\pdfoutput=1
\documentclass[sigconf,authorversion=true,screen]{acmart}
\usepackage[english]{babel} 
\usepackage[T1]{fontenc} 
\usepackage[utf8]{inputenc} 

\usepackage{amsmath,amsfonts,amsthm,amssymb,bm}
\usepackage[ruled,vlined,linesnumbered]{algorithm2e}
\usepackage{multirow}
\usepackage{stmaryrd}
\usepackage[caption=false,font=footnotesize,hangindent=10pt]{subfig}
\usepackage{enumitem}
\usepackage[capitalise]{cleveref}

\usepackage{graphicx}
\graphicspath{{figs/}}

\AtBeginDocument{%
  \providecommand\BibTeX{{%
    \normalfont B\kern-0.5em{\scshape i\kern-0.25em b}\kern-0.8em\TeX}}}
    
\setcopyright{acmcopyright}

\def\BibTeX{{\rm B\kern-.05em{\sc i\kern-.025em b}\kern-.08emT\kern-.1667em\lower.7ex\hbox{E}\kern-.125emX}}

\newtheorem{theorem}{\textbf{Theorem}}

\SetVlineSkip{2pt}

\SetCommentSty{myAlgComment}
\SetKwProg{Fn}{Function}{:}{}

\SetAlFnt{\small}
\SetAlCapFnt{\small}
\SetAlCapNameFnt{\small}

\newcommand{\E}{\mathbb{E}}
\newcommand{\Z}{\mathbb{Z}}
\newcommand{\F}{\mathbb{F}}
\newcommand{\var}{\operatorname{Var}}
\newcommand{\indr}{\mathbf{1}}
\newcommand{\Rand}{\texttt{Rand}}
\newcommand{\RandExp}{\texttt{RandExp}}
\newcommand{\ZeroTest}{\texttt{ZeroTest}}

\newcommand{\header}[1]{\noindent\textbf{#1}}
\newcommand{\bullethdr}[1]{\smallskip\noindent\textbullet\,\textbf{#1}}

\newcommand{\share}[1]{\llbracket{#1}\rrbracket}

\begin{document}

\title{An Effective and Differentially Private Protocol for Secure Distributed Cardinality Estimation}


\author{Pinghui Wang}
\authornote{ Pinghui Wang is the corresponding author.}
\affiliation{%
  \institution{Xi’an Jiaotong University}
  \city{Xi’an}
  \country{China}
}
\email{wangpinghui369@gmail.com}

\author{Chengjin Yang}
\affiliation{%
  \institution{Xi’an Jiaotong University}
  \city{Xi’an}
  \country{China}
}
\email{cjyang2643@stu.xjtu.edu.cn}

\author{Dongdong Xie}
\affiliation{%
  \institution{Xi’an Jiaotong University}
  \city{Xi’an}
  \country{China}
}
\email{xdddd1@stu.xjtu.edu.cn}

\author{Junzhou Zhao}
\affiliation{%
  \institution{Xi’an Jiaotong University}
  \city{Xi’an}
  \country{China}
}
\email{junzhou.zhao@xjtu.edu.cn}

\author{Hui Li}
\affiliation{%
  \institution{Xidian University}
  \city{Xi’an}
  \country{China}
}
\email{hli@xidian.edu.cn}

\author{Jing Tao}
\affiliation{%
  \institution{Xi’an Jiaotong University}
  \city{Xi’an}
  \country{China}
}
\email{jtao@mail.xjtu.edu.cn}

\author{Xiaohong Guan}
\affiliation{%
  \institution{Xi’an Jiaotong University}
  \city{Xi’an}
  \country{China}
}
\email{xhguan@mail.xjtu.edu.cn}

%
\begin{abstract}
  Counting the number of distinct elements distributed over multiple data holders is a fundamental problem with many real-world applications ranging from crowd counting to network monitoring.
Although a number of space and computationally efficient sketch methods (e.g., the Flajolet-Martin sketch and the HyperLogLog sketch) for cardinality estimation have been proposed to solve the above problem, 
these sketch methods are insecure when considering privacy concerns related to the use of each data holder's personal dataset.
Despite a recently proposed protocol that successfully implements the well-known Flajolet-Martin (FM) sketch on a secret-sharing
based multiparty computation (MPC) framework for solving the problem of private distributed cardinality estimation (PDCE),
we observe that this MPC-FM protocol is not differentially private.
In addition, the MPC-FM protocol is computationally expensive, which limits its applications to data holders with limited computation resources.
To address the above issues, in this paper we propose a novel protocol \emph{DP-DICE}, which is computationally efficient and differentially private for solving the problem of PDCE.
Experimental results show that our DP-DICE achieves orders of magnitude speedup and reduces the estimation error by several times in comparison with state-of-the-arts under the same security requirements.

\end{abstract}
%
%
%
\maketitle

\pdfoutput=1
\section{Introduction} \label{sec:introduction}

Distributed cardinality computing (DCC), the task of counting the number of
distinct elements in the union of multiple datasets is fundamental for many
real-world applications.
Consider the following motivating examples.

\header{Logistics Monitoring.}
RFID systems have been increasingly deployed in logistics and transportation~\cite{zhengxia2010modern}.
An RFID system typically places RFID readers in distributed locations, where an
RFID reader collects information about its nearby goods through RFID tags.
During a time period, RFID readers located at different places may observe goods
in common when goods (e.g., vehicles) move over time.
Computing logistic statistics such as the number of distinct vehicles observed by
RFID readers in an area during a period, which reflects the severity of congestion
in the area.

\header{Disease Epidemiology.}
Researchers from medical institutions, government agencies, and insurance
companies may want to know the number of patients in a country having a specific
disease such as diabetes~\cite{DBLP:journals/corr/abs-2003-14412}.
Medical records are distributed over hospitals and patients often get medical care
from multiple hospitals.
Therefore, the above problem 
can be formulated as a DCC task.

\header{Audience Reach Reporting.}
On the Internet, an advertiser often conducts a campaign across several
publishers, which show advertisements on behalf of the advertiser.
The campaign's reach is defined as the number of distinct users exposed to the
campaign by at least one publisher, which is a critical metric for evaluating the
campaign's efficacy~\cite{ulbrich2021tracking}~\cite{GhaziKKMPSWW22}.
Different publishers may reach overlapping sets of individuals.
Thus, the reach computation can be formulated as a DCC task.

\header{Collection of Internet Traffic Statistics.}
Counting the number of distinct online clients is a fundamental task for network
measurement and monitoring.
Sometimes, the traffic of interest may be distributed over multiple devices
belonging to different holders (e.g., routers held by different Internet service
providers and anonymity network egress nodes in anonymous communication networks~\cite{DBLP:conf/uss/DingledineMS04})
and a client may occur over more than one of these devices.

The cardinality calculation task is simple to complete when all data are collected
and stored centrally.
However, sometimes data may be distributed over multiple data holders (DHs), and
it is prohibitive to collect the entire data due to the sheer size of data and the
limited resources of DHs.
To solve the problem of distributed cardinality estimation, a straightforward
solution is to locally compute the cardinality of each DH's dataset and then sum
all DHs' cardinalities together.
However, using this aggregation to approximate the union dataset's cardinality may
exhibit a large estimation error because different DHs' datasets may have many
elements in common.
To address this challenge, a number of sketch methods~\cite{harmouch2017cardinality} have been proposed such as
the Flajolet-Martin (in short FM) sketch~\cite{flajolet1985probabilistic} and the
HyperLogLog (in short HLL) sketch~\cite{HeuleNH13}.
The key to these methods is the construction of a compact and mergeable sketch
(i.e., a data summary) on each DH's dataset.
Given the mergeable sketches on multiple DHs' datasets, one can easily merge these
sketches into a sketch on the union of all DHs' datasets.
Then, the union's cardinality can be estimated based on the merged sketch.
These properties significantly reduce the memory, computation, and network costs
for the task of distributed cardinality estimation.

Besides efficiency and accuracy, data privacy is also critical for many
applications.
In the above examples, it is well known that goods tags collected by RFID systems
as well as personal records collected by hospitals and Internet publishers may
cause severe privacy risks, as these might reveal confidential information (e.g.
traces) of individuals.
Therefore, it is desirable to have a solution that can effectively solve the problem of distributed cardinality estimation without revealing the private
information of individuals collected by each DH.

When privacy issues are taken into account, the above sketch methods all fail because their
generated sketches contain confidential information linked to personal privacy.
Although~\cite{smith2020flajolet,DickensArxiv2203} found that many popular sketch methods, including FM and HLL preserve differential privacy as long as $n$ (the cardinality of the set of interest) is larger than a constant, they assume that the functions employed in the procedure should also be kept private,
which is non-trivial to be satisfied in the setting of private distributed cardinality estimation (PDCE).

To solve the above issue,
recently, Hu et al.~\cite{Hu0LGWGLD21} designed a protocol that successfully
implements the well-known FM sketch on a secret-sharing based multiparty
computation (MPC) framework, which significantly improves the performance of
existing methods.
Although this MPC-FM protocol uses secure computation to protect against attacks when collecting and merging FM sketches collected from DHs, 
honest but curious DHs can infer confidential information linked to personal privacy from the protocol's output (refer to~\cref{subsec:challenges} for details).
In addition, the MPC-FM protocol exerts a heavy computational burden over the DHs, which limits its usage on edge devices (e.g., devices for the
Internet of things) with limited computational resources.
To address the above challenges, in this paper we propose a novel protocol \emph{DP-DICE} for solving the problem of PDCE. We demonstrate our DP-DICE protocol is secure and differentially private.
Moreover, it is more accurate and computationally efficient than the
state-of-the-art protocol i.e., MPC-FM~\cite{Hu0LGWGLD21}.
Our main contributions can be summarized as follows.
\begin{itemize}[leftmargin=3ex]
\item We reveal that the state-of-the-art protocol MPC-FM~\cite{Hu0LGWGLD21} is
  not differentially private, which is inconsistent with the claim in the original
  paper~\cite{Hu0LGWGLD21}.
  To guarantee differential privacy, one can inject a certain amount of noise
  during the procedure of computing the cardinality estimation.
  Unfortunately, it results in large estimation errors.

\item We propose a novel protocol \emph{DP-DICE} for solving the problem of PDCE.
  Our DP-DICE is designed based on a fast and accurate sketch method, \emph{FMS},
  which is proposed in this paper.
  Compared with sketch methods FM and HLL, \emph{FMS} exhibits
  significantly smaller estimation errors when combing with additive noise
  mechanisms for differential privacy.
  To implement FMS for solving the PDCE problem, we design a new distributed
  additive noise mechanism to guarantee that the DP-DICE protocol is
  differentially private.

\item We conduct extensive experiments on a variety of datasets.
  Compared with the state-of-the-art methods, our DP-DICE accelerates computation
  speed on DHs by orders of magnitudes and decreases estimation errors by several
  times to achieve the same security requirements.
\end{itemize}


\pdfoutput=1
\section{Problem Formulation}
\label{sec:problem}

\subsection{PDCE Problem}
Given $d$ \emph{data holders} (DHs), where each DH $j\in \{1,\ldots, d\}$ holds a
set $S_j$ consisting of elements from a universal set $U$.
Let $S$ be the union of all sets $S_j$, i.e., $S=\bigcup_{j=1}^dS_j$, and the
cardinality of set $S$ is denoted by $n$, i.e., $n=|S|$.
For example, a DH may represent a hospital, and $S_j$ is a set of patients in
hospital $j$, then $n$ is the number of distinct patients appearing in all these
hospitals.

Our goal is to estimate $n$ without revealing that an element $e$ is a
member of set $S_j$.
In many real-world applications, such membership information is
confidential and typically related to personal privacy.
For example, leaking the information that a person is a patient of some hospital
could harm the person's healthy privacy.

We refer to this task as the {\em Private Distributed Cardinality Estimation}
(PDCE) problem.
This problem is challenging because different sets among sets $S_1, \ldots,
S_d$ may have elements in common and so we cannot obtain $n$ by simply summing up
the cardinality of each set $S_j$, $j=1, \ldots, d$.



\subsection{Threat Model and Privacy Goals}

In real-world applications, there may exist a large number of DHs with limited
computational and network resources.
Therefore, it is prohibitive to directly perform secure multi-party computation
(MPC) over these DHs.
Following the framework of~\cite{Hu0LGWGLD21}, we require the following
assumptions.

\begin{itemize}[leftmargin=3ex]
\item \textbf{Assumption 1.}
  There exist $c$ {\em computation parties} (CPs), where at least one CP is
  trusted.
  The untrusted CPs are modeled as corrupted and controlled by a single adversary.
  The adversary can behave arbitrarily on these untrusted CPs (e,g., tamper with
  data on any untrusted CP).

\item \textbf{Assumption 2.}
  Each DH $j$ is responsible for collecting set $S_j$ and securely sharing $S_j$
  (or a summary of $S_j$) among the CPs.
  On the contrary, all the CPs are responsible for collaboratively and securely approximating $n$ based on their collected data. 

\item \textbf{Assumption 3.}
  The estimation result of cardinality $n$ is publicly available to all the DHs,
  the CPs, and the adversary.
  For many real-world applications, this assumption is practical because it may
  reveal the cardinality estimation to the public for research and business
  purpose.

\item \textbf{Assumption 4.}
  The DHs will faithfully execute the protocol but may try to learn all possible
  information legally.
  Notably, \cite{Hu0LGWGLD21} has assumed each DH is honest and
    \textit{incurious}, which is impractical for some real-world applications.
  For example, a DH (e.g., a hospital) may be curious about whether
  an element $e\in U$ (e.g., a person with citizen ID $e$) is held by other DHs.
\end{itemize}

Based on the above assumptions, we aim to design a secure protocol to estimate the
cardinality $n$ of the union set $S$ that achieves:

\begin{itemize}[leftmargin=3ex]
\item \textbf{Target 1.}
  The adversary learns nothing from executing the protocol except the protocol's
  output and is unable to change it.

\item \textbf{Target 2.}
  The adversary fails to destroy the correctness of the computation without being
  detected.

\item \textbf{Target 3.}
  The protocol complies with differential privacy.
  Specifically, given the protocol's output, the adversary is unable to infer
  whether any element $e\in U$ is in set $S$ or not.
  In addition, a DH also fails to infer whether any element $e\in U$ is held by
  any other DHs;

\item \textbf{Target 4.}
  The computation on all CPs and DHs as well as the communication between these
  parties are as small as possible.

\end{itemize}

\pdfoutput=1
\section{Preliminaries}
\label{sec:preliminary}

In this section, we first introduce classical cardinality estimation methods:
the Flajolet-Martin (FM) sketch~\cite{flajolet1985probabilistic} and the
HyperLogLog (HLL) sketch~\cite{flajolet2007hyperloglog}\cite{HeuleNH13}.
Then, we describe the definition of differential privacy. 
At last, we briefly review the state-of-the-art protocol in~\cite{Hu0LGWGLD21} for
solving the PDCE problem and discuss its weaknesses,
which will be addressed in this paper.

\pdfoutput=1
\subsection{FM Sketch}
\label{sec:FM}

\begin{figure*}[htbp]
  \centering
  \includegraphics[width=\textwidth]{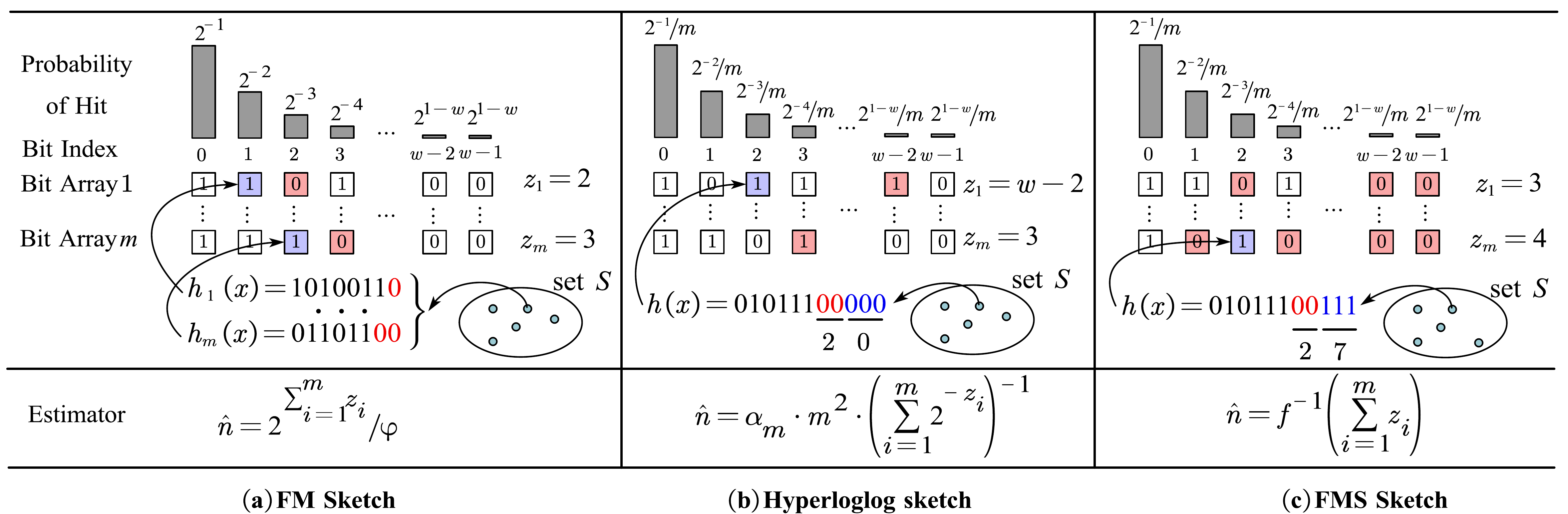}
  \caption{Illustration of our FMS sketch in comparison with the FM
    sketch~\cite{flajolet1985probabilistic} and the HLL sketch~\cite{HeuleNH13}.
    (a) The FM sketch.
    The blue boxes are the bits updated to 1; the red box in each array is the
    zero with the least bit index and is denoted as $z_i$, $i=1,\ldots, m$.
    The red numbers represent trailing zeros.
    (b) The HLL sketch.
    The blue boxes are the bits updated to 1; the red box in each array is the one
    with the greatest bit index and is denoted as $z_i$, $i=1,\ldots, m$.
    The red numbers represent trailing zeros and the blue numbers represent bit array indices in binary form.
    (c) Our FMS sketch.
    The blue boxes are the bits updated to 1; the red boxes are all the zero bits;
    the red numbers represent trailing zeros and the blue numbers represent the
    selected bit array's index in binary form.
    The number of zero bits in array $i$ is denoted as $z_i$, $i=1,\ldots,m$.}
  \label{fig:sketches}
\end{figure*}

The Flajolet-Martin (FM) sketch is a classic data structure to efficiently
estimate the cardinality (i.e., the number of distinct elements) of a large set
$S$, which may given as a sequence including duplicated elements.
As shown in the first column of \cref{fig:sketches}, an FM sketch $\{B_i,
h_i\}_{i=1, \ldots, m}$ consists of $m$ bit arrays $B_1, \ldots, B_m$ and $m$
independent hash functions $h_1, \ldots, h_m$.
Each bit array $B_i$ has $w$ bits, which are all initialized to zero.
Each hash function $h_i$ used for updating array $B_i$ maps an element $e\in U$ to
a $(w-1)$-bit string $h_i(e)$ uniformly selected from set $\{0,1\}^{w-1}$ at
random.
The FM sketch of set $S$ is computed as: enumerate each element $e\in S$ and set
each bit $B_i[\rho(h_i(e))]=1$, $i=1, \ldots, m$, where function $\rho(h_i(e))$
returns the number of trailing zeros of $h_i(e)$.\footnote{For example,
  $\rho(1011)=0$, $\rho(1010)=1$, $\rho(1100)=2$, $\rho(1000)=3$, $\rho(0000)=4$.}
The cardinality $n$ of set $S$ can be estimated as
$\hat{n}\approx 2^\frac{Z^*}{m}\varphi^{-1}$,
where $\varphi$ is about $0.77351$ which is a correction factor and variable $Z^*$
is defined as:
\begin{equation}\label{eq:fmz}
  Z^*\triangleq\sum_{i=1}^m z_i.
\end{equation}



The FM sketch can be applied to solve the problem of distributed cardinality
estimation, i.e., approximate the cardinality of the union of sets $S_1, \ldots,
S_d$.
Given the FM sketch $\{(B_i^{(j)}, h_i)\}_{i=1, \ldots, m}$ of each set $S_j$,
$j=1, \ldots, d$.
Note that all these sets use the same set of hash functions $h_1, \ldots, h_m$.
Then, the FM sketch $\{(B_i, h_i)\}_{i=1, \ldots, m}$ of the union set $S=S_1\cup
\cdots\cup S_d$ is computed as
$B_i = B_i^{(1)} \lor \cdots \lor B_i^{(d)}$, $i=1, \ldots, m$,
where $\lor$ is the bitwise OR operation.

\pdfoutput=1
\subsection{HLL Sketch}
\label{sec:HLL}
set cardinality.
Compared with the earlier FM sketch, the HLL sketch requires less memory
space and computation operations.
The updating procedure is displayed in the second column in \cref{fig:sketches}.
It finds that the largest index $z_i$ with $B_i[z_i]=1$ can also be
used to predict the cardinality.
Without directly storing array $B_i$ in memory, the HLL sketch uses a register to
keep tracking of $z_i$ for each array $B_i$, which significantly decreases the
memory usage from $w\times m$ bits to $\lceil \log_2 w \rceil \times m $ bits.
In addition, the HLL sketch applies a stochastic averaging technique to reduce the
complexity of processing each element from $O(m)$ to $O(1)$.
Formally, the HLL sketch of a set $S$ consists of $m$ registers $R[1], \ldots,
R[m]$ initialized to 0.
For simplicity, $m$ is set to $2^{r}$, where $r$ is a positive integer.
In addition, a hash function $h$ is used to map an element $e\in S$ to a $(r +
w-1)$-bit string $h(e)$ uniformly selected from set $\{0,1\}^{r+w-1}$ at random.
For each element $e\in S$, the HLL sketch splits its hash string $h(e)$ into two
parts $h^{(\text{L})}(e)$ and $h^{(\text{H})}(e)$, which consist of the lowest $r$
bits and the highest $w-1$ bits of string $h(e)$, respectively.
Then, the HLL sketch updates one specific register $R\left[h^{(\text{L})}(e) + 1\right]$ as:
\[
  R\left[h^{(\text{L})}(e) + 1\right] \gets \max\left(R\left[h^{(\text{L})}(e) +
      1\right], \rho(h^{(\text{H})}(e))\right),
\]
where $\rho(h^{(\text{H})}(e))$ returns the number of trailing zeros of $h^{(\text{H})}(e)$.

Given the HLL sketch of set $S$, the cardinality of $S$ is estimated as
$\hat{n}=\alpha _m m^2 \left( \sum_{i=1}^{m}{2^{-R[i]}} \right) ^{-1}$,
where $\alpha_m=\left(m\int_{0}^{\infty} \left(\log \frac{2+u}{1+u}\right)^m du
  \right)^{-1}$.
When $\alpha_m m^2 \left(\sum_{i=0}^{m-1} 2^{-R_{v, a}[i]}\right)^{-1} <
\frac{5}{2}m$, the above estimation $\hat n$ exhibits a large estimation
error.
For this case, the cardinality $n$ is more accurately estimated as
$\hat{n} = -m \ln \frac{E}{m}$,
where $E$ is defined as the number of registers among $R[1], \ldots, R[m]$ that
equal $0$.

Similar to the FM sketch, the HLL sketch is also mergeable.
Specifically, given the HLL sketch $R^{(j)}$ of each set $S_j$, $j=1, \ldots, d$.
Again, all merged sketches share the same hash function $h$.
Then, the HLL sketch $R$ of the union set $S=S_1\cup \cdots\cup S_d$ is computed
as
$R[i] = \max_{j=1,\ldots,d} R^{(j)}[i]$, $i=1, \ldots, m$.

\pdfoutput=1
\subsection{Differential Privacy}\label{subsec:DP}
Differential privacy is an elegant mathematical definition of privacy, which takes the following form:

\textbf{Definition 1. Differential Privacy (DP)~\cite{DworkKMMN06}.} A randomized algorithm $M:
\mathcal{D}\rightarrow \mathcal{Y}$ satisfies $(\varepsilon, \delta)$-differential
privacy if and only if for all pairs of $x, x'\in \mathcal{D}$ differ in at most
one element, for all events $E\subseteq \mathcal{Y}$, we have
\[
  P(M(x)\in E)\le \exp(\varepsilon) P(M(x')\in E) +\delta.
\]

\textbf{Definition 2. Concentrated Differential Privacy (CDP)~\cite{dwork2016concentrated,bun2016concentrated}.}
A randomized algorithm $M: \mathcal{D}\rightarrow \mathcal{Y}$ satisfies
$\frac{1}{2}\varepsilon^2$-concentrated differential privacy if and only if for all
pairs of $x, x'\in \mathcal{D}$ differ in at most one element, we have
\[
  \sup_{\alpha\in (1, \infty)} \frac{1}{\alpha} D_{\alpha} (M(x)||M(x')) \le
  \frac{1}{2}\varepsilon^2,
\]
where $D_{\alpha} (M(x)||M(x'))$ is the $\alpha$-R\'enyi divergence of $M(x)$'s
distribution with respect to $M(x')$'s distribution.
Formally, we define $D_{\alpha} (M(x)||M(x'))$ as:
\[
  D_{\alpha} (M(x)||M(x')) = \frac{1}{\alpha-1} \ln \textup{E}_{Y\sim Q}
  \left(\left(\frac{Q(Y)}{Q'(Y)}\right)^{\alpha-1}\right),
\]
where functions $Q(\cdot)$ and $Q'(\cdot)$ refer to the distributions of $M(x)$ and $M(x')$, respectively.

\textbf{Relation Between Differential Privacy and Concentrated Differential
  Privacy.}
Kairouz et al.~\cite{KairouzL021} find that a randomized algorithm $M$ satisfies
$(\varepsilon, 0)$-differential privacy, then it satisfies
$\frac{1}{2}\varepsilon^2$-concentrated differential privacy.
If algorithm $M$ satisfies $\frac{1}{2}\varepsilon^2$-concentrated differential
privacy, then for any $\delta>0$, $M$ satisfies $(\varepsilon_\delta^*,
\delta)$-differential privacy where $\varepsilon_\delta^*$ is defined as:
\[
  \varepsilon_\delta^*
  =\inf_{\alpha>1} 0.5\varepsilon^2\alpha + \frac{\ln(1/\alpha\delta)}{\alpha-1} +
  \ln(1-1/\alpha)
  \le 0.5\varepsilon(\varepsilon + 2\sqrt{-2\ln\delta}).
\]

\textbf{Definition 3. Discrete Gaussian Distribution~\cite{canonne2020discrete}.} 
Adding noise drawn from Gaussian distribution ~\cite{dwork2014algorithmic} is widely used to preserve differential privacy. However, finite computers cannot represent continuous noise precisely, and using finite precision approximations may incur privacy destruction ~\cite{mironov2012significance}. Therefore, Canonne et al.~\cite{canonne2020discrete} proposed the discrete Gaussian distribution $\mathcal{N}_{\mathbb{Z}}(\mu, \sigma^2)$ defined as:
\[
 P(X=x) =
 \frac{\exp\left(-\frac{(x-\mu)^2}{2\sigma^2}\right)}
 {\sum_{y\in\Z}\exp\left(-\frac{(y-\mu)^2}{2\sigma^2}\right)},
 \quad x\in\Z.
\]
and find the following theorem.
\begin{theorem}~\cite{canonne2020discrete} \label{theorem:discrete_gaussian_CDP}
Let $f: \mathcal{D}\rightarrow\mathcal{\mathbb{Z}}$ be a function with sensitivity $\Delta$.
The sensitivity $\Delta$ is defined as:  $\max|f(x)-f(y)|$, where the maximum is over all pairs of datasets $x$ and $y$ in $\mathcal{D}$ differing in at most one element.
Define a randomized algorithm $M: \mathcal{D}\rightarrow \mathbb{Z}$ as $M(x)=f(x) + X$, where $X\leftarrow \mathcal{N}_{\mathbb{Z}}(0, \varDelta^2/\varepsilon^2).$ Then, $M$ satisfies $\frac{1}{2}\varepsilon^2$-concentrated differential privacy.
\end{theorem}
Furthermore, the following theorem given in~\cite{KairouzL021} reveals that the sum of  independent discrete Gaussian variables can also be used as additive noise for achieving differential privacy.
\begin{theorem} \label{theorem:distributed_discrete_gaussian_CDP}
Define a randomized algorithm $M: \mathcal{D}\rightarrow \mathbb{Z}$ as $M(x)=f(x) + \sum^{d}_{i=1}X_i$, where $X_i\leftarrow \mathcal{N}_{\mathbb{Z}}(0, \sigma^2), i=1,\ldots, d$ and $\sigma>\frac{1}{2}$. Then, $M$ satisfies $\frac{1}{2}\varepsilon^2$-differential privacy, where $\varepsilon$ equals
$$\min \left\{ \sqrt{\frac{\Delta ^2}{d\sigma ^2}+5\sum\nolimits_{k=1}^{d-1}{e^{-2\pi ^2\sigma ^2\frac{k}{k+1}}}}, \frac{|\Delta |}{\sqrt{d}\sigma}+10\sum\nolimits_{k=1}^{d-1}{e^{-2\pi ^2\sigma ^2\frac{k}{k+1}}} \right\},
$$
and $\Delta$ is the sensitivity of function $f$.
\end{theorem}

\pdfoutput=1
\subsection{MPC-FM Protocol}

Hu et al.~\cite{Hu0LGWGLD21} propose the state-of-the-art protocol for solving the
problem of Private Distributed Cardinality Estimation (PDCE).
We name their protocol MPC-FM.
MPC-FM works as follows: Each DH $i=1, \ldots, d$ builds the FM sketch $F_i$ of
its private set $S_i$.
Then, each DH $i$ splits its sketch $F_i$ (a secret) into $c$ pieces and each CP
holds one piece.
Each piece here is called a share.
Individual shares are of no use on their own and the secret sketch $F_i$ can be
reconstructed only when all of the $c$ CPs collude.
All the DHs use the same set of hash functions to generate FM sketches, therefore
all $F_1, \ldots, F_d$ are mergeable.
In other words, one can compute the FM sketch $F$ of the union set $S$ according
to the values of sketches $F_1, \ldots, F_d$.
Based on this property, all the CPs compute the shares of sketch $F$ by
collaboratively and securely merging the shares of all sketches $F_1, \ldots, F_d$
sent from the DHs.
Note that each CP holds only a share of sketch $F$ and so an adversary cannot
reconstructs the secret $F$ when at least one CP is not corrupted by the
adversary.
At last, based on the shares of sketch $F$, all the CPs securely compute the
variable $Z^*$ (defined in \cref{eq:fmz}), which is the result for cardinality estimation under MPC-FM.
To implement the above procedure, MPC-FM adopts the framework of
SPDZ~\cite{DamgardPSZ12}.
SPDZ is a secret-sharing-based multiparty computation (MPC) scheme that supports
secure computation over a finite field and is secure in the presence of an
adversary statically making arbitrarily much corruption.

\pdfoutput=1
\subsection{Challenges}
\label{subsec:challenges}

In this paper, we aim to address the following unresolved challenges.

\noindent\textbullet\,\textbf{Challenge 1.~MPC-FM is expensive for DHs.}
For the MPC-FM protocol, we see that each DH $j$ needs to compute $m\times |S_j|$
hash operations for generating the FM sketch of set $S_j$.
Its cost is expensive for a large set $S_j$ because $m$ is typically set to
thousands.

\noindent\textbullet\,\textbf{Challenge 2.~MPC-FM leaks personal privacy when DHs are honest but curious.}
We first give the following example to demonstrate that FM leaks privacy when its hash functions and the entire final value of the sketch are available to the public. Consider an FM sketch SK1 of a set $S$ after protocol aggregation. An attacker can find an element $e$ after an attempt to make $S \cup \{e\}$ generate a different FM sketch SK2, i.e., $e$ fills at least one zero bit of SK1. Then, the adversary can easily deduce that $e$ is definitely not in $S$. Of course, if the adversary does not know the hash function, it cannot infer the membership of $e$. 
In our problem setting, all DHs know the same hash function, and a DH may be curious about privacy information linked to other DHs. If a protocol like MPC-FM simply implements the FM sketch on the secure computation platform SPDZ to solve our problem, privacy leakage appears. For example, denote $Z$ as the estimator (i.e., the sum of indexes of first zero bits in FM sketches) of the union set $S$. When a DH $i$ (holding set $S_i$) finds that the estimator $Z_e$ of a set $S_i \cup \{e\}$ is greater than $Z$, then DH $i$ can deduce that $e$ is not in any other data holders’ sets. More generally, when a DH $i$ finds that the estimator $Z_e$ of set $S_i \cup E$  equals $Z + t$, where $t$ is a positive integer, then DH $i$ can deduce that at least $t$ elements in set $E$ are not in the sets holding by any other DHs.

Take the private collection of Internet traffic statistics discussed in Section~\ref{sec:introduction} as an instance.
Internet network devices such as routers have limited computation resources and network traffic collected on these devices is private.
One may attempt to give a modification MPC-HLL to solve Challenge 1, which
replaces the FM sketch with the HLL sketch.
Clearly, the HLL sketch will significantly reduce the complexity of computing the
FM sketch $F_i$ from $O(m|S_i|)$ to $O(|S_i|)$.
Unfortunately, merging HLL sketches require MAX operations, which are complex and
expensive to be implemented on CPs at the top of MPC frameworks such as SPDZ.
In addition, similar to MPC-FM, we easily find that MPC-HLL is not differentially
private either.
To address Challenge 2, one may enhance MPC-FM with additive noise
mechanisms.
The way of merging local data using MPC and adding an additive noise to the aggregation output is standard for satisfying differential privacy and has also been applied in other areas such as federated learning~\cite{JayaramanW0G18,KairouzL021}.
Unfortunately, in our later experimental results (shown in Fig.~\ref{fig:central} and Fig.~\ref{fig:local}), we observe that MPC-FM (and
also MPC-HLL) with additive noise mechanisms exhibits significant estimation
errors.

\pdfoutput=1

\section{Our Method}
\label{sec:method}

In this section, we first introduce a new method, \emph{FMS sketch}, which is used
as a building block of our protocol \emph{DP-DICE}.
Then, we elaborate on our protocol DP-DICE and discuss its performance.

\pdfoutput=1
\subsection{Our FMS Sketch}\label{sec:FMS}
\header{Data Structure}.
As shown in ~\cref{fig:sketches}, our FMS sketch $\{B_i\}_{i=1, \ldots, m}$
consists of $m$ bit arrays $B_1, \ldots, B_m$ with $m = 2^{r}$, where $r$ is a
positive integer.
Each $B_i$, $i=1, \ldots, m$ is a one dimensional array of $w$ bits, where all
bits $B_i[0], \ldots, B_i[w-1]$ are initialized to 0.
In addition, the FMS sketch uses a single hash function $h(\cdot)$ that maps an
element $e\in U$ into a random $(r + w - 1)$-bit string $h(e)\in \{0,1\}^{r+w-1}$.

\header{Sketch Generation}.
For each element $e$ in set $S$, similar to the HLL sketch~\cite{HeuleNH13},
according to the hash string $h(e)$, we first compute
\[
  (i, j)\gets \left(h^{(\text{L})}(e) + 1,
    \rho\left(h^{(\text{H})}(e)\right)\right),
\]
where $h^{(\text{L})}(e)$ and $h^{(\text{H})}(e)$ consist of the lowest $r$ bits
and the highest $w-1$ bits of string $h(e)$ respectively, and
$\rho(h^{(\text{H})}(e))$ returns the number of trailing zeros of $h^{(\text{H})}(e)$.
Then, we update the bit array $B_{i}$ as: $ B_{i}[j] \gets 1$.

\header{Cardinally Estimation}.
Given FMS sketches $B_1, \ldots, B_m$, next we describe our method to estimate $n$
(i.e., the cardinality of set $S$).
Define $Z$ as the number of zero bits in all the bit arrays $B_1,\ldots,B_m$,
that is,
\begin{equation}\label{eq:Z}
  Z
  = mw - \sum_{i=1}^{m} \sum_{x=0}^{w-1} B_i[x]
  = \sum_{i=1}^{m}\sum_{x=0}^{w-1} (1-B_i[x]).
\end{equation}
\begin{theorem} \label{theorem:expandvariance}
Define $
  p_x\triangleq\begin{cases}
                 2^{-x-1}/m,  & 0\le x \le w-2,\\
                 2^{-w+1}/m, &x=w-1
               \end{cases}
$ and $V\triangleq\frac{Z}{mw}$.
Then, the expectation is $\E(V) = \frac{1}{w}\sum_{x=0}^{w-1} (1-p_{x}^{})^n$ and variance of $V$ is
\begin{align*}
  &\var(V) =\frac{1}{m w^2} \sum_{x=0}^{w-1}{\left( 1-p_x \right) ^n} \left( 1-m\sum_{x=0}^{w-1}{\left( 1-p_x \right) ^n} \right) \\
        &+\frac{2}{mw^2}\sum_{x=0}^{w-1}{\sum_{y=x+1}^{w-1}{\left(1-p_x-p_y\right)^n}} +\frac{m-1}{mw^2}\sum_{x=0}^{w-1}{\sum_{y=0}^{w-1}{\left(1-p_x-p_y\right)^n}}.
\end{align*}
\end{theorem}
\pdfoutput=1
\proof We easily find that
\begin{equation} \label{eq:phe}
  P\left(\rho(h^{(\text{H})}(e))= x\right) = p_x.
\end{equation}
Define variables
$
  \indr_{j, x} \triangleq 1-B_j[x].
$
After updating all elements in set $S$, from \cref{eq:phe}, we have
\[
  P(\indr_{j, x}=1) = P(B_j[x]=0) = (1-p_x)^n,
\]
Therefore, the expectation of $Z$ is computed as:
\begin{equation*}
  \E(Z)
  = \sum_{j=0}^{m-1}\sum_{x=0}^{w-1} \E(\indr_{j, x})
  = m\sum_{x=0}^{w-1}(1-p_x)^n.
\end{equation*}
Similarly, we define
$
  Z_j \triangleq \sum_{x=0}^{w-1}{\indr_{j,x}}.
$
Then we have
\begin{align*}
  \E(Z^2) &= \sum_{i=0}^{m-1}\sum_{j=0}^{m-1}\E(Z_iZ_j)\\ 
        &= \sum_{i=0}^{m-1}\E(Z_iZ_i) + 2\sum_{i=0}^{m-1}\sum_{j=i+1}^{m-1}\E(Z_iZ_j) \\
          &= m\sum_{x=0}^{w-1}\E(\indr_{j,x}) +2m\sum_{x=0}^{w-1}\sum_{y=x+1}^{w-1}\E(\indr_{j,x}\indr_{j,y}) \\
          &\quad + 2\sum_{i=0}^{m-1}\sum_{j=i+1}^{m-1}\sum_{x=0}^{w-1}\sum_{y=0}^{w-1}\E(\indr_{i,x}\indr_{j,y}) \\
          &= m\sum_{x=0}^{w-1}(1-p_x)^n
            + 2m\sum_{x=0}^{w-1}\sum_{y=x+1}^{w-1}(1-p_x-p_y)^n \\
          &\quad + m(m-1)\sum_{x=0}^{w-1}\sum_{y=0}^{w-1}(1-p_x-p_y)^n,
\end{align*}
where the last equation holds because for any two different tuples $(i,x)$ and
$(j,y)$ we have
\[
  P(\indr_{i, x} \indr_{j, y} =1) = P(B_i[x]=B_j[y]=0) = (1-p_x-p_y)^n.
\]
Then, we have
\begin{equation*}
\begin{split}
  &\var(Z) = \E(Z^2)  - \E(Z)^2 \\
  &= 2m\sum_{x=0}^{w-1}\sum_{y=x+1}^{w-1}(1-p_x-p_y)^n
    + m(m-1)\sum_{x=0}^{w-1}\sum_{y=0}^{w-1}(1-p_x-p_y)^n \\
  &\quad + m\sum_{x=0}^{w-1}(1-p_x)^n\left(1 - m\sum_{x=0}^{w-1}(1-p_x)^n\right).
\end{split}
\end{equation*}
Then, we easily derive the formulas of $\E(V)$ and $\var(V)$ as $\E(V) =
\frac{\E(Z)}{m w}$ and $\var(V) = \frac{\var(Z)}{ m^2 w^2}$.
\endproof

Based on the above theorem, we define a function $f(n)$ as:
\begin{equation} \label{eq:fn}
  f(n) \triangleq \frac{1}{w}\sum_{x=0}^{w-1}(1-p_x)^n \approx \frac{1}{w}\sum_{x=0}^{w-1} \text{EXP}(-n p_x),
\end{equation}
where the last approximation equation holds because $p_x$ is smaller than $\frac{1}{2m}$,
which is typically a small number.

We then estimate $n$ by $\hat n$ such that $f(\hat n)=V$.
It is not difficult to find that $f(n)$ decreases as $n$ increases.
We can use a binary search method to compute $\hat n  = f^{-1}(V)$, where $f^{-1}(V)$ is the inverse function of function $f(n)$ defined in
\cref{eq:fn}.

\header{Error Analysis.}
From Theorem~\ref{theorem:expandvariance} and Eq.~(\ref{eq:fn}), we have 
\[
f(n)=\E(V)=\frac{1}{w} \sum_{x=0}^{w-1} (1-p_x)^n.
\]
Since $f(n)$ is a monotonic function, we undo the operation of function $f(n)=\E(V)$. After that, we have 
\[ 
n = f^{-1}(\E(V)).
\]
Given the complex forms of $\E(V)$, $\text{Var}(V)$, and $f(\cdot)$, it is not easy to derive $\text{Var}(\hat n) = \text{Var}(f^{-1}(V))$.

To address this issue, inspired by the analysis of HLL~\cite{HeuleNH13},
we give an asymptotic analysis under the Poisson model.
Recall that the $n$ elements in $S$ into sketches $B_1, \ldots, B_m$ independently using hash function $h(\cdot)$.
Let $n_i$ denote the number of elements in $S$ hashed into $B_i$, $i=1,\ldots, m$.
We see that $\sum_{i=1}^m n_i = n$, therefore $n_1, \ldots, n_m$ are not independent, which makes the analysis of $\text{Var}(\hat n)$ challenging.
When $m$ is large, $n_1, \ldots, n_m$ can be approximated as identical and independent Poisson variables generated according to $\text{Pois}(\frac{n}{m})$~\cite{Mitzenmacherbook}.
Define $V_i = \frac{Z_i}{w}$, where
\begin{equation}\label{eq:Zi}
  Z_i
  = w - \sum_{x=0}^{w-1} B_i[x]
  = \sum_{x=0}^{w-1} (1-B_i[x]).
\end{equation}

Therefore, we have
$
V=\frac{\sum_{i=1}^m V_i}{m}.
$
Suppose that $n_1, \ldots, n_m$ are identical and independent Poisson variables generated according to the distribution $\text{Pois}(\frac{n}{m})$.
Then, all $V_1, \ldots, V_m$ are identical and independent random variables with the same expectation $\mu_\text{pois}$ and variance $\sigma^2_\text{pois}$.
For each of the $n_i\sim \text{Pois}(\frac{n}{m})$ elements hashed into the sketch $B_i$,
the probability that it is hashed into $B_i[x]$ is $p_x \times m$.
Then, it is not hard to find that the number of elements hashed into $B_i[x]$ is a random variable according to the distribution $\text{Pois}(np_x)$.
Therefore, we have that $B_i[x]$ equals zero with probability $\text{EXP}(-n p_x)$ and one with probability $1-\text{EXP}(-n p_x)$.
Based on the above analysis, it is not difficult to obtain the expectation $\mu_\text{pois}$ of each $V_1, \ldots, V_m$ as:
\begin{equation}\label{eq:mu}
\mu_\text{pois} = \frac{1}{w}\sum_{x=0}^{w-1} \text{EXP}(-n p_x)\approx f(n),
\end{equation}
where function $f(n)$ is defined in Eq.~(\ref{eq:fn}).
Similarly, the variance $\sigma^2_\text{pois}$ of each $V_1, \ldots, V_m$ is computed as:
\begin{equation}\label{eq:sigma2}
\begin{split}
\sigma^2_\text{pois} &= \frac{1}{w^2} \sum_{x=0}^{w-1} \text{EXP}(-n p_x) - \text{EXP}(-2n p_x)\\
&\approx \frac{1}{w^2} \left(\text{EXP}\left(-\frac{n}{2^w m}\right) - \text{EXP}\left(-\frac{n}{ m}\right)\right).
\end{split}
\end{equation}

From the Lindeberg–L\'evy central limit theorem~\cite{Bill86}, as $m$ approaches infinity, we have that the random variables $\sqrt{m} (V-\mu_\text{pois})$ converge in distribution to a normal distribution $\mathcal{N}(0, \sigma^2_\text{pois})$, i.e.,
\begin{equation}\label{eq:CLT}
\sqrt{m} (V-\mu_\text{pois}) \overset{D}{\rightarrow} \mathcal{N}(0, \sigma^2_\text{pois}).
\end{equation}
To derive $\text{Var}(f^{-1}(V))$ (i.e., $\text{Var}(\hat n)$),
using ~\emph{the delta method}~\cite{Hoef2012WID}, we then have
\[
\sqrt{m} (g(V)-g(\mu_\text{pois})) \overset{D}{\rightarrow} \mathcal{N}(0, \sigma^2_\text{pois}\times (g'(\mu_\text{pois}))^2),
\]
where function $g(\cdot) = f^{-1}(\cdot)$, i.e., 
the inverse function of function $f(n)$ defined in
\cref{eq:fn}.
Then, we have $g(\mu_\text{pois})\approx n$ because $\mu_\text{pois} \approx f(n)$ and $g'(\mu_\text{pois}) \approx \frac{1}{f'(n)}\approx -\frac{w}{\sum_{x=0}^{w-1} p_x \text{EXP}(-n p_x)}$,
where $f'(n)$ is the derivative of function
$f(n)$, therefore the mean square error of our estimation $\hat n$ can be approximated as $\frac{\sigma^2_\text{pois}}{m\times (f'(n))^2)} = \frac{\text{EXP}\left(-\frac{n}{2^w m}\right) - \text{EXP}\left(-\frac{n}{m}\right)}{m(\sum_{x=0}^{w-1} p_x \text{EXP}(-n p_x))^2}$. 
Besides, we have $\sum_{x=0}^{w-1} p_x \text{EXP}(-n p_x)\approx \sum_{x=0}^{w-1} \frac{n}{m} 2^{-x-1} \text{EXP}{(-\frac{n}{m} 2^{-x-1})}\approx\int_{0}^{w}\frac{n}{m} 2^{-x-1} \text{EXP}{(-\frac{n}{m} 2^{-x-1})}\text{d} x
=\frac{1}{\ln 2}(\text{EXP}\left(-\frac{n}{2^w m}\right) - \text{EXP}\left(-\frac{n}{m}\right))$. When setting $w\ge \lceil \log_2 (n/m)+6 \rceil$, we have $\text{EXP}\left(-\frac{n}{2^w m}\right)$ approximates 1 closely. Based on the above observations, we have  
$
\text{STDErr}(\hat{n}) \approx\frac{\ln 2}{\sqrt{m}}\cdot\frac{1}{\sqrt{1 - \text{EXP}\left(-\frac{n}{m}\right)}}
$.
When $\frac{n}{m} \ge 3$, we have $\text{STDErr}(\hat{n}) \approx \frac{\ln 2}{\sqrt{m}} \approx \frac{0.69}{\sqrt{m}}$. The standard error fits well with the experiments in \cref{fig:pure_sketch}.

As we mentioned in Section~\ref{subsec:challenges},
revealing the value of $Z=mwV$ (defined in Eq.~(\ref{eq:Z})) leaks privacy in the setting of PDCE studied in this paper.
To address this issue, we add a random variable 
$
N\sim \mathcal{N}\left(0, \sigma^2_\text{noise}\right)
$
to the variable $Z$ to achieve differential privacy.
That is, we define
$
V_\text{diff} \triangleq \frac{Z + N}{mw} = V + \frac{N}{mw},
$
and from Eq.~(\ref{eq:CLT}) then have
\[
\sqrt{m} (V_\text{diff} -\mu_\text{pois}) \overset{D}{\rightarrow} \mathcal{N}\left(0, \sigma^2_\text{pois} + \frac{\sigma_\text{noise}^2}{m w^2}\right).
\]
Again, using the delta method, we have 
\[
\sqrt{m} (g(V_\text{diff})-g(\mu_\text{pois})) \overset{D}{\rightarrow} \mathcal{N}\left(0, \left(\sigma^2_\text{pois} + \frac{\sigma_\text{noise}^2}{m w^2}\right) \times (g'(\mu_\text{pois}))^2\right).
\]
Thus, the mean square error of estimation $\hat n^*$ inferred from $V_\text{diff}$ can be approximated as $\frac{\sigma^2_\text{pois} + \frac{\sigma_\text{noise}^2}{m w^2}}{m\times (f'(n))^2)} = \frac{\text{EXP}\left(-\frac{n}{2^w m}\right) - \text{EXP}\left(-\frac{n}{m}\right) + \frac{\sigma_\text{noise}^2}{m w^2}}{m(\sum_{x=0}^{w-1} p_x \text{EXP}(-n p_x))^2}$.
Similar to the procedure of approximating $\text{STDErr}(\hat{n})$, we have $\text{STDErr}(\hat{n}^*)\approx \frac{0.69}{\sqrt{m}}\cdot\sqrt{1+\frac{\sigma_\text{noise}^2}{m w^2}}$.

\pdfoutput=1
\subsection{Our DP-DICE Protocol}\label{subsec:dp-dice}

\subsubsection{\textbf{Overview}}

\begin{figure}[t]
  \centering
	\includegraphics[width=\linewidth]{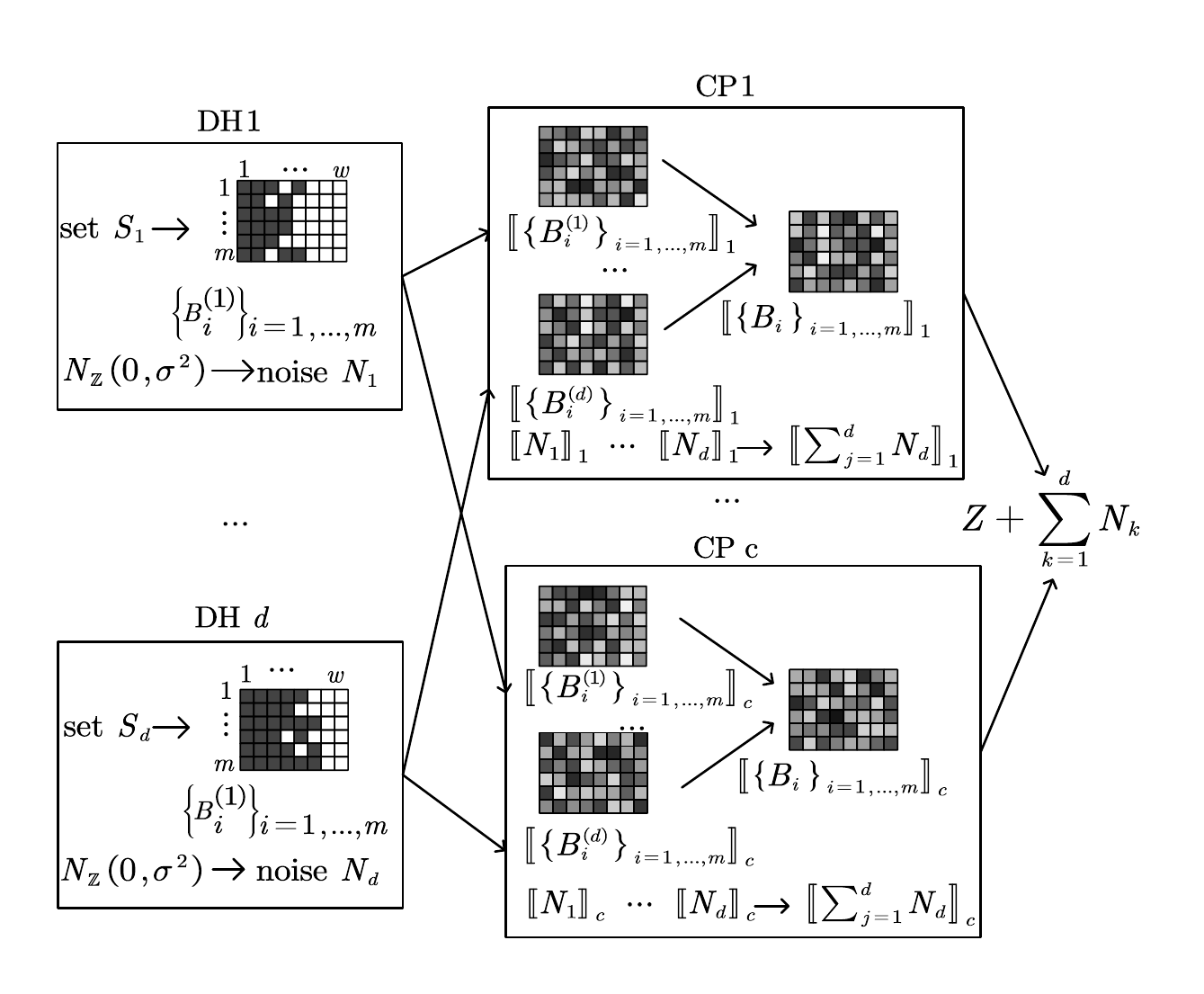}
  \caption{Overview of our DP-DICE.
    $\{B_i^{(k)}\}_{i=1, \ldots, m}$, $k=1, \ldots, d$ is the FMS sketch of set
    $S_k$ holding by DH $k$ and $N_k$ is a random variable according to discrete
    Gaussian distribution $\mathcal{N}_\Z (0, \sigma^2)$.
    $\{\llbracket{B_i^{(k)}}\rrbracket_j\}_{i=1, \ldots, m}$, $j=1, \ldots, c$ and
    $\llbracket{N_k}\rrbracket_j$ holding by CP $j$ are the additive shares of
    $\{B_i^{(k)}\}_{i=1, \ldots, m}$ and $N_k$, respectively.}
	\label{fig:framework}
\end{figure}

Our DP-DICE is designed to securely estimate the cardinality of set
$S=\cup_{j=1,\ldots, d} S_j$, where set $S_j$ is locally held by DH $j$, which is
honest but curious.
The overview of our DP-DICE is shown in \cref{fig:framework}.
Each DH $j$ is responsible for collecting its set $S_j$ and building the FMS
sketch $\{B_i^{(j)}\}_{i=1, \ldots, m}$ of $S_j$.
In addition, it also generates a variable $N_j$ used for generating additive noise
to satisfy differential privacy.
Then, DH $j$ securely sends both $\{B_i^{(j)}\}_{i=1, \ldots, m}$ and $N_j$ to all
the CPs.
After receiving all the DHs' FMS sketches, all the CPs collaboratively merge them
to compute the FMS sketch $\{B_i\}_{i=1, \ldots, m}$ of set $S$ in a secure
manner.
To avoid privacy leakage, each CP only holds a share of FMS sketch $\{B_i\}_{i=1,
  \ldots, m}$ of the union set $S$.
At last, all the CPs securely compute the variable $Z$ (defined in \cref{eq:Z}) of
FMS sketch $\{B_i\}_{i=1, \ldots, m}$ and then output $Z+\sum_{j=1}^d N_j$ to the
public for estimating the cardinality $n$ of set $S$, where $\sum_{j=1}^d N_j$ is
noise added to meet the requirement of differential privacy.
Our DP-DICE protocol consists of three phases: offline preparation phase, data
collection phase, and data aggregation phase.
We build our DP-DICE on the SPDZ framework~\cite{KellerPR18}, which utilizes
authenticated shares.
Next, we introduce each phase in detail.

\subsubsection{\textbf{Offline Preparation Phase}}

This phase mainly initializes some parameters used in the protocol and generates
random numbers for subsequent online steps according to the MPC protocol.

\header{Parameter Initialization.}
All the DHs and CPs negotiate parameters to be used in the protocol during the
parameter initialization.
First, they agree on a finite field $\F_p=\{0, 1, \ldots, p-1\}$, which
will be used as the basis of data representation, secret sharing, and all
computation.
The modulus $p$ is a $(\lambda+\tau)$-bit prime number, where $\lambda$ (e.g.
$\lambda=40$) is a statistical security parameter and $\tau$ (e.g.
$\tau=32$) is determined by the plaintext domain.
Next, all the CPs and DHs agree on the FMS sketch's parameters $m$ and $w$ and all
the CPs run the setup protocol of SPDZ to obtain their parameters and keys.
In addition, all the DHs choose a keyed hash function $H$ and establish a secret
key $\mathbb{k}$ using the authenticated group key exchange protocol
in~\cite{KatzY03}.
Then, the hash function $h$ implemented by the FMS sketch is defined as $h(e) =
H(\mathbb{k}||e)$, where $||$ is the string concatenation operation.

\header{Random Number Generation.}
All the CPs run several offline protocols to generate a number of random numbers
which will be used later in the data collection and aggregation phases.
Note that each CP only holds a piece (i.e., a share) of any generated random
number and the adversary can obtain the random number only when all the CPs are
corrupted and controlled.
The offline protocols to generate random numbers we used include \Rand,
\Rand2, and \RandExp, which all are available on the framework of SPDZ.

\subsubsection{\textbf{Data Collection Phase}}

This phase is for DHs, which are responsible for completing the collection of
datasets and covertly sharing their generated FMS sketches with all the CPs.
Each DH also sends a variable to facilitate the generation of additive noise on the CPs, which is used to satisfy differential privacy.

\header{FMS Sketch Generation.}
Using the same hash function $h(e)=H(\mathbb{k}||e)$, each DH $j$ computes the FMS
sketch $\{B_i^{(j)}\}_{i=1,\ldots,m}$ of set $S_j$ using the method in \cref{sec:FMS}.
The FMS sketch $\{B_i^{(j)}\}_{i=1, \ldots, m}$ will be securely sent to all the
CPs to compute the union FMS sketch.

\header{Distributed Discrete Noise Generation.}
MPC frameworks such as SPDZ are more convenient and efficient to handle integers than floating-point numbers.
In addition, we notice that the variable $Z$ of the union FMS sketch $\{B_i\}_{i=1, \ldots, m}$ (defined in \cref{eq:Z}), which will be revealed to the public for computing an estimate of the
cardinality of the union set $S$, is a positive integer.
To add controlled discrete noise to $Z$, therefore, we adopt the state-of-the-art secure aggregation method, \emph{distributed discrete Gaussian noise
mechanism}~\cite{KairouzL021}, which is originally designed for securely aggregating model updates of clients in federated learning.
Specially, each DH $j$ generates a random variable $N_j\in\Z$ (i.e.
$N_j$ is a random integer) to protect against differential attacks.
$N_j$ is a random variable sampled from the discrete Gaussian distribution
$\mathcal{N}_\Z (0, \sigma^2)$, that is,
$
  P(N_j=x) =
  \frac{\exp\left(-\frac{x^2}{2\sigma^2}\right)}
  {\sum_{y\in\Z}\exp\left(-\frac{y^2}{2\sigma^2}\right)},
$
$x\in\Z$.

Each DH $j$ will covertly send its noise variable $N_j$ to all the CPs and the sum
of all the DHs' noise variables $N_j$, i.e., $\sum_{j=1}^d N_j$ will be later
added to variable $Z$.

\header{Secure Data Sharing.}
In SPDZ, an integer $x$ in the shared form is defined as
$
  \share{x}\triangleq
  (x_1,\ldots,x_c, m_1^{(x)},\ldots,m_c^{(x)}, \Delta_1,\ldots,\Delta_c),
$
and each CP $j$ holds a tuple $\share{x}_j = (x_j, m_j^{(x)}, \Delta_j)$ where the
three parts in the tuple are the additive shares of $x$, the MAC, and the MAC key,
respectively.
The MAC and the MAC key are used to verify the authenticity of variable $x$, which
makes us aware when the adversary tampers with $x$.
To securely share a variable $x$ held by a DH to all the CPs, we adopt the method
used in~\cite{Hu0LGWGLD21}.
Specifically, all the CPs reveal a random number
$
\share{a} = (a_1, \ldots, a_c,
m_1^{(a)}, \ldots, m_c^{(a)}, \Delta_1, \ldots, \Delta_c)
$
in the shared form to
the DH, where $a\in\F_p$ is generated offline by function \Rand{} provided by
SPDZ.
In detail, each CP $j$ sends its $a_j$ to the DH, then the DH obtains
$a=\sum_{j=1}^c a_j$.
Given $a$, the DH computes $x-a$ and broadcasts it to all the CPs.
Then, all the CPs compute
$
  \share{x} = \share{a} + x - a.
$
That is, each CP $j$ computes
\[
  \share{x}_j =
  \begin{cases}
    {(a_j + x - a, m_j^{(a)} + (x-a)\Delta_j, \Delta_j)} & \text{if} \, j =1, \\
    {(a_j, m_j^{(a)} + (x-a)\Delta_j, \Delta_j)}         & \text{otherwise.}
	\end{cases}
\]
We easily see that the value of $a$ is unknown to the adversary which controls up
to $c-1$ CPs.
Therefore, the above procedure is secure.
Each random variable $a\in\F_p$ is generated by the \Rand{} function provided by
the SPDZ framework and against potential attacks we discard it once it is used.
In other words, different random numbers need to be used for securely sharing
different variables from the DH to all the CPs.
Using the above method, each DH $j$ securely sends and shares each entry
$B_i^{(j)}[l]$, $i=1, \ldots, m$, $l=1, \ldots, w$ of its FMS sketch as well as
the noise variable $N_j$ to all the CPs.

\subsubsection{\textbf{Data Aggregation Phase}}
This phase is for CPs.
CPs are responsible for securely computing the FMS sketch $\{B_i\}_{i=1, \ldots,
  m}$ of the union set $S$ based on the shares of $B_i^{(j)}$ for $1\leq i\leq
m, 1\leq j\leq d$, and calculating the variable $Z$ defined in \cref{eq:Z}.
Before revealing the value of $Z$ to the public, all the CPs securely insert a
noise $\sum_{j=1}^d N_j$ to $Z$ to meet the requirement of differential privacy.

\header{Merge FMS Sketches.}
To compute the secret shares of the union set's FMS sketch $\{B_i\}_{i=1, \ldots,
  m}$, we first define variables $B_i^*[l]$ as:
\[
  B_i^*[l] = \sum_{j=1}^d B_i^{(j)}[l], \quad i=1,\ldots,m, \quad l=1,\ldots,w.
\]
We easily see that $B_i[l] = 1$ when $B_i^*[l] > 0$ and $B_i[l] = 0$ otherwise.
In the framework of SPDZ, for any integers $x$ and $y$, we have
\[
  \share{x + y}_j = \share{x}_j + \share{y}_j
  = (x_j + y_j, m_j^{(x)} + m_j^{(y)}, \Delta_j).
\]
Therefore, given all tuples $\share{B_i^{(1)}[l]}_j, \ldots,
\share{B_i^{(d)}[l]}_j$ held by CP $j$, CP $j$ computes $\share{B_i^*[l]}_j$ as:
$
  \share{B_i^*[l]}_j = \sum_{k=1}^d \share{B_i^{(k)}[l]}_j.
$
To securely compute the value of $\share{B_i[l]}_j$ for CP $j$, we apply the
protocol \ZeroTest{} proposed in~\cite{LipmaaT13} on $\share{B_i^*[l]}$, which can
securely compare whether $B_i^*[l]$ equals zero.
The pseudo-code of \ZeroTest{} is shown in \cref{alg:zerotest}.

\pdfoutput=1
\begin{algorithm}[t]
	\SetKwInOut{Input}{Input}
	\SetKwInOut{Output}{Output}

	\Input{$\share{x}$, $x\in\F_p$}
	\Output{$\share{b}$, where $b=0$ when $x=0$ and $b=1$ otherwise}

  \vspace*{2ex}
	\textbf{Offline phase:} \\
	\ForEach{$t=0,1,\ldots,L-1$}{\label{ln:a1}
    $\share{a_t}\gets\Rand2()$\tcp*{generate shares of a random bit}
	}
  $\share{a}\gets\sum_{t=0}^{L-1}2^{t}\share{a_t}$\tcp*{$a_t$ is the $t$-th bit of $a$}
  \label{ln:a2}

	$(\beta_0, \ldots, \beta_\tau)\gets\mathtt{Interpolate}(\phi(x))$
  \tcp*{$\phi(x)=\sum_{t=0}^\tau \beta_t x^t$}
  \label{ln:phi}

	$(\share{R^{-1}},\share{R},\share{R^2},\ldots,\share{R^\tau})\gets\RandExp(\tau)$\;
  \label{ln:R}

  \vspace*{2ex}
	\textbf{Online phase:}\\
	$\share{r}\gets\share{a}+\share{x}$ and reveal $r$ to all the CPs\;\label{ln:r}
  $\share{h}\gets \sum_{t=0}^{L-1}(\share{a_t}+r_t-2r_t\cdot\share{a_t})$\tcp*{$r_t$
    is the $t$-th bit of $r$}
  $\share{\gamma}\gets\share{R^{-1}}\cdot\share{1+h}$ and reveal $\gamma$ to all
  the CPs\;\label{ln:gamma}
  $\share{b}\gets\beta_0$\;
  \ForEach{$t=1, 2,\ldots, \tau$}{
		$\share{b}\gets\share{b}+\beta_t\cdot\gamma^{t}\share{R^t}$\tcp*{note
      that $\gamma^{t}\cdot\share{R^t}=\share{(1+h)^t}$}
	}
  \Return $\share{b}$\;
	\caption{Function $\ZeroTest(\share{x})$}\label{alg:zerotest}
\end{algorithm}

To test whether an integer $x\in\F_p$ is zero or not, \ZeroTest{} first computes
$\share{r} = \share{a + x} = \share{a} + \share{x}$ and reveals $r$ to all the CPs
(\cref{ln:r}), where $a\in\F_p$ is a random integer generated in the offline
preparation phase (Lines~\ref{ln:a1}--\ref{ln:a2}).
Since no CP knows $a$, therefore, revealing $r$ does not leak information of $x$
but facilitates computing the Hamming distance $h$ between $r$ and $a$ in the
shared form.
Specifically, we have
$
  \share{h} = \sum_{t=0}^{L-1}(\share{a_t} + r_t - 2r_t\cdot\share{a_t}),
$
where $L=\lceil \log_2 p\rceil$ is the bit length of the modulus $p$ used for the
SPDZ framework and $a_t$ and $r_t$ are the $t$-th bits of random variables $a$ and
$r$ respectively, i.e., $a=\sum_{t=0}^{L-1}2^ta_t$ and $r=\sum_{t=0}^{L-1}2^tr_t$.
Note that $a_t$ is computed offline via function \Rand2 provided by SPDZ.

We easily see that $h=0$ when $x=0$ and $h\in \{1, \ldots, \tau\}$ otherwise,
where $\tau$ (e.g., $\tau=32$) is the bit length of the plaintext domain setting
for the SPDZ.
\ZeroTest{} uses a lookup function that is a polynomial $\phi(\cdot)$ such that
$\phi(0) = 0 $ and $\phi(x) = 1$ for all other $x\in\{1, \ldots, \tau\}$.
The lookup polynomial will be interpolated accordingly, say, using the Lagrange
Interpolation (\cref{ln:phi}).
Given $\share{h}$, it is inefficient to directly compute $\share{h^t}$, which is
included in the lookup polynomial.
To solve this issue, \ZeroTest{} uses a function $\RandExp(\tau)$ on SPDZ to
generate $(\share{R^{-1}}, \share{R}, \share{R^2}, \ldots, \share{R^\tau})$, the
shares of a nonzero random number $R\in\F_p\backslash\{0\}$, as well as the shares
of its $t$-th power, $t\in\{-1, 1, 2, \ldots, \tau\}$, which are all generated in
the offline phase (\cref{ln:R}).
Then, it computes $\share{\gamma}=\share{R^{-1}}\cdot\share{1+h}$ and reveals
$\gamma$ to all the CPs (\cref{ln:gamma}).
Here we use $h+1$ instead of $h$ because it is easy to infer $h=0$ when
$R^{-1}h=0$, which leaks information of $h$.
Because $\gamma^t=R^{-t}(1+h)^t$, we have $\gamma^t\cdot\share{R^t}=\share{(1+h)^t}$.
Formally, we define $\phi(x)=\sum_{t=0}^\tau\beta_tx^t$ such that $\phi(1)=0$ and
$\phi(x)=1$ for $x\in\{2,\ldots,\tau+1\}$.
Given $\gamma, \beta_0,\ldots,\beta_\tau$, as well as $\share{R},\share{R^2},
\ldots, \share{R^\tau}$, we then the result of \ZeroTest{} as:
$
  \share{b} = \share{\phi(1+h)} = \sum_{t=0}^\tau\beta_t\gamma^t\cdot\share{R^t}.
$
Applying \ZeroTest{} for each $\share{B_i^*[l]}$, we finally obtain
$\share{B_i[l]}$, for each $1\leq i\leq m$.

\header{Merge Noise Variables.}
Given all tuples $\share{N_1}_j, \ldots, \share{N_d}_j$ held by CP $j$, CP $j$
computes $\share{N}_j$ in share form as:
$
  \share{N}_j = \sum_{k=1}^d \share{N_k[l]}_j,
$
where $N=\sum_{k=1}^dN_k[l]$ is the final additive noise used for differential
privacy protection.

\header{Estimate Cardinality.}
Given $\share{B_i[l]}_j$ and $\share{N}_j$, each CP $j$ first computes
$
  \share{Z + N}_j = \share{N}_j + \sum_{i=1}^m\sum_{l=1}^w\share{B_i[l]}_j.
$
Before the SPDZ framework reveals the value $x=Z+\sum_{k=1}^dN_k$ in the shared
form $ \share{x}=(x_1, \ldots, x_c, m_1^{(x)}, \ldots, m_c^{(x)}, \Delta_1,
\ldots, \Delta_c), $ it first checks the authenticity of $x$ as follows: let each
CP $j$ compute $\sigma_j^{(x)}= m_j^{(x)} - x \Delta_j$ and broadcast
$\sigma_j^{(x)}$, then check whether $\sum_{j=1}^c \sigma_j^{(x)} = 0$.
It aborts if $\sum_{j=1}^c \sigma_j^{(x)} \ne 0$, and outputs $x$ otherwise.
Given $x=Z + \sum_{k=1}^c N_k$, we use the method in \cref{sec:FMS} to estimate
the cardinality $n$ of the union set $S$.

\pdfoutput=1
\subsection{Protocol Security and Differential Privacy}
We consider the protocol's security properties including confidentiality and
correctness.
As we mentioned, the adversary can corrupt all but one CPs.
However, the adversary is unable to tamper with the correctness of computing
$Z+\sum_{j=1}^d N_j$ without being detected, which is the salient feature of the
SPDZ framework.
In addition, it learns nothing from executing the protocol except the protocol's
output $Z+\sum_{j=1}^d N_j$, which is differentially private discussed later.
Following the security proof of MPC-FM given in~\cite{Hu0LGWGLD21}, we also prove
the security of our DP-DICE in the Universally Composable
framework~\cite{Canetti20}, which enables our protocol to serve as a component of
a larger system without losing its security properties.

Next, we discuss the differential privacy of our protocol.
To set a proper scale $\sigma$ for the discrete Gaussian distribution
$\mathcal{N}_\mathbb{Z} (0, \sigma^2)$ used to guarantee the differential privacy
for the protocol's output $Z+N$, we first compute the sensitivity of variable $Z$
defined as 
$\Delta_Z = \max |Z(S) - Z(S')|$,
where the maximum is over all pairs of sets $S$ and $S'$, which are two subsets of
the universal set $U$ differing in at most one element, and $Z(S)$ and $Z(S')$ are
the values of $Z$ for sets $S$ and $S'$, respectively.
For the definition of $Z$ given in \cref{eq:Z}, we find that $\Delta_Z=1$.
Define a variable 
$
  \varepsilon_d = \min \left(\sqrt{\frac{1}{d\sigma^2}+\frac{\tau_d}{2}},
    \frac{1}{\sqrt{d}\sigma }+\tau_d\right),
$
where $\tau_d=10\sum_{k=1}^{d-1} \exp\left(-\frac{2k\pi^2\sigma^2}{k+1}\right)$.
From Theorem ~\ref{theorem:distributed_discrete_gaussian_CDP} in  Section~\ref{subsec:DP}, we then find that the result of adding
$d$ independent discrete Gaussian variables $N_1, \ldots, N_d\sim
\mathcal{N}_\mathbb{Z} (0, \sigma^2)$ to variable $Z$ satisfies
$\frac{1}{2}\varepsilon_d^2$-concentrated differential privacy when $\sigma\ge 0.5$.
From Section~\ref{subsec:DP}, we find that $Z$ achieves $(\varepsilon_\delta^*,
\delta)$-differential privacy where $\varepsilon_\delta^*
    \le 0.5\varepsilon_d(\varepsilon_d + 2\sqrt{-2\ln\delta})$.
Besides the adversary, a curious DH $i$ may subtract its local variable $N_i$ from
the protocol's output $Z+\sum_{j=1}^d N_j$.
However, it still learns nothing except the output $Z + \sum_{1\le j\le d \wedge
  j\ne i} N_j$.
Similarly, we see that the variable $Z + \sum_{1\le j\le d \wedge
  j\ne i} N_j$ satisfies $\frac{1}{2}\varepsilon_{d-1}^2$-concentrated
differential privacy and $(\varepsilon_\delta^*,
\delta)$-differential privacy where $\epsilon_\delta^*
    \le 0.5\varepsilon_{d-1}(\varepsilon_{d-1} + 2\sqrt{-2\ln\delta})$.

\pdfoutput=1
\begin{figure*}[htp]
  \centering
  \subfloat[AARE vs. $m$, where $n$ = $10^3$]{%
    \includegraphics[width=.25\linewidth]{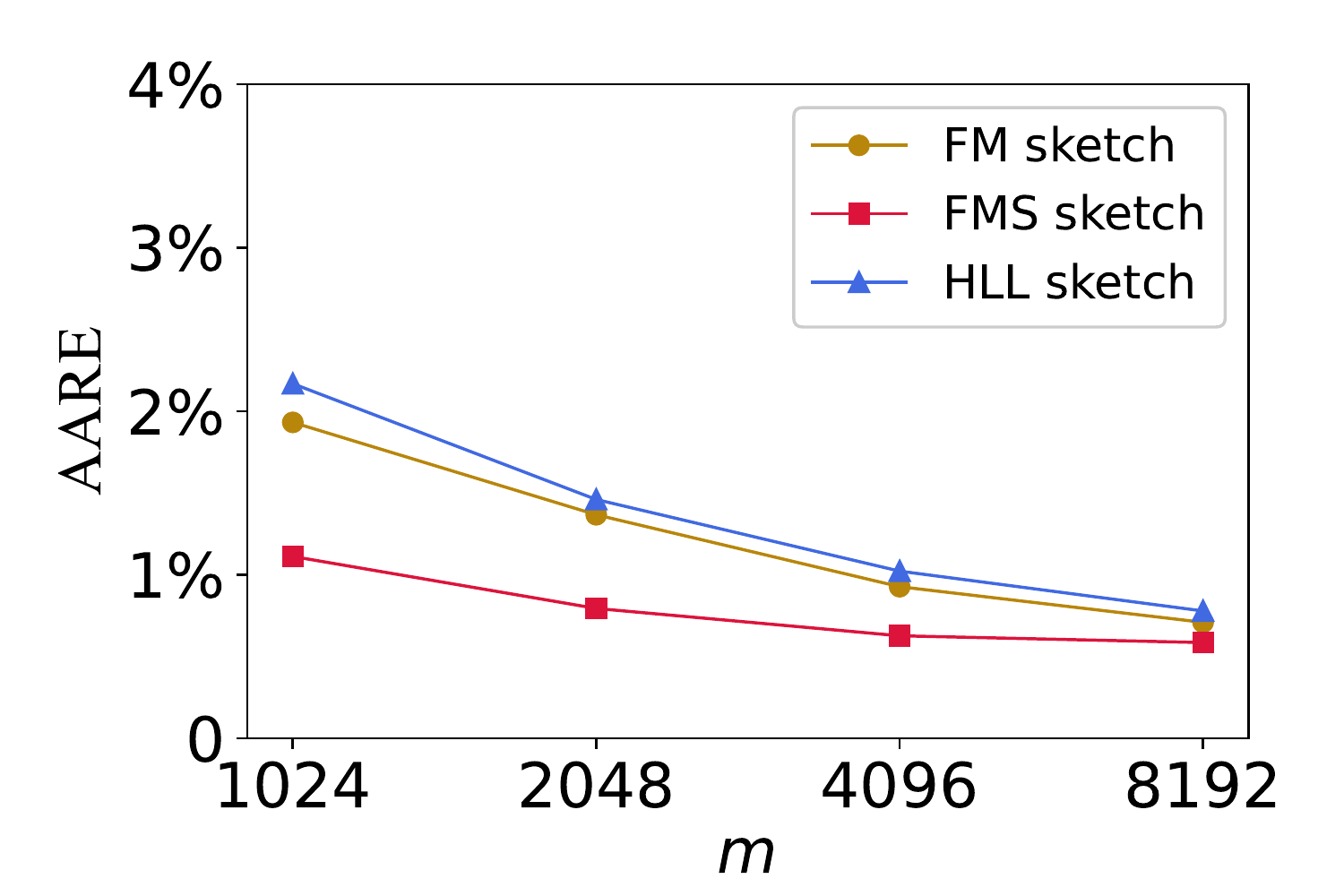}}
  \subfloat[AARE vs. $m$, where $n$ = $10^5$]{%
    \includegraphics[width=.25\linewidth]{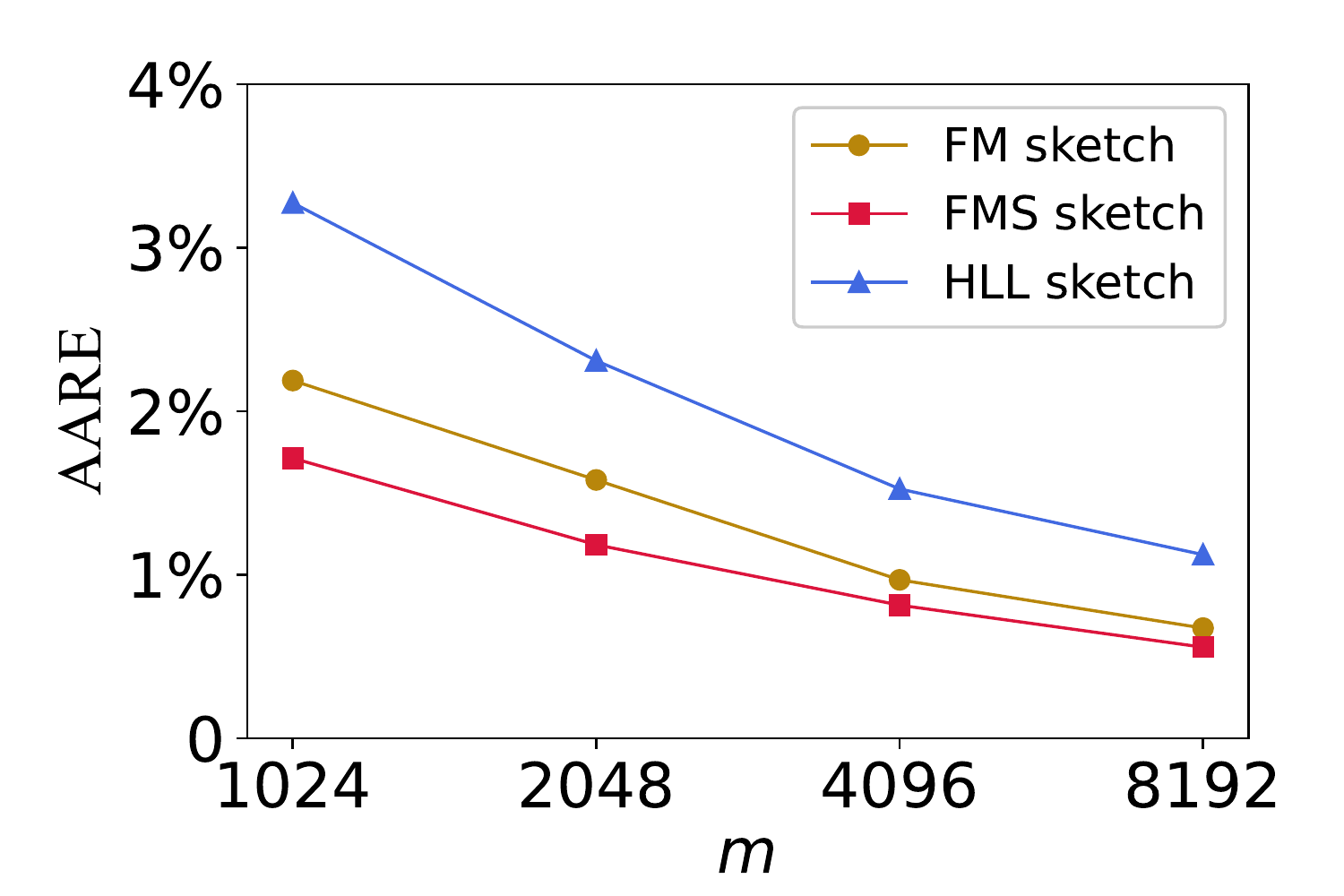}} 
  \subfloat[AARE vs. $m$, where $n$ = $10^7$]{%
    \includegraphics[width=.25\linewidth]{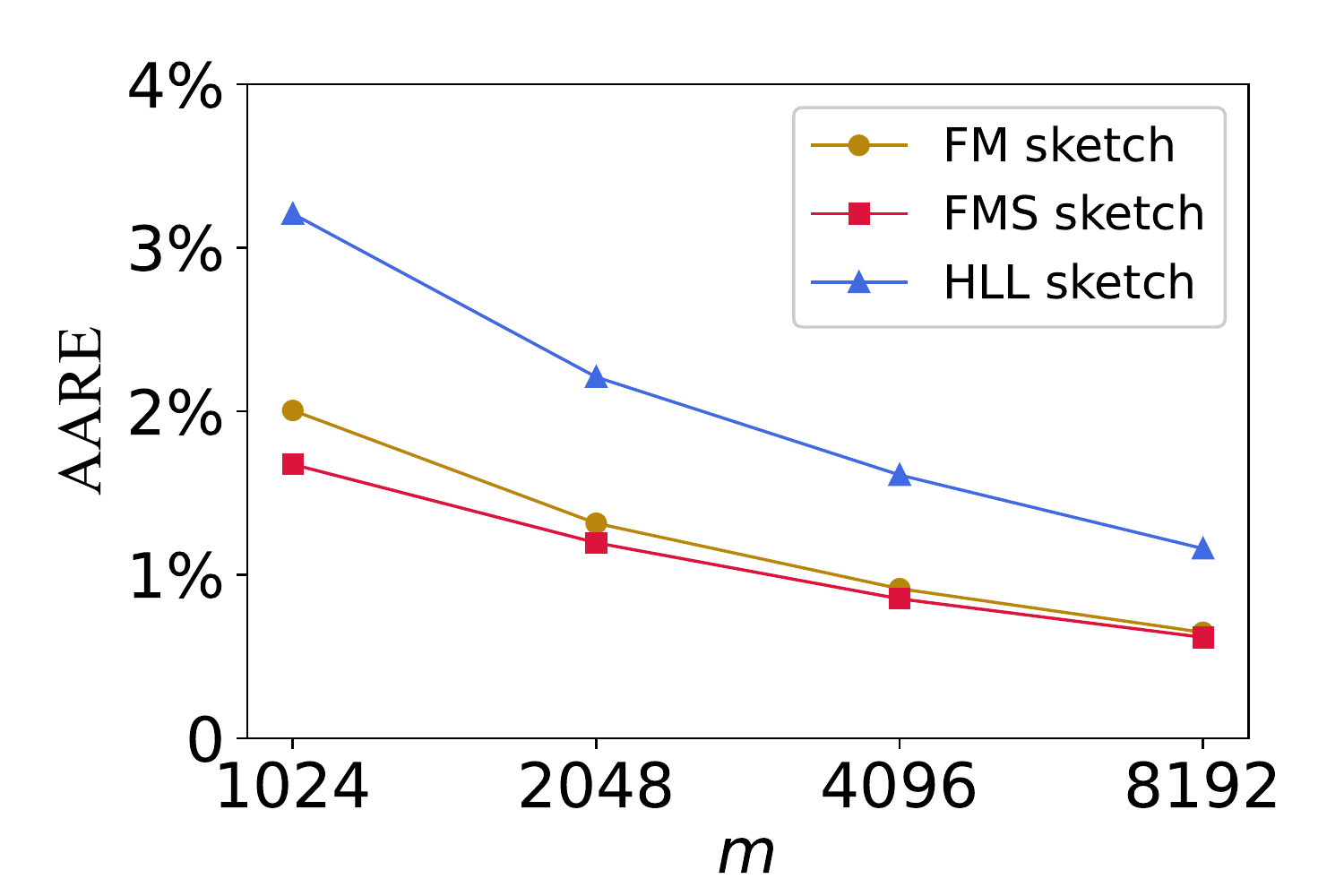}}
  \subfloat[AARE vs. $m$, where $n$ = $10^9$]{%
    \includegraphics[width=.25\linewidth]{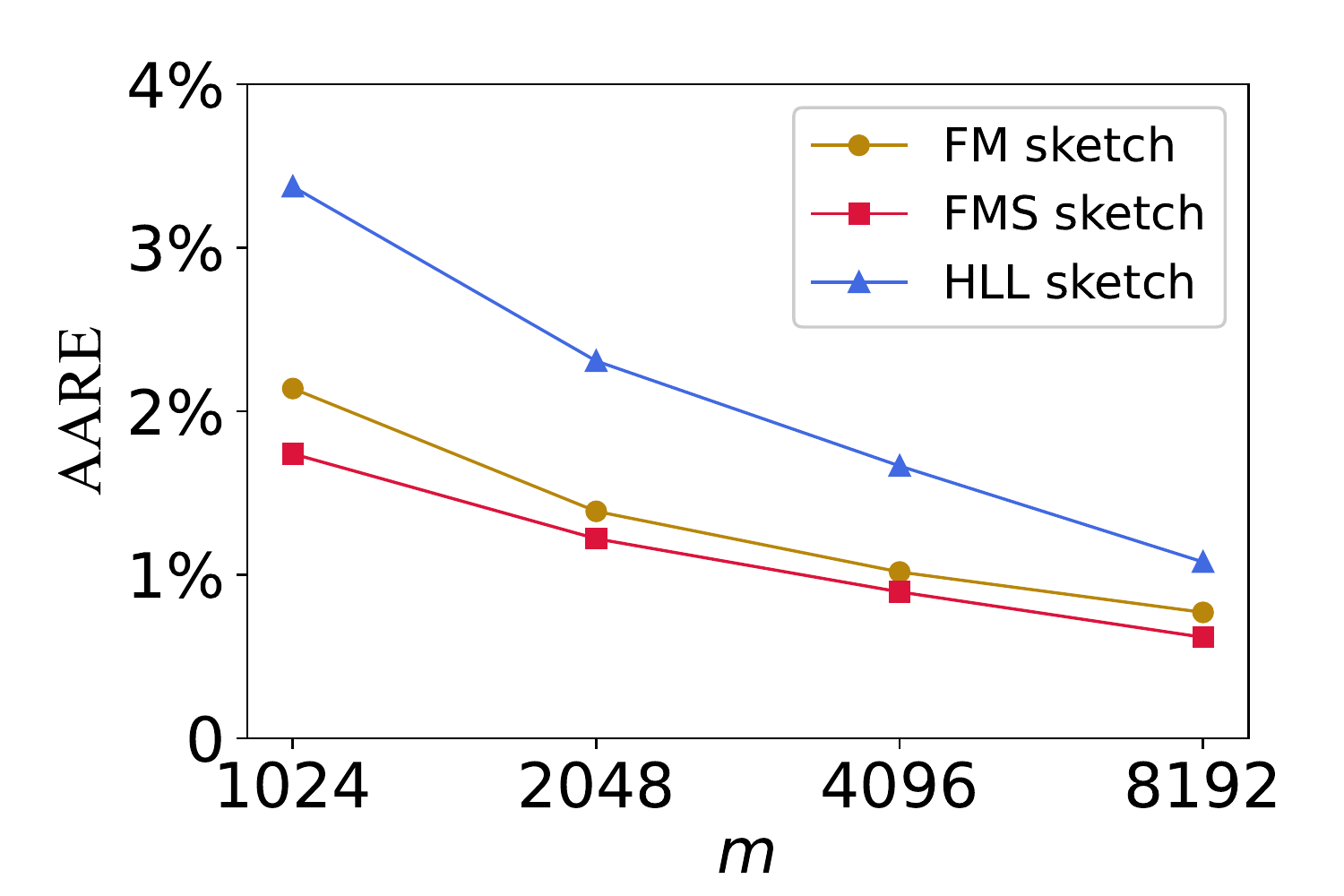}}
  \caption{Accuracy of our FMS sketch compared with FM sketch and HLL sketch when privacy is not considered.}
  \label{fig:pure_sketch}
\end{figure*}

\begin{figure*}[htp]
  \centering
  \subfloat[AARE vs. $\varepsilon$, where $n$ = $10^3$]{%
    \includegraphics[width=.25\linewidth]{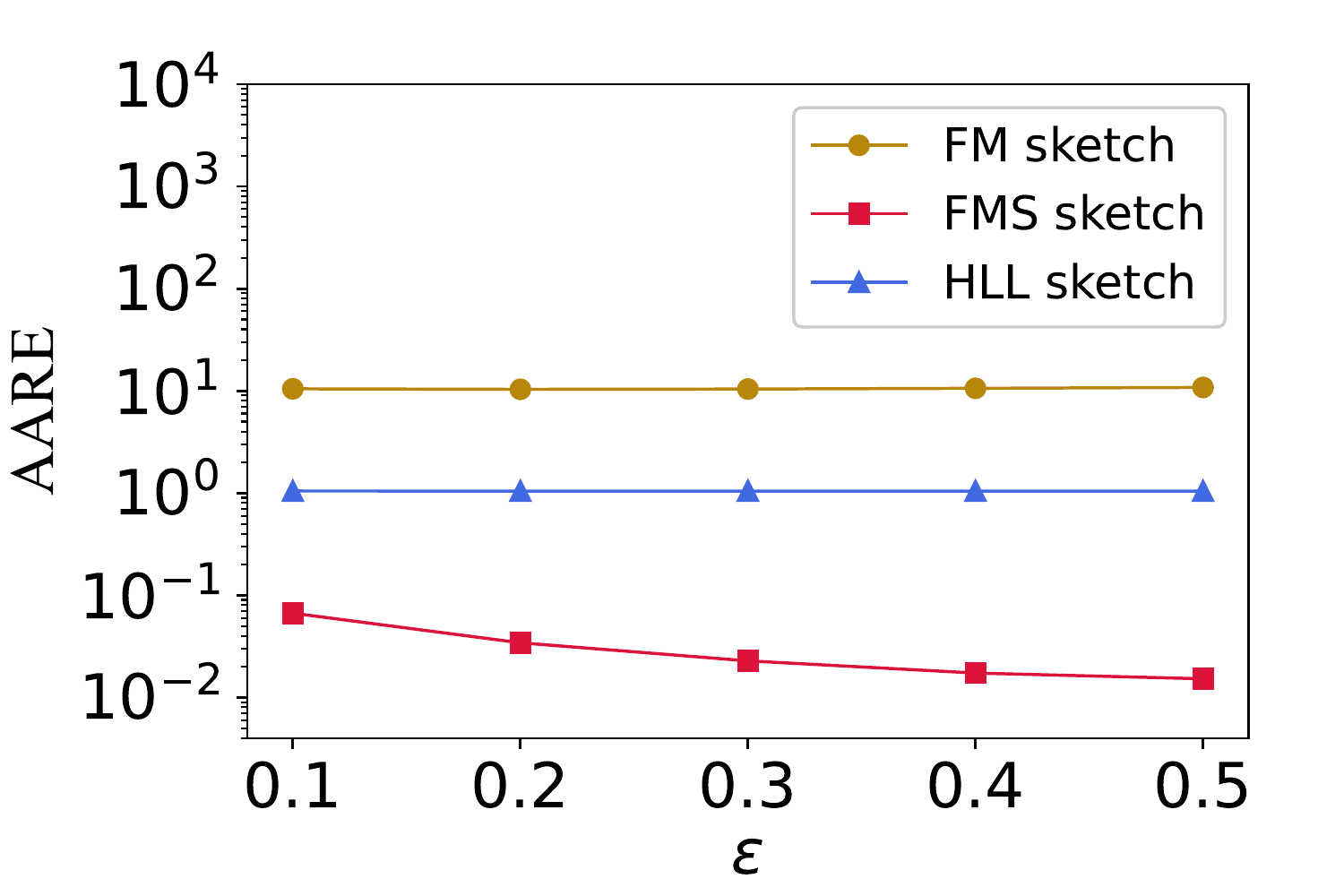}}
  \subfloat[AARE vs. $\varepsilon$, where $n$ = $10^5$]{%
    \includegraphics[width=.25\linewidth]{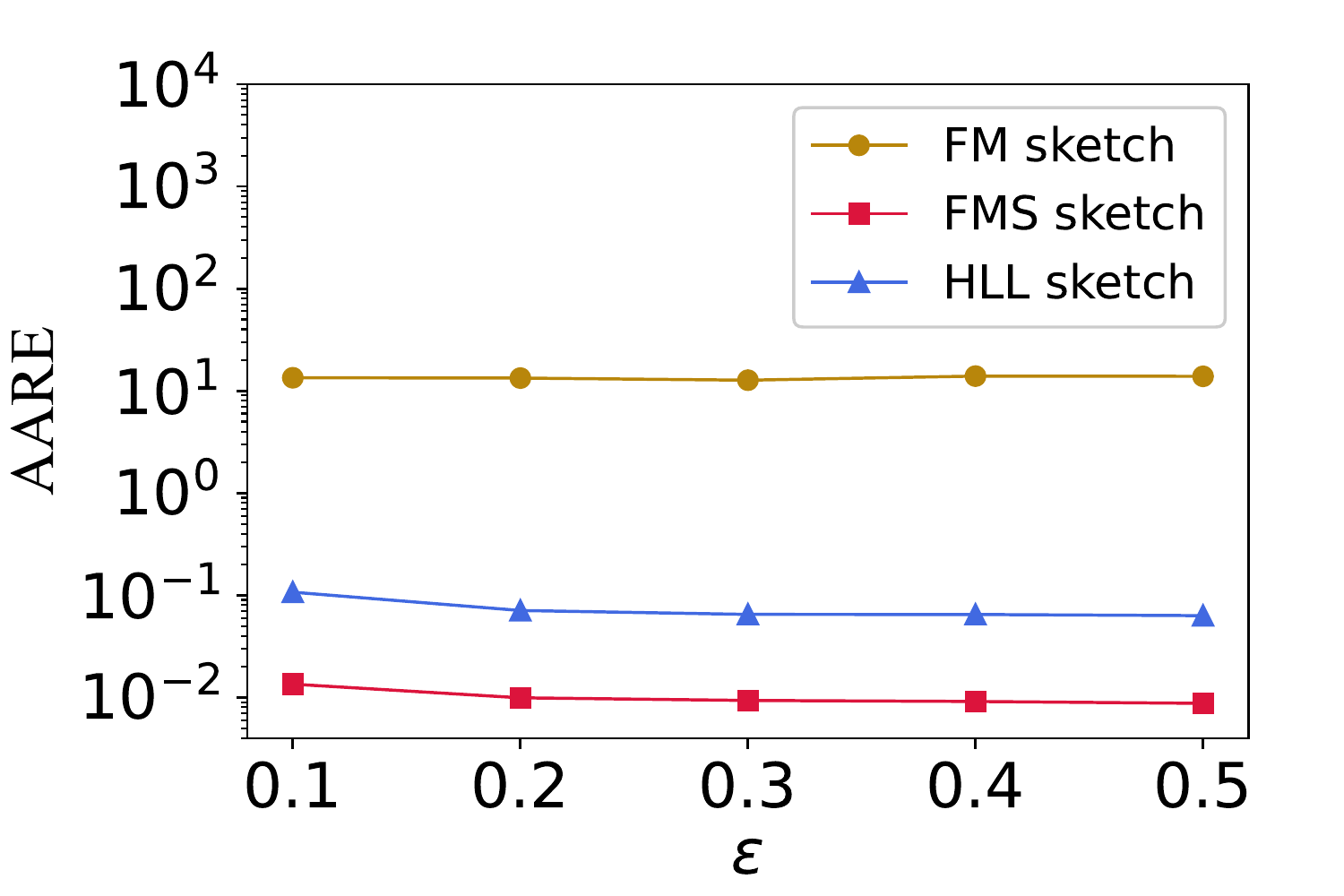}} 
  \subfloat[AARE vs. $\varepsilon$, where $n$ = $10^7$]{%
    \includegraphics[width=.25\linewidth]{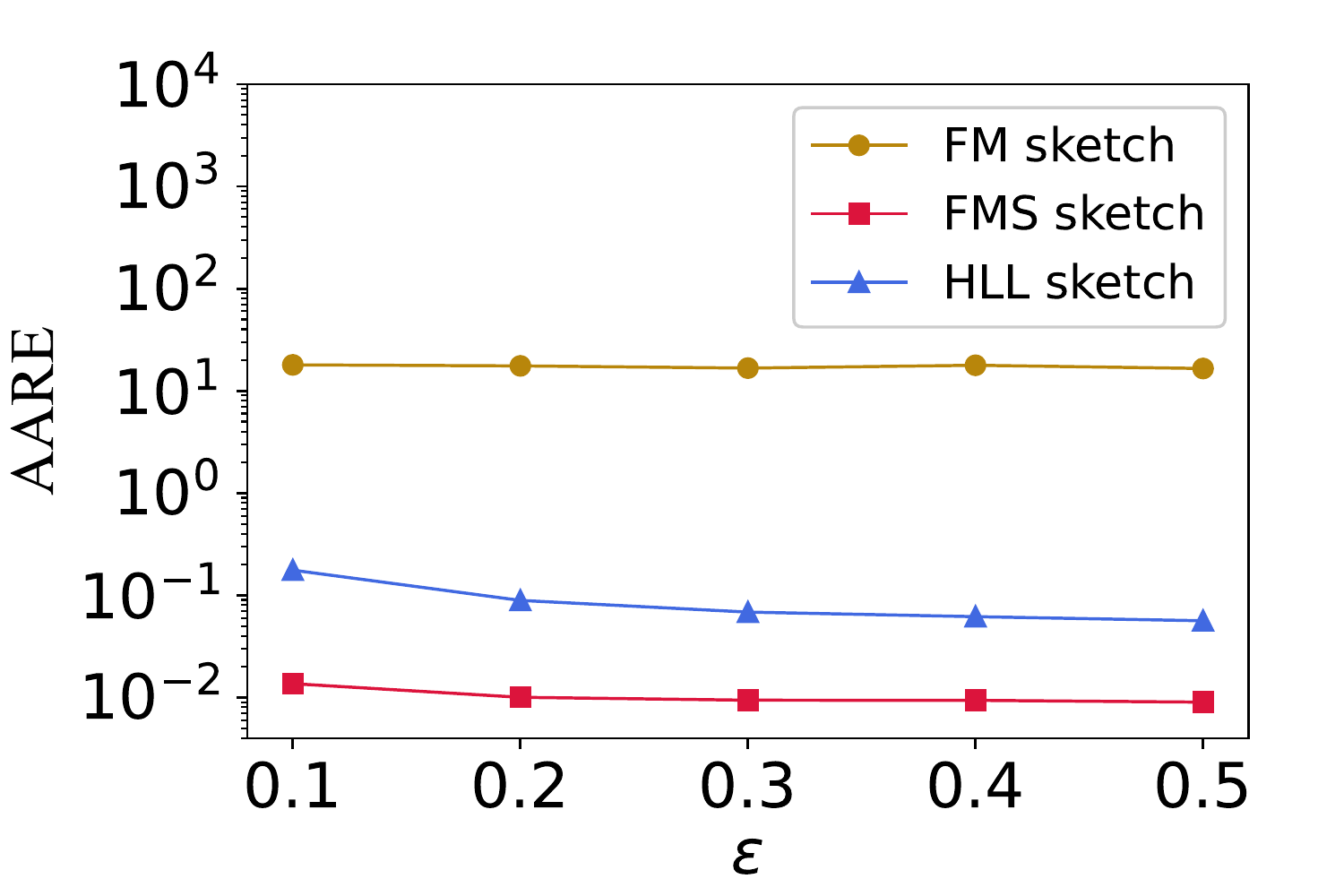}}
  \subfloat[AARE vs. $\varepsilon$, where $n$ = $10^9$]{%
    \includegraphics[width=.25\linewidth]{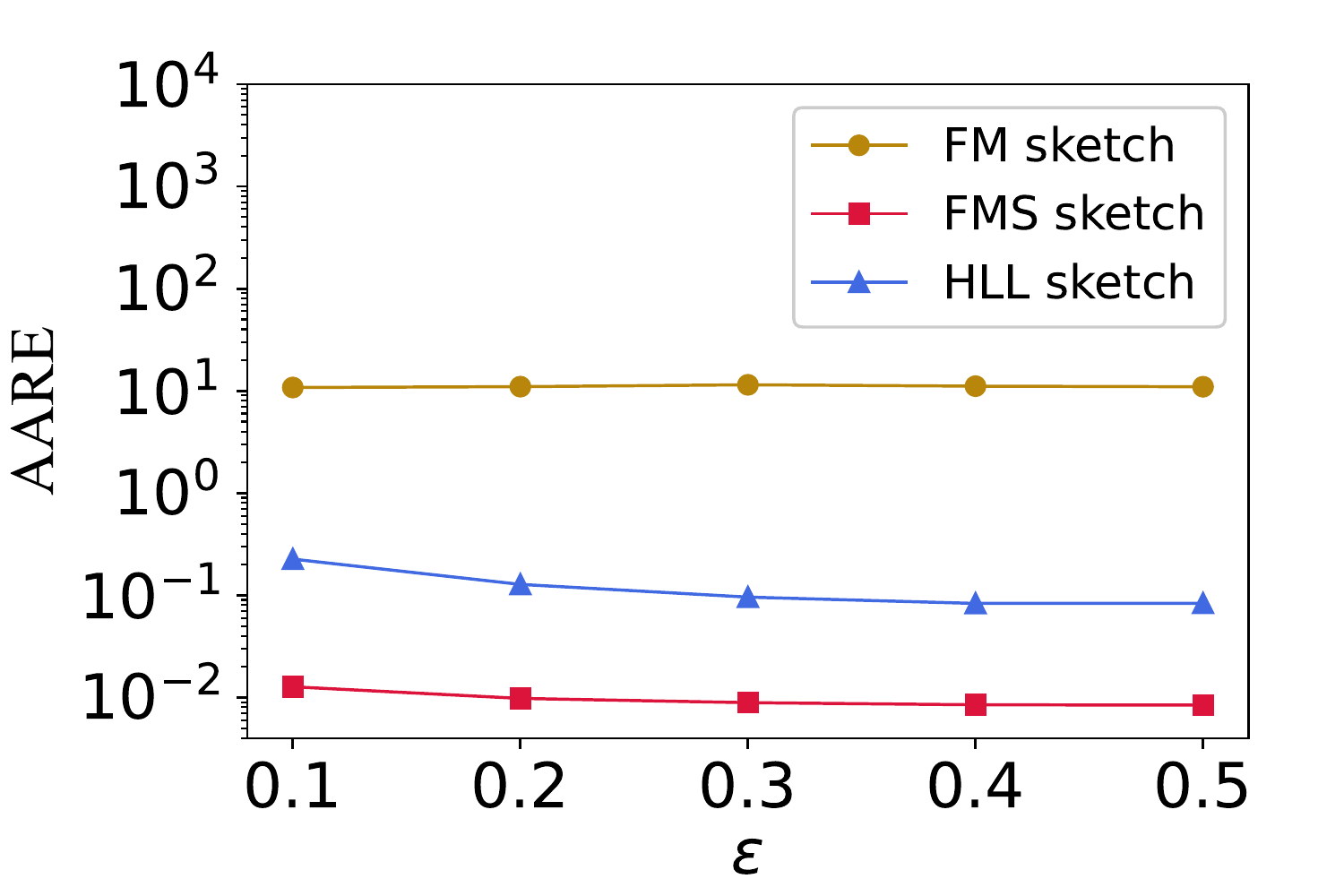}} \\
  \subfloat[AARE vs. $m$, where $n$ = $10^3$]{%
    \includegraphics[width=.25\linewidth]{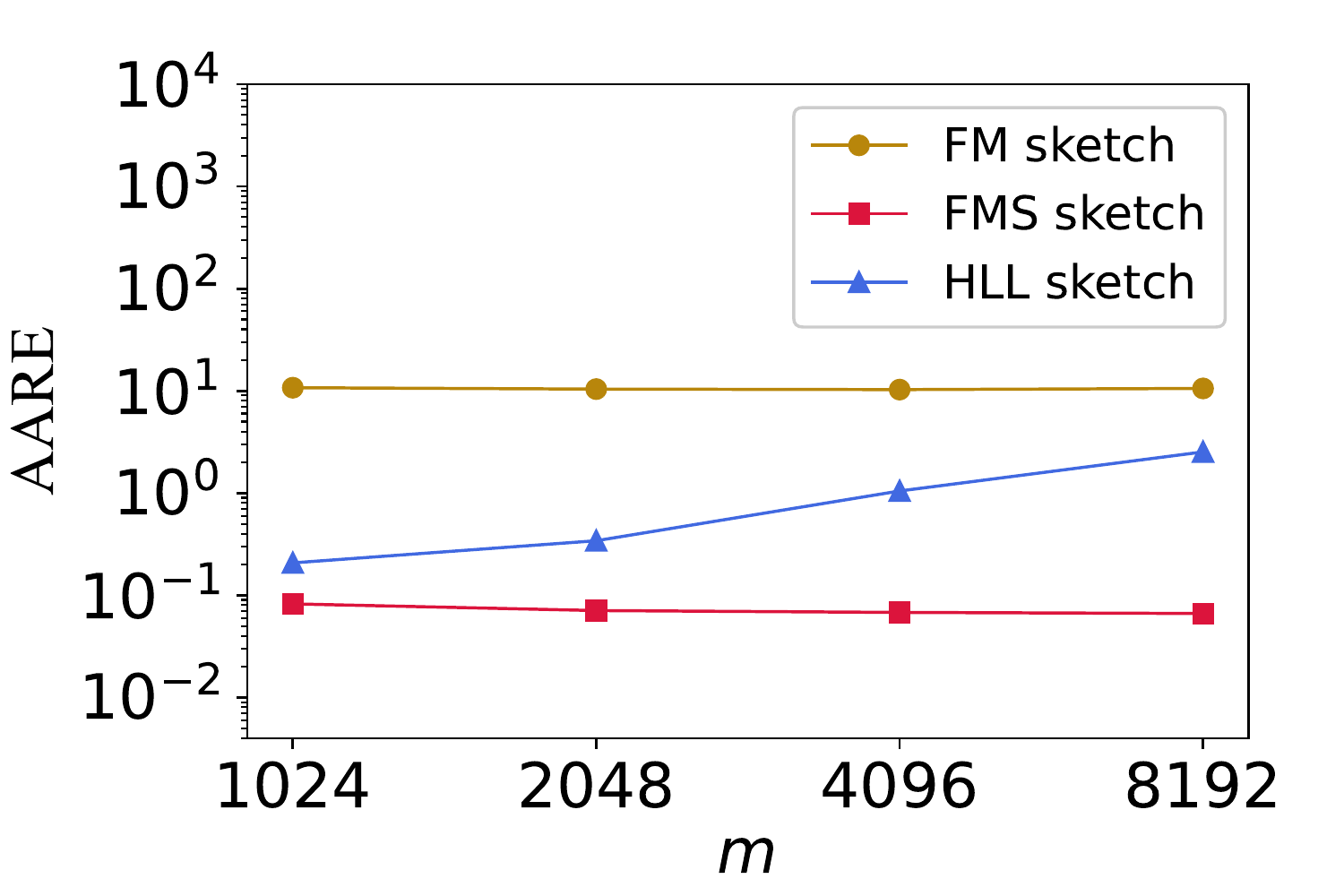}}
  \subfloat[AARE vs. $m$, where $n$ = $10^5$]{%
    \includegraphics[width=.25\linewidth]{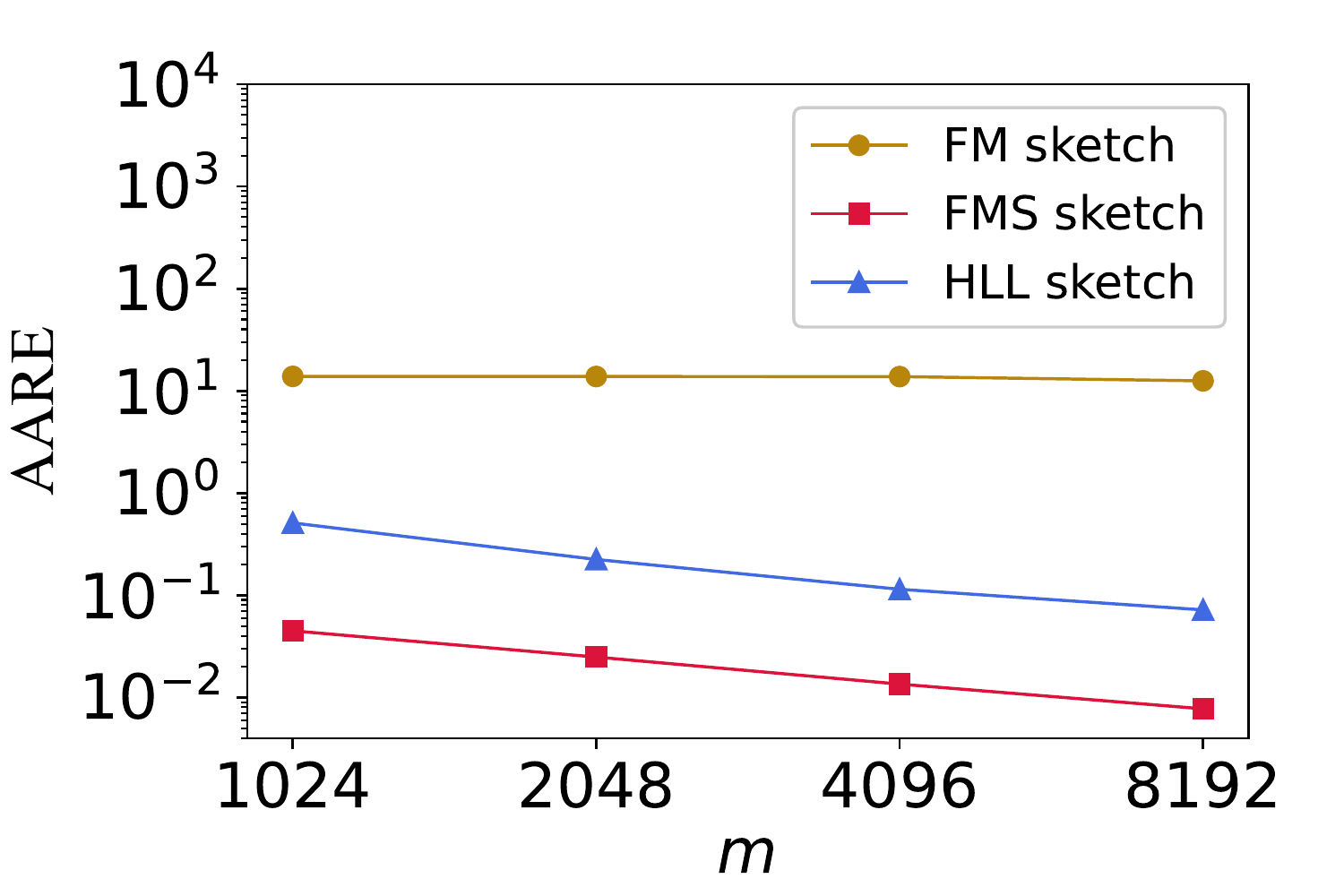}} 
  \subfloat[AARE vs. $m$, where $n$ = $10^7$]{%
    \includegraphics[width=.25\linewidth]{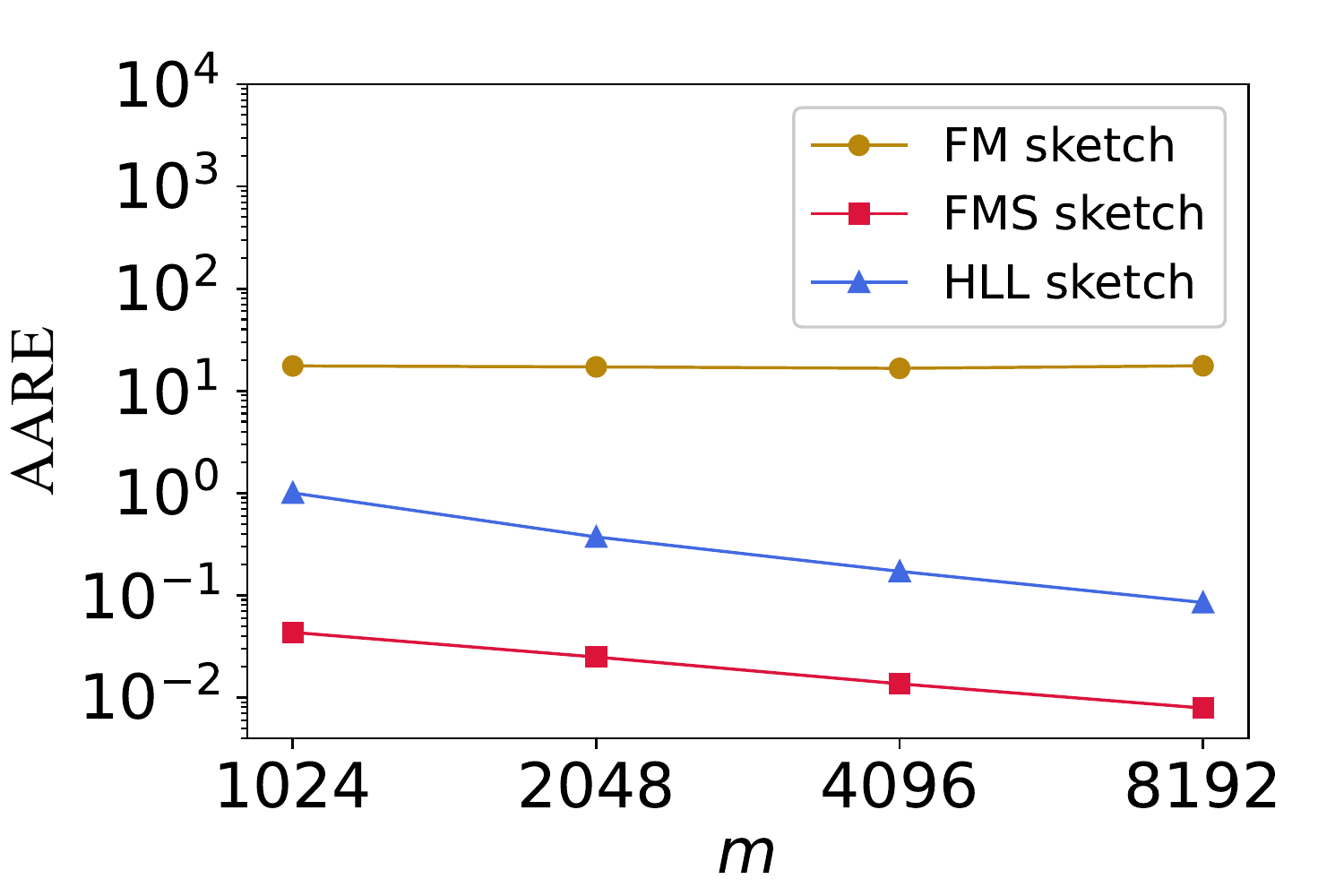}}
  \subfloat[AARE vs. $m$, where $n$ = $10^9$]{%
    \includegraphics[width=.25\linewidth]{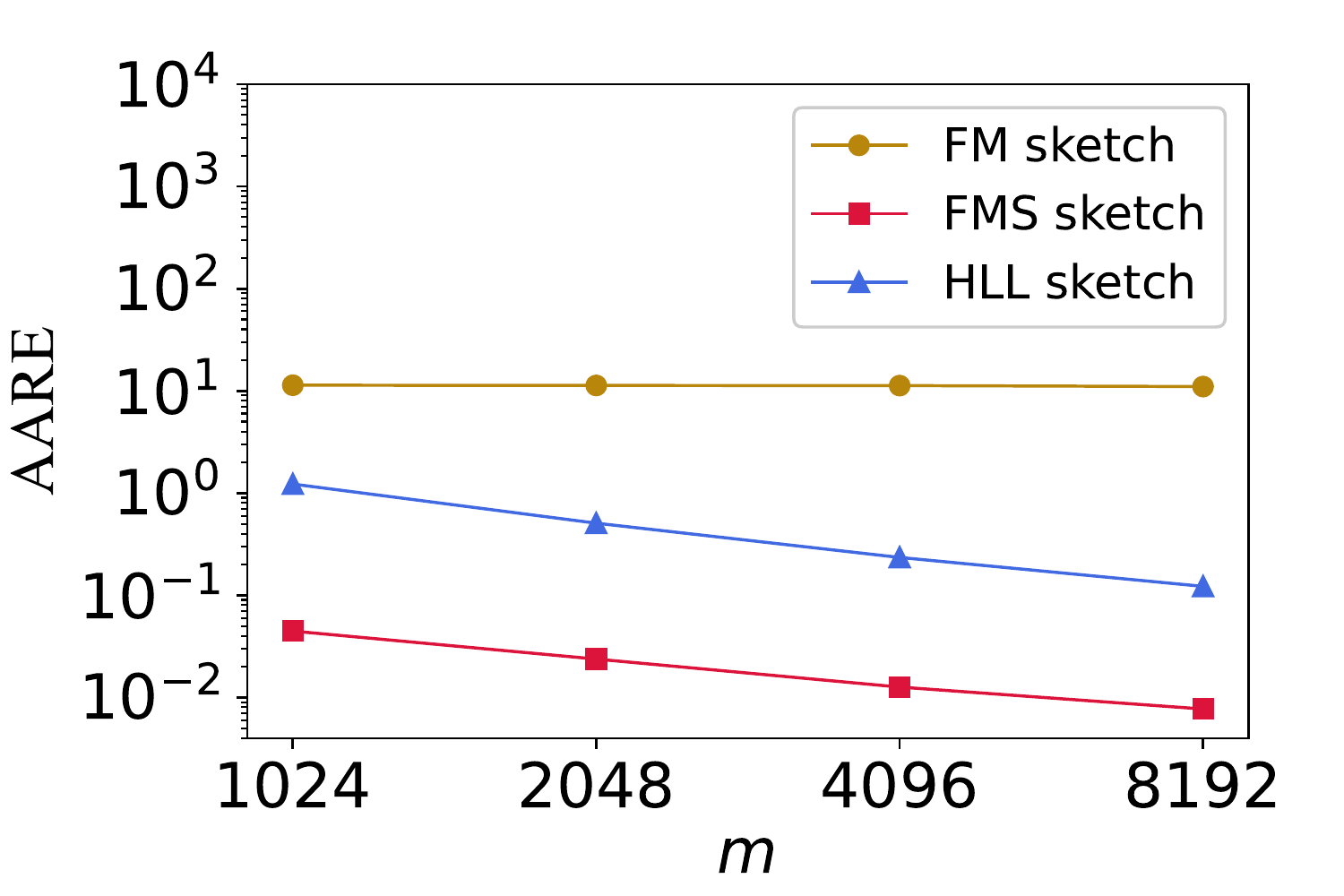}}
  \caption{Accuracy of our FMS sketch compared with FM sketch and HLL sketch in central
    differential privacy settings.}
  \label{fig:central}
\end{figure*}

\subsection{Protocol Complexities}

Compared with the-state-of-art MPC-FM~\cite{Hu0LGWGLD21}, our DP-DICE uses the FMS
sketch while the protocol of MPC-FM uses the FM sketch.
Therefore, our DP-DICE requires the same memory usage on each DH as the MPC-FM
protocol, while it reduces the DP's computational complexity of processing an
element from $O(m)$ to $O(1)$, where $m$ is often set to thousands.
In addition, our DP-DICE also reduces the computation on the CPs because the
variable $Z$ (defined in \cref{eq:Z}) used by our DP-DICE is simpler to compute on
the MPC framework than the variable $Z^*$ (defined in \cref{eq:fmz}) required for
MPC-FM.
From our later experiments, we observe that the network communion cost is also
significantly reduced due to this simplification.

\pdfoutput=1

\section{Evaluation} \label{sec:results}
We perform experiments to evaluate the performance of our
protocol DP-DICE in comparison with state-of-the-art MPC-FM~\cite{Hu0LGWGLD21}
and PSC~\cite{fenske2017distributed} of which codes are publicly available.
Following MPC-FM,
we implemented a prototype of our protocol DP-DICE in C++.
For facilitating future research, we will release our code to the public.

\pdfoutput=1
\subsection{Experimental Setup}

\bullethdr{Parameter Settings}.
We compare FMS sketch with FM sketch and HyperLogLog sketch in the same number of bit arrays (or registers) $m$. As discussed in ~\cite{flajolet1985probabilistic}, it would be enough to set $w$ for the FM sketch to be $\lceil \log_2 n+6 \rceil $. For the HLL sketch and FMS sketch, it would suffice to set $w$ to be $\lceil \log_2(n/m)+6 \rceil $ since each register (or bit array) is expected to receive $n/m$ elements.
In evaluating the accuracy of FMS sketch, we vary the differential privacy parameter $\varepsilon \in \{0.1, 0.2, 0.3, 0.4, 0.5 \}$, the bit arrays number $m \in \{1024, 2048, 4096, 8192 \}$ and the number of DHs $d \in \{5, 10, 15, 20, 25 \}$.
In evaluating the running time and communication cost of our DP-DICE, We vary the number of CPs, i.e., $c\in \{2, 3, \dots, 7\}$.
We set the default the number of DHs $d=20$, the number of CPs $c=5$,
the cardinality value $n=10^6$, the number of sketches $m=4096$, and the
privacy parameters $\varepsilon$ and the corresponding $\varepsilon_d$ to be 0.1 and 0.012,
respectively.
We set these defaults as experimental parameters unless otherwise stated.

\bullethdr{Computing and Network Environment.} We performed extensive experiments in both LAN (Local Area Network) and WAN (Wide Area Network) environments. In the WAN environment, CPs and DHs are evenly distributed in three different Huawei data centers located in Beijing, Guangzhou, and Shanghai, China, respectively. Huawei cloud servers use ECS instances of type c7.8xlarge.2, where each instance has 32 vCPUs (16 physical cores) based on Intel Ice Lake 3.0GHz series CPUs, 64GB of RAM, and a maximum bandwidth of 30 Gbit/s per network interface. 
In the LAN environment, each CP works on a Linux server with 2 Quad-Core Intel(R) Xeon(R) Gold 6140 CPUs @ 2.30GHz processors,
and each DH runs on a desktop with typical hardware configurations of Intel Quad-Core i7-10400F CPUs and 16GB of RAM.

\bullethdr{Datasets and Metrics}.
We randomly generate a variety of independent sets with sizes in $\{10^3,
10^5, 10^7, 10^9\}$, which are used as the union set $S$.
We use a metric \emph{Average Absolute Relative Error} (AARE) to measure the
accuracy of estimates $\hat n$ with respect to the true cardinality value $n$.
Formally, we define the AARE of $\hat n$ as $\mathbb{E}\left(\frac{|\hat n -
    n|}{n}\right)$.
Besides, we evaluate the running time as well as communication costs between the
DHs and CPs.
By default, all experimental results are calculated from the average of 100
independent runs.

\pdfoutput=1
\subsection{Results}

\begin{figure*}[htp]
  \centering
  \subfloat[AARE vs. $\varepsilon$, where $n = 10^3$]{%
    \includegraphics[width=.25\linewidth]{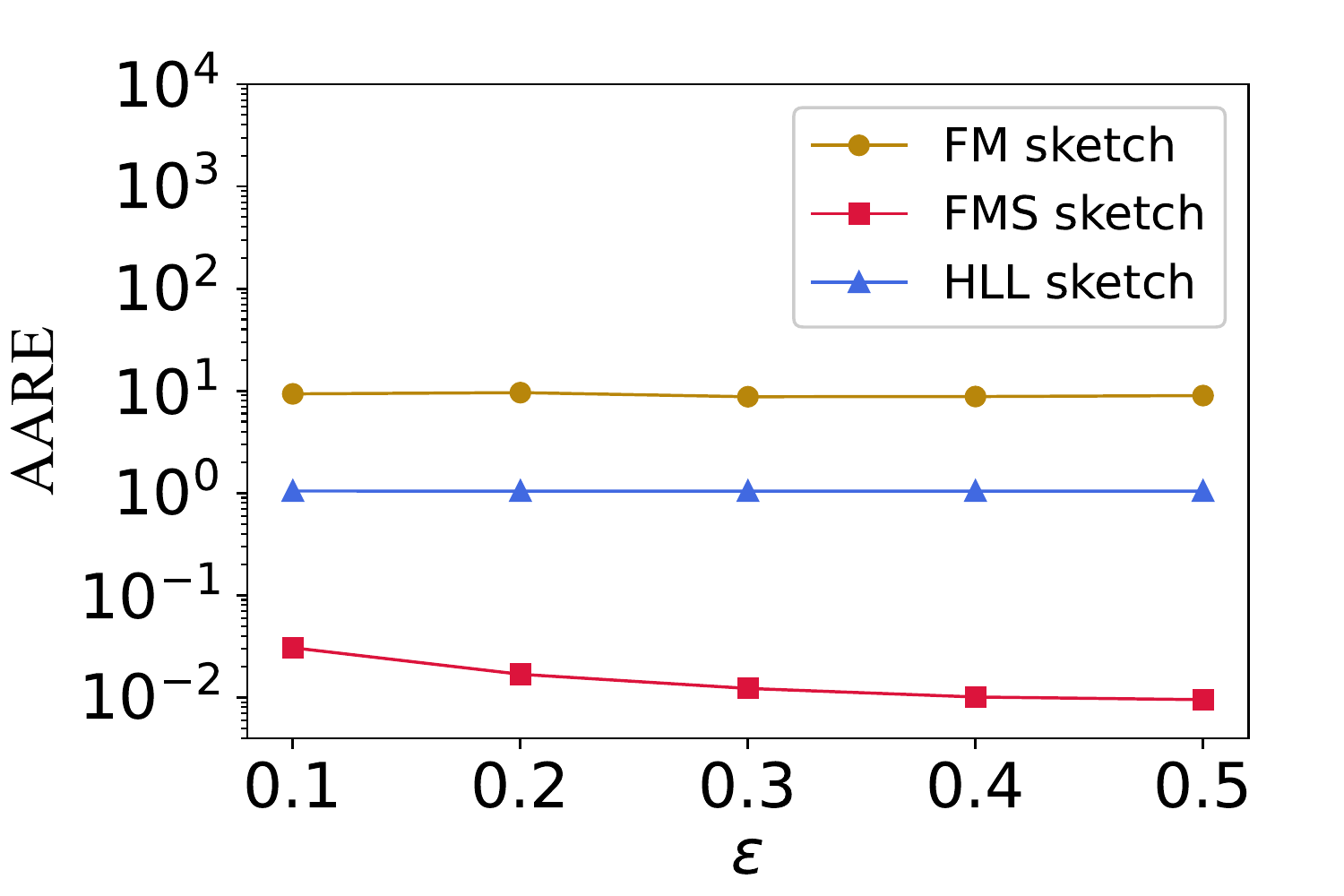}}
  \subfloat[AARE vs. $\varepsilon$, where $n = 10^5$]{%
    \includegraphics[width=.25\linewidth]{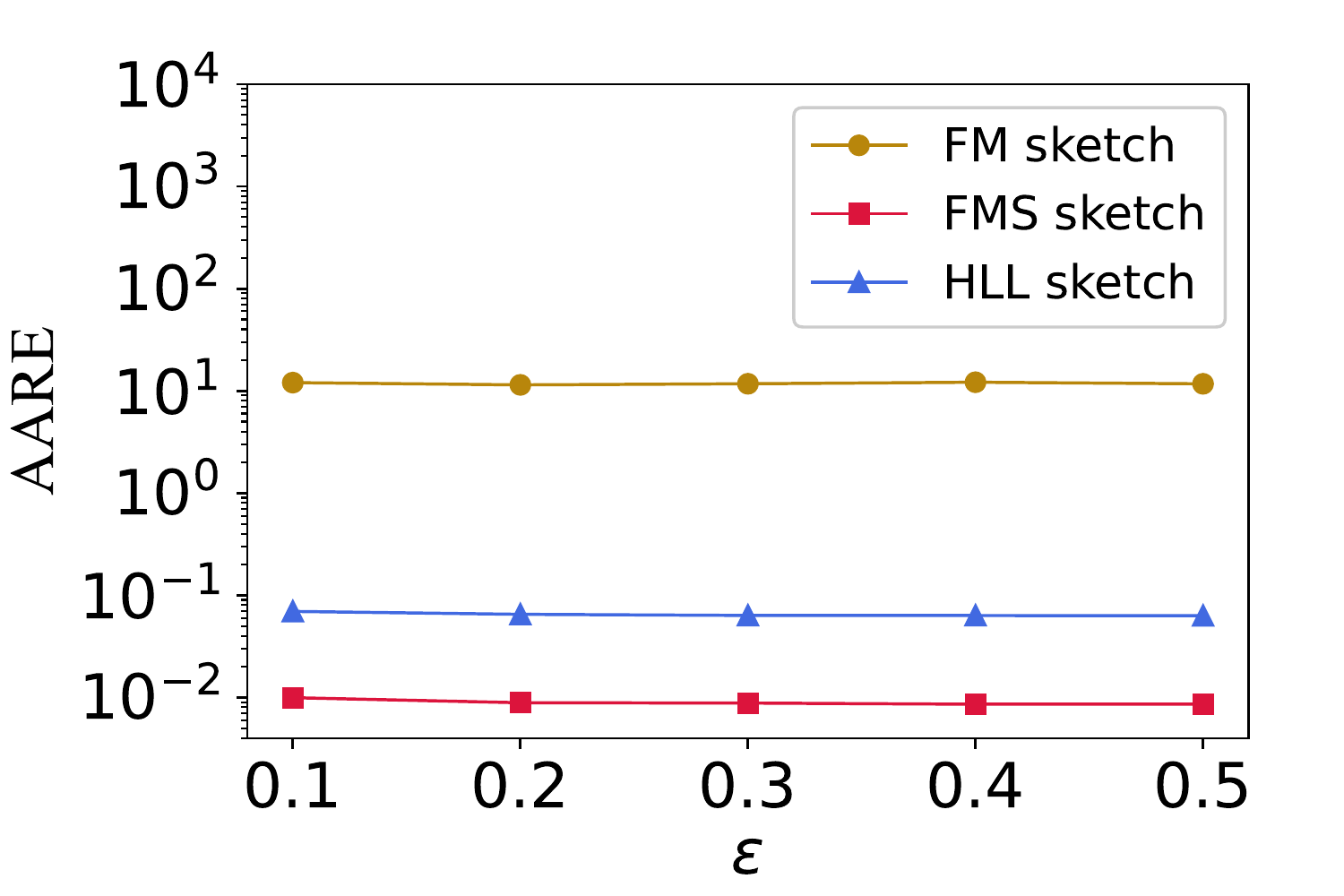}}
  \subfloat[AARE vs. $\varepsilon$, where $n = 10^7$]{%
    \includegraphics[width=.25\linewidth]{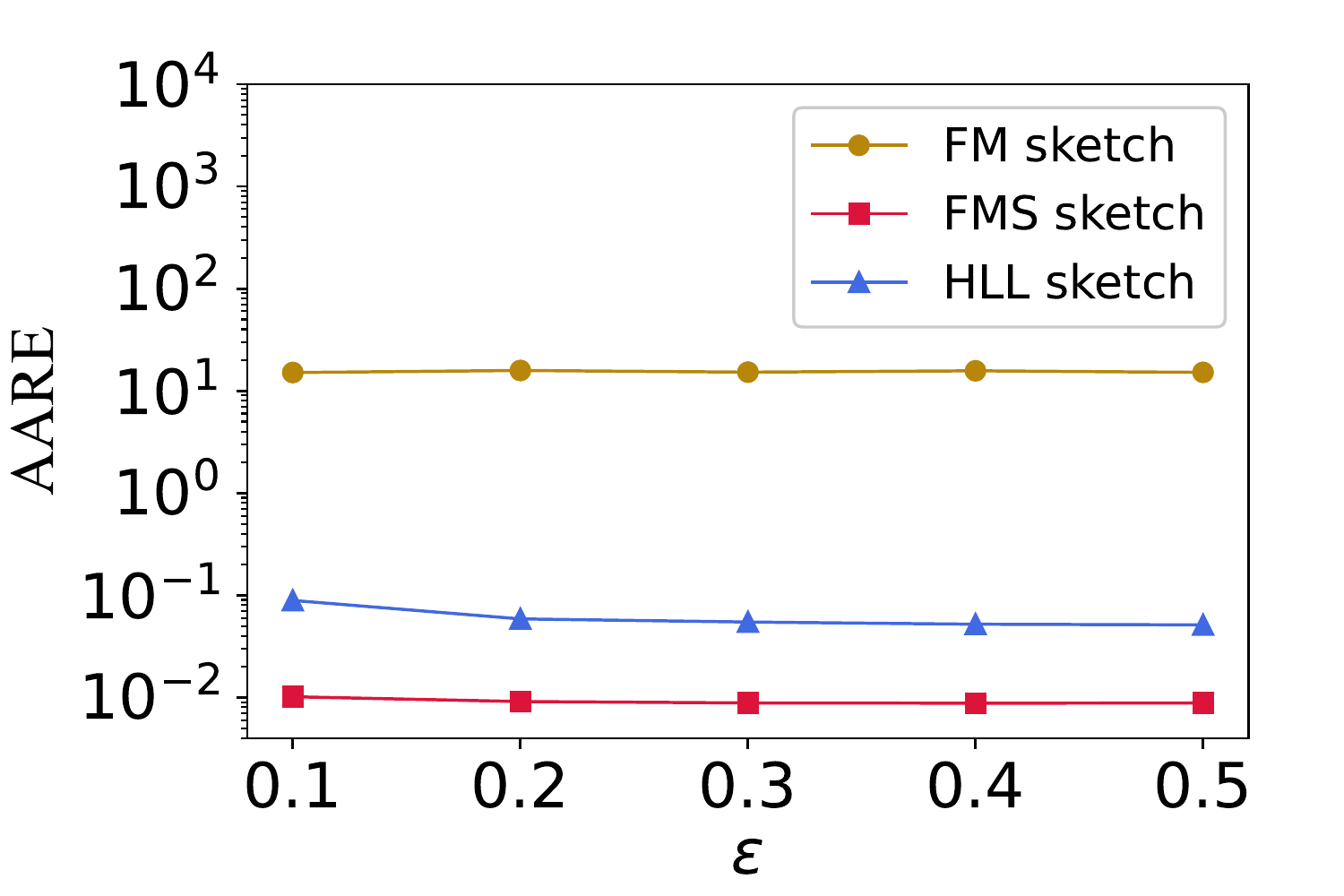}}
  \subfloat[AARE vs. $\varepsilon$, where $n = 10^9$]{%
    \includegraphics[width=.25\linewidth]{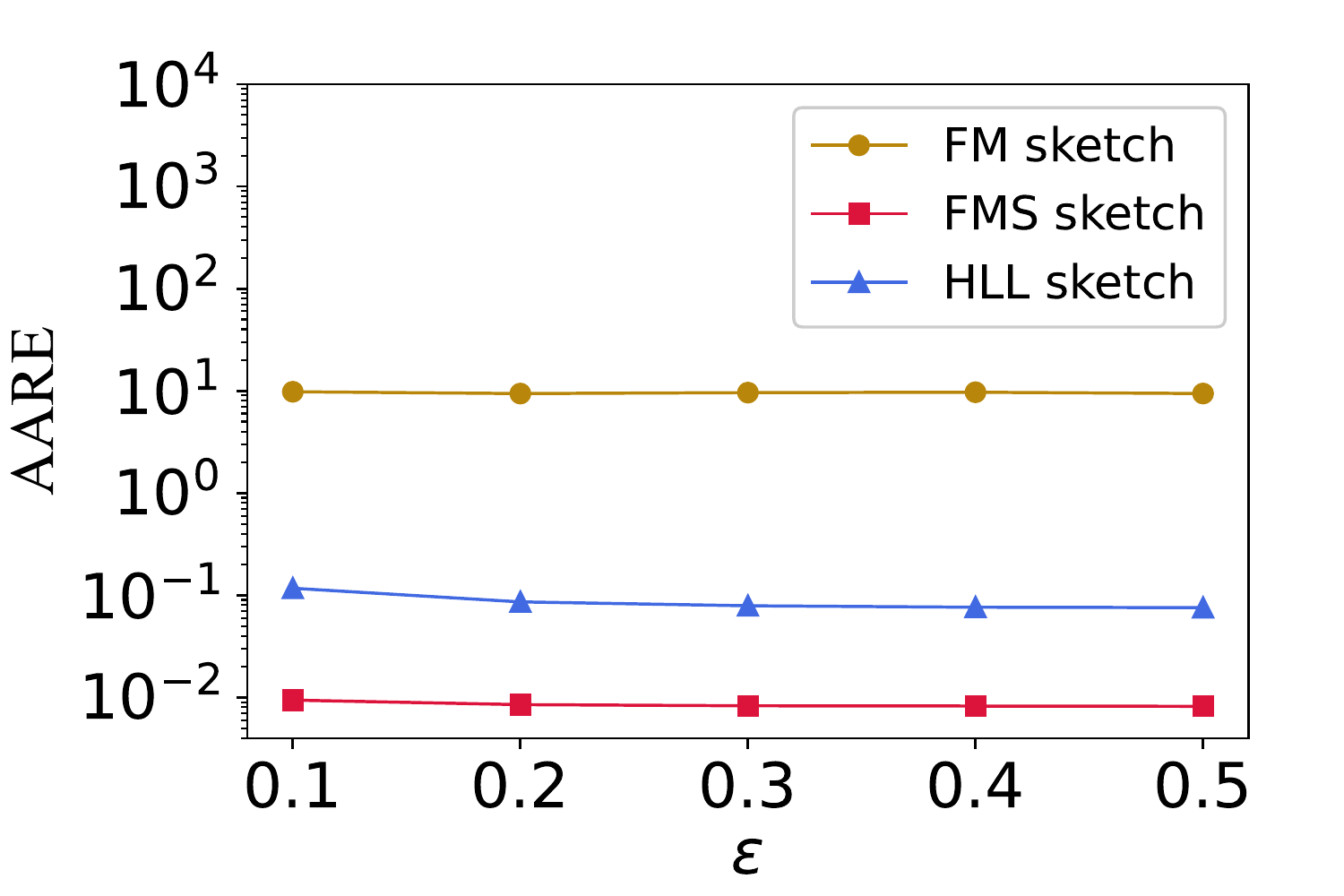}}\\
  \subfloat[AARE vs. $m$, where $n = 10^3$]{%
    \includegraphics[width=.25\linewidth]{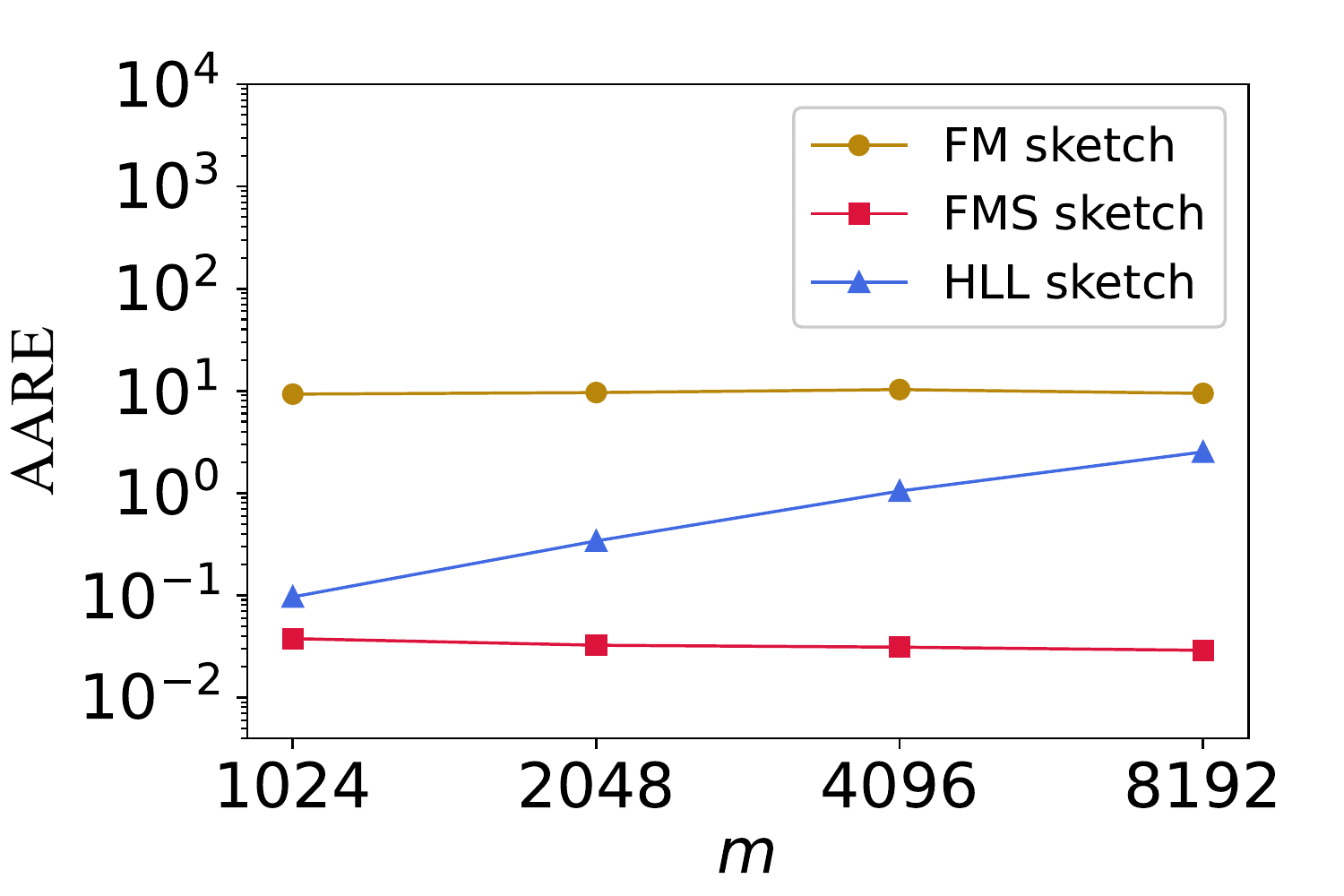}}
  \subfloat[AARE vs. $m$, where $n = 10^5$]{%
    \includegraphics[width=.25\linewidth]{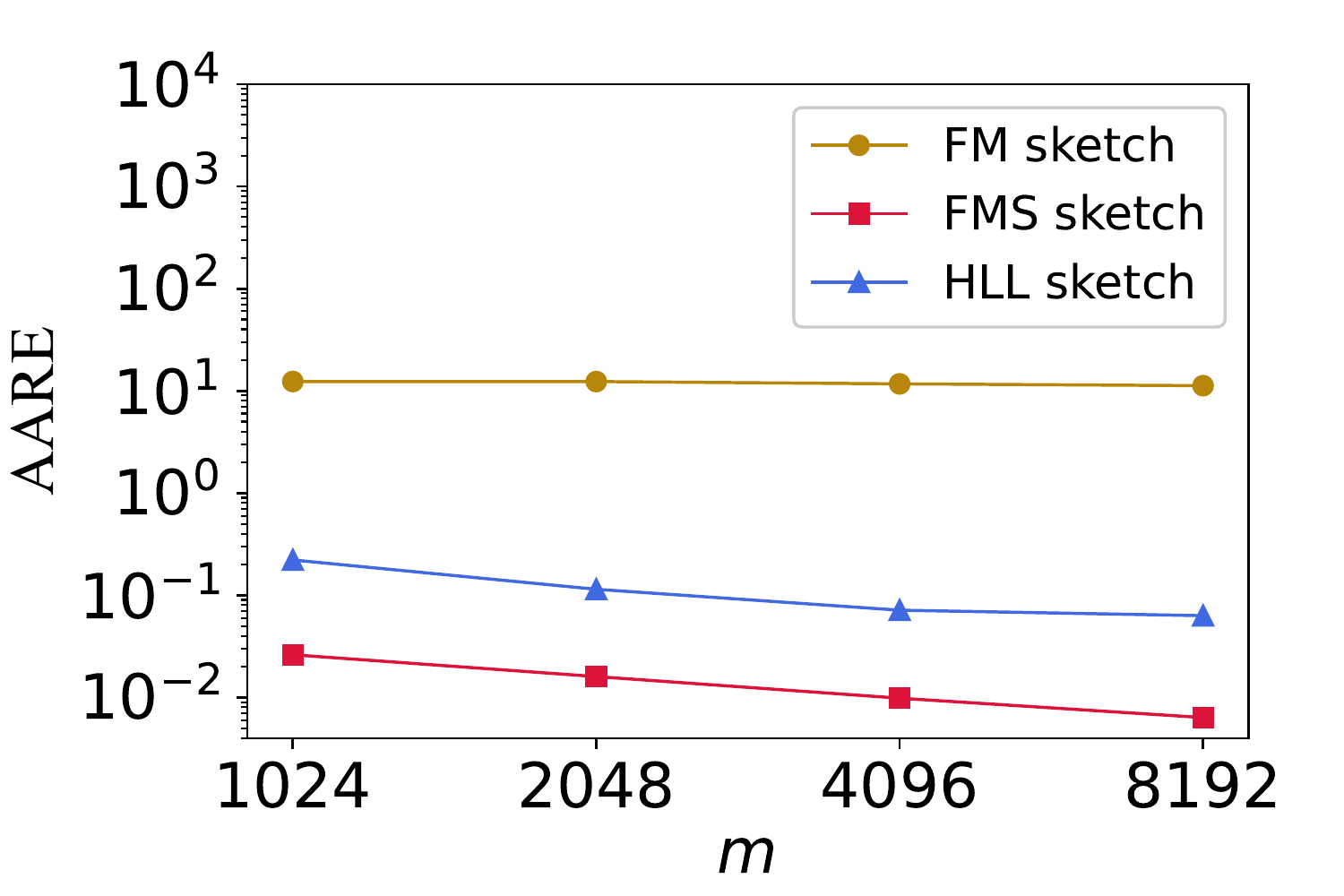}}
  \subfloat[AARE vs. $m$, where $n = 10^7$]{%
    \includegraphics[width=.25\linewidth]{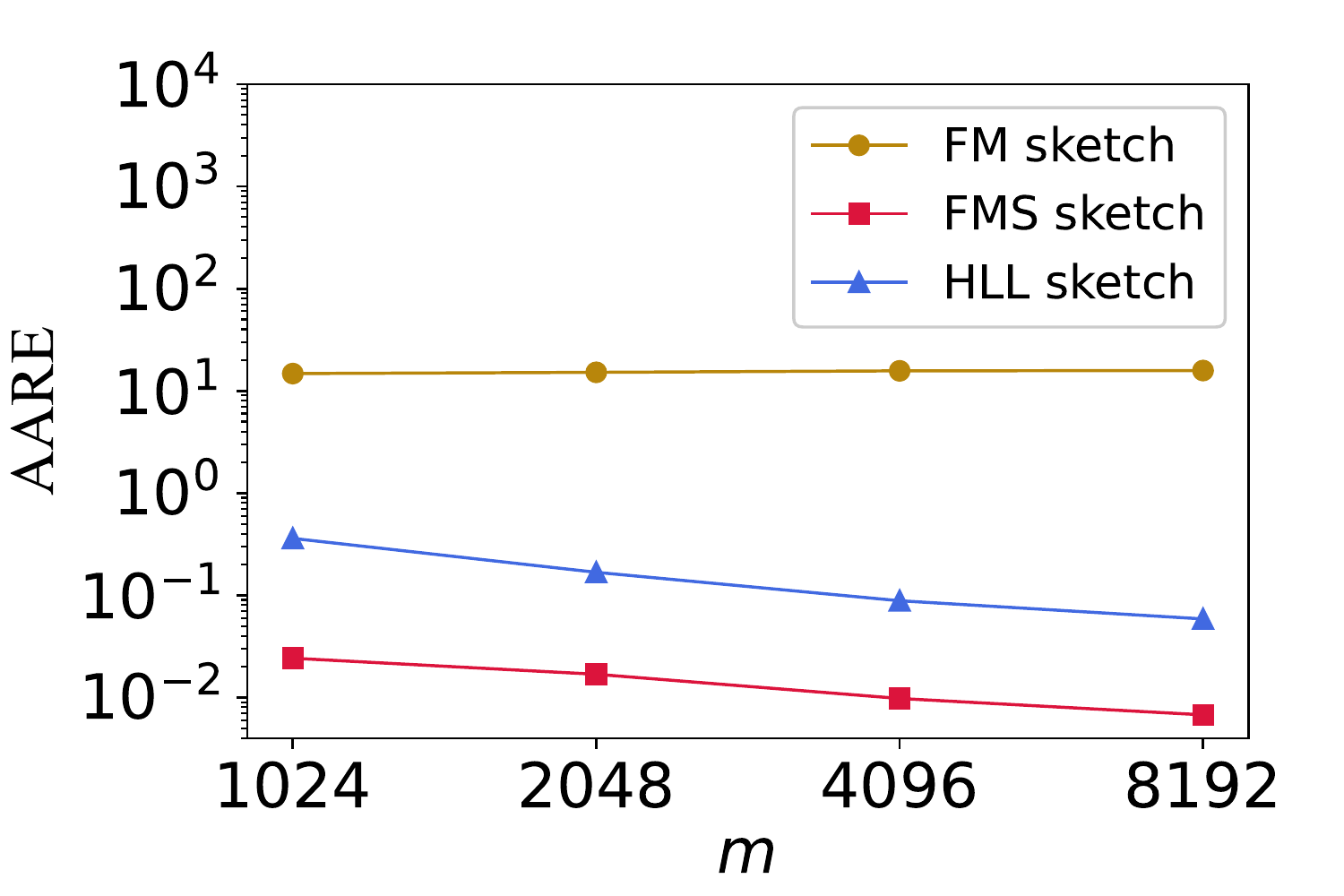}}
  \subfloat[AARE vs. $m$, where $n = 10^9$]{%
    \includegraphics[width=.25\linewidth]{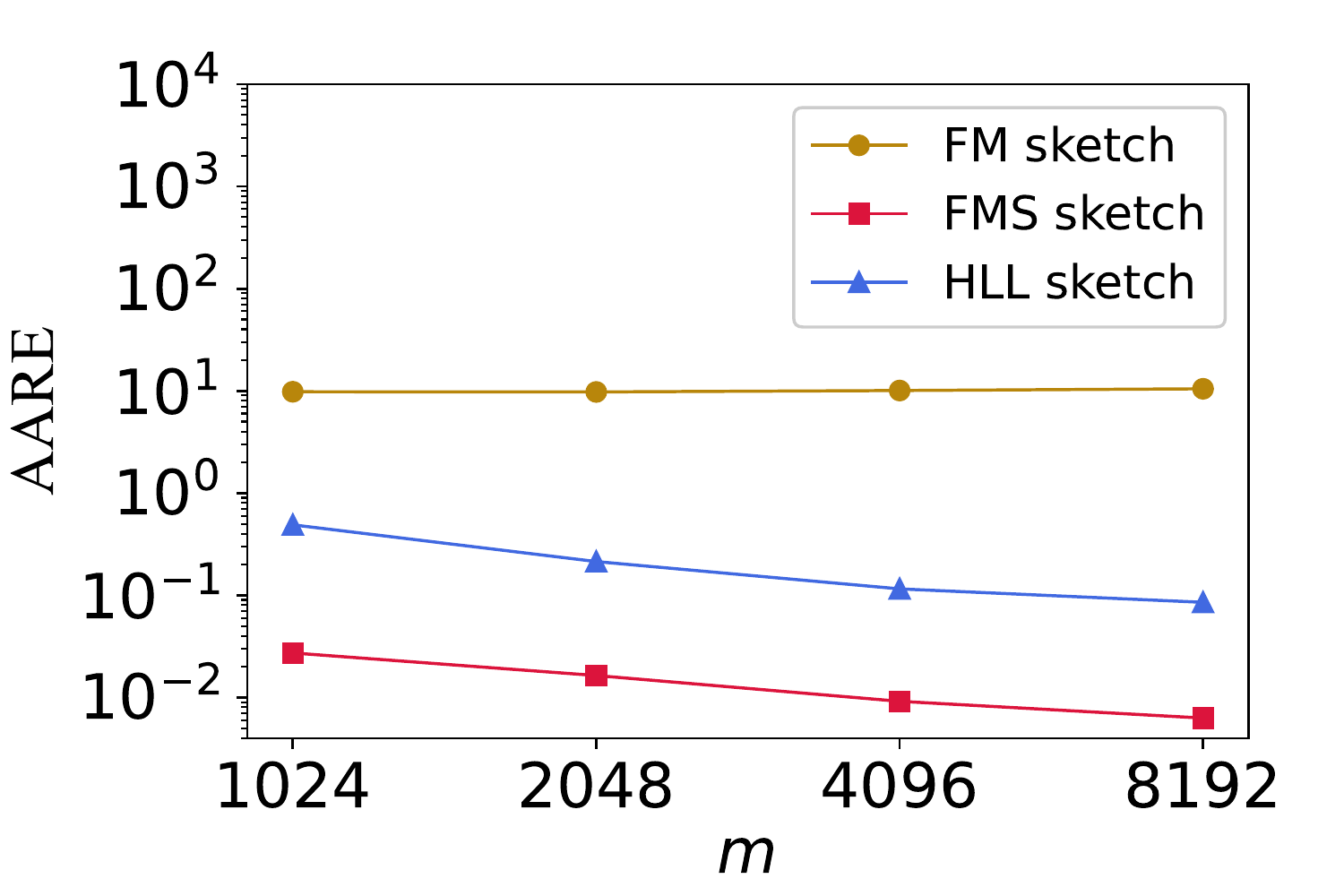}}\\
  \subfloat[AARE vs. No.DHs, where $n = 10^3$]{%
    \includegraphics[width=.25\linewidth]{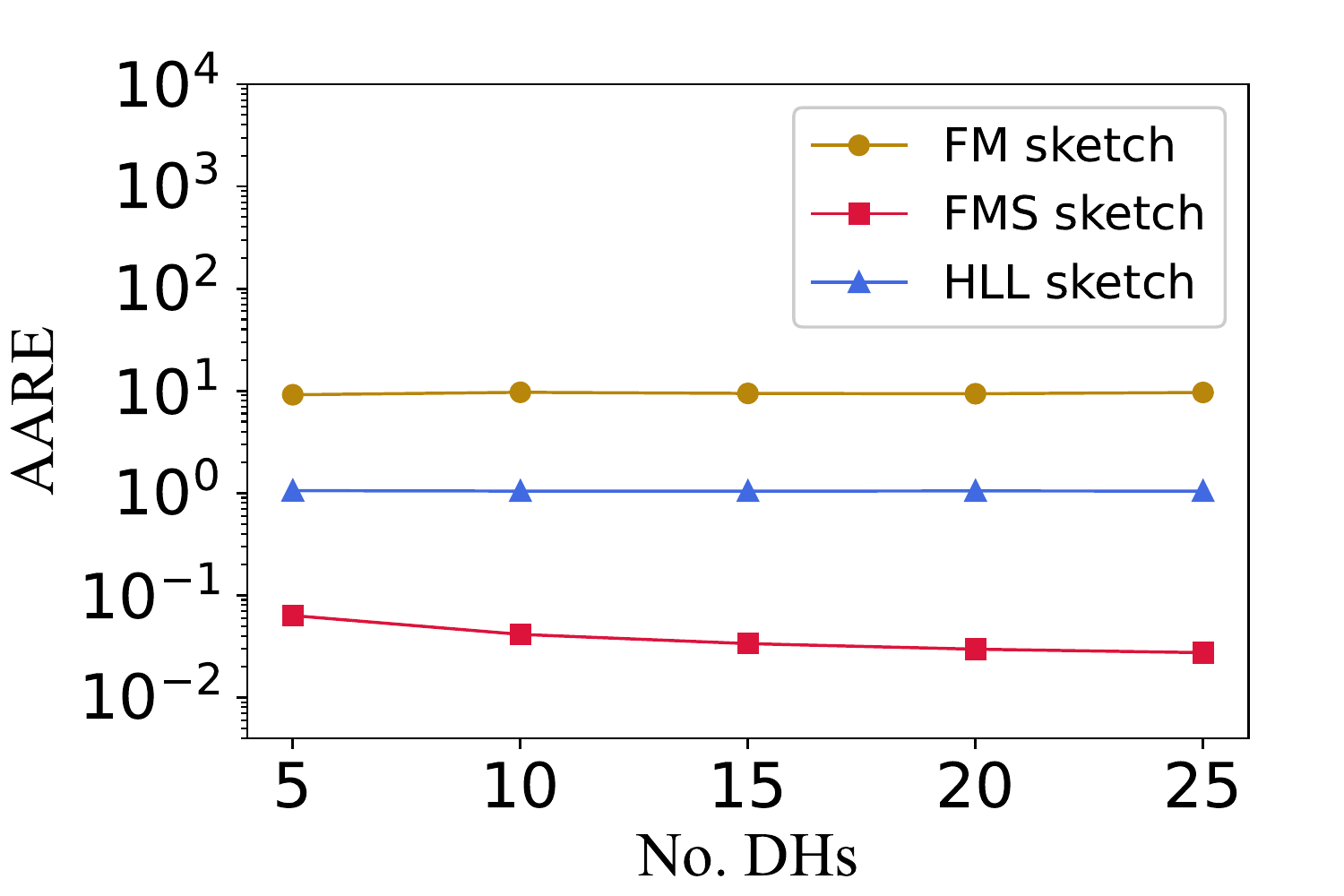}}
  \subfloat[AARE vs. No.DHs, where $n = 10^5$]{%
    \includegraphics[width=.25\linewidth]{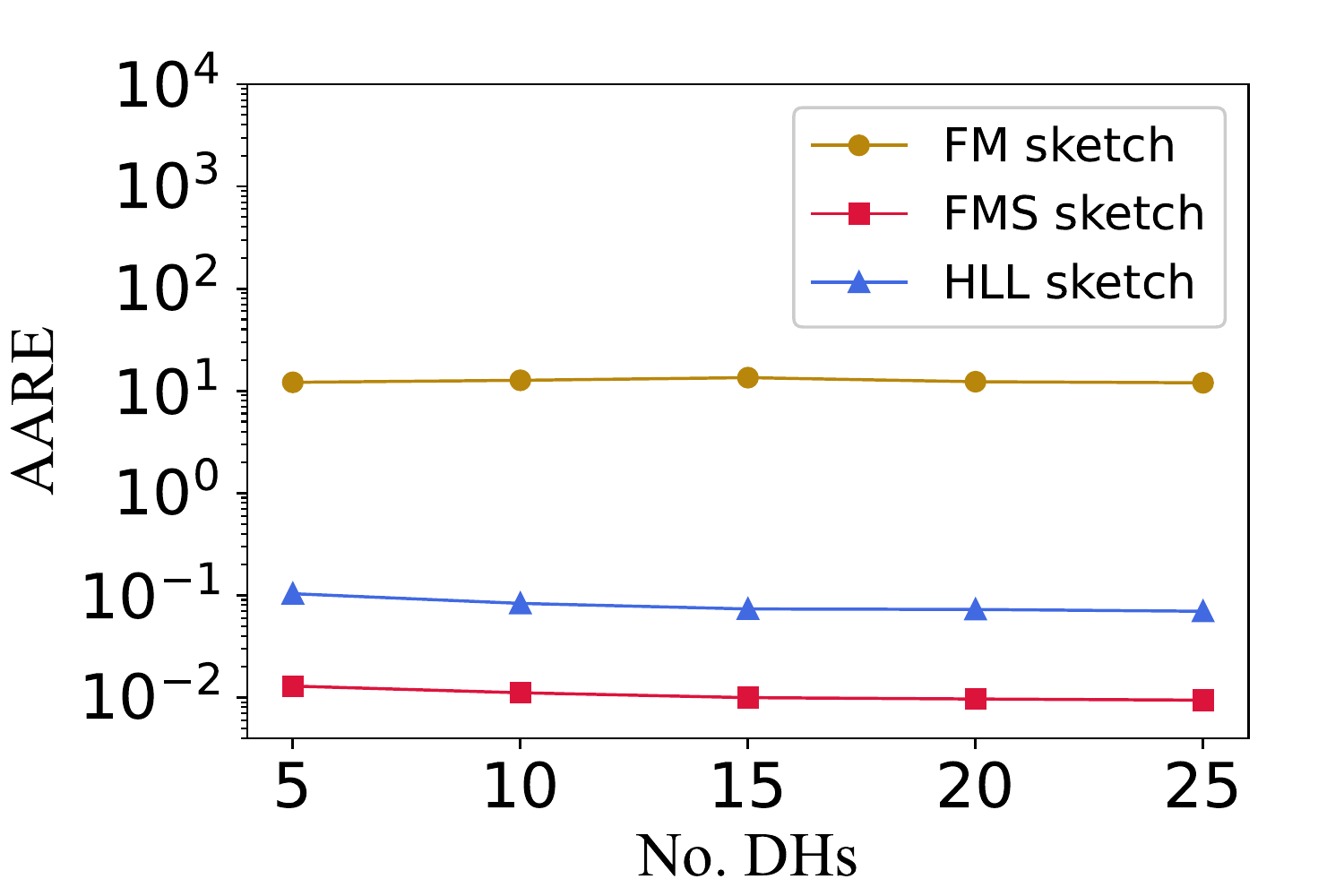}}
  \subfloat[AARE vs. No.DHs, where $n = 10^7$]{%
    \includegraphics[width=.25\linewidth]{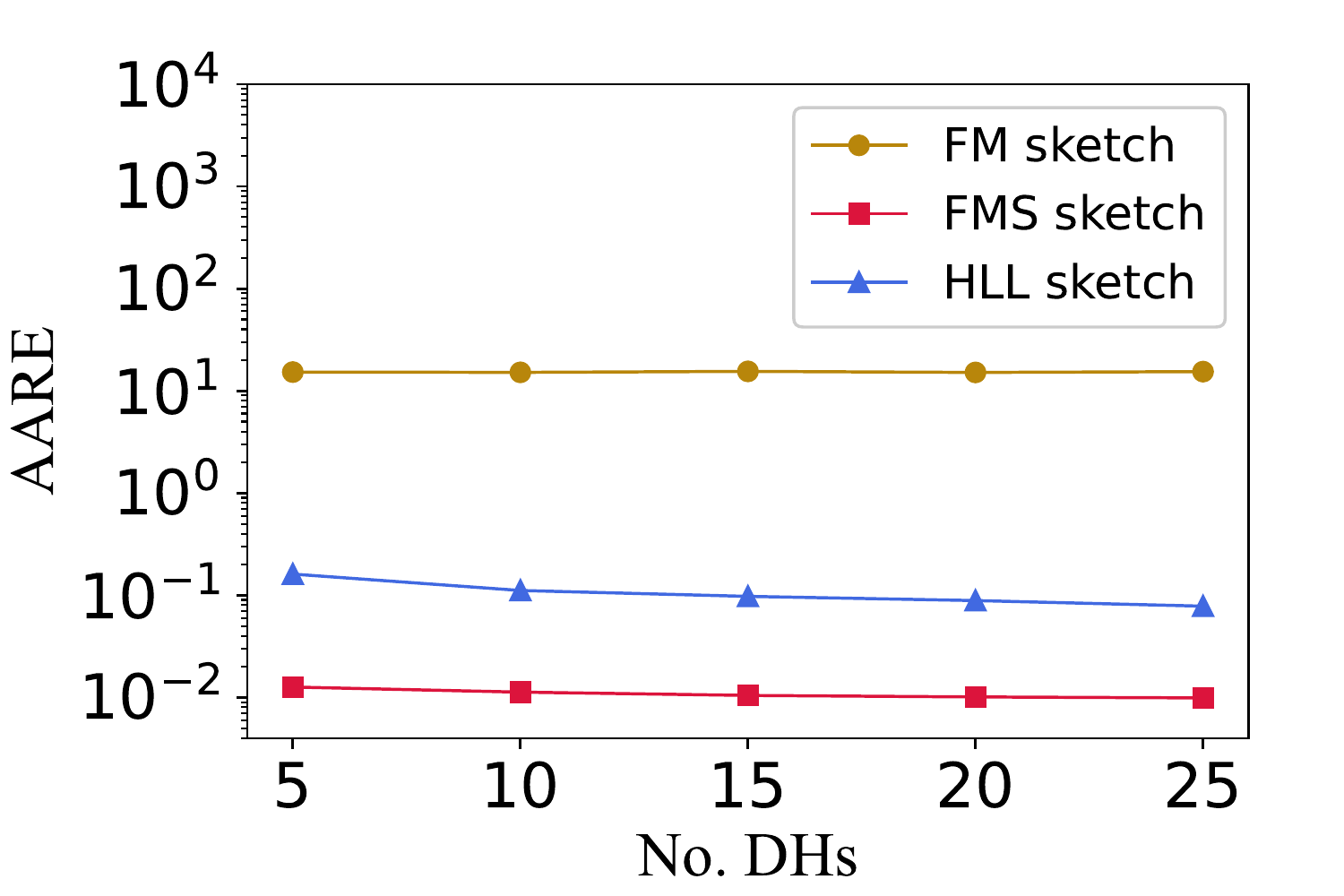}}
  \subfloat[AARE vs. No.DHs, where $n = 10^9$]{%
    \includegraphics[width=.25\linewidth]{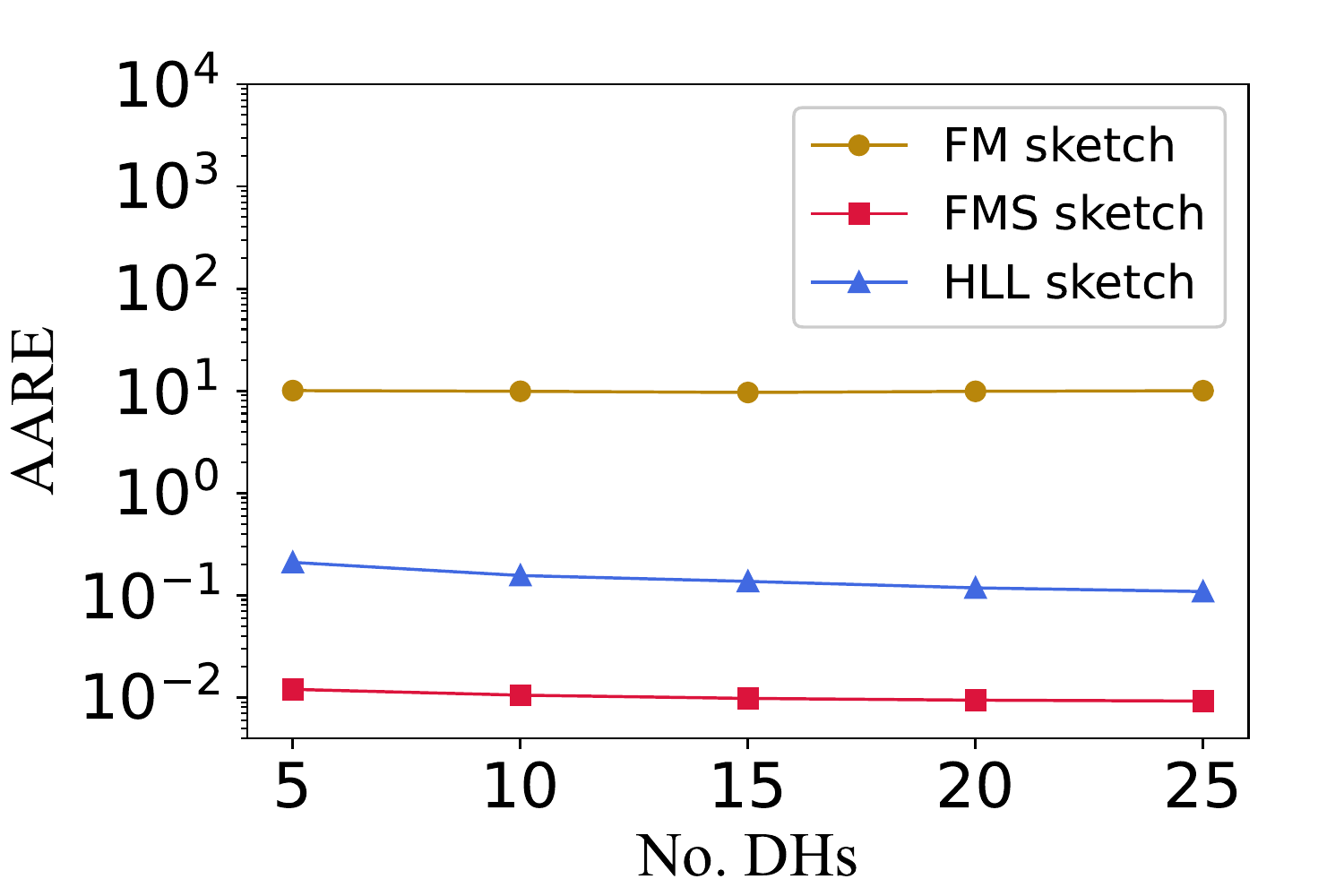}}
  \caption{Accuracy of our FMS sketch compared with FM sketch
    and HLL sketch in distributed differential privacy settings.}
  \label{fig:local}
\end{figure*}

\subsubsection{FMS vs. FM \& HLL}
In \cref{fig:pure_sketch}, we show the accuracy of our FMS sketch in comparison with the FM and HLL sketches when privacy is not considered. We compare sketch methods in different $m$ (the number of registers, chosen from $\{1024, 2048, 4096, 8192\}$) and different cardinality scale $n \in \{10^3, 10^5, 10^7, 10^9\}$. As we discussed in Section~\ref{sec:FM},~\ref{sec:HLL}, and~\ref{sec:FMS}, these sketch methods all have standard errors proportional to $1/\sqrt{m}$. When the cardinality scale is fixed, their AAREs decrease as $m$ increases, which corresponds well with the theoretical analyses. 
Our FMS sketch always has lower AAREs than the others.
In \cref{fig:pure_sketch}(a), $n=10^3$, when $m$ is $1024, 2048, 4096, 8192$, the AARE of our FMS sketch is  1.7, 1.7, 1.5, 1.2 times smaller than that of the FM sketch and is 2.0, 1.8, 1.6, 1.3 times smaller than that of the HLL sketch. When $n=10^5$ and $m$ is $1024, 2048, 4096, 8192$ respectively, the AARE of the FMS sketch is 1.3, 1.3, 1.2, 1.2 times smaller than the FM sketch and is 1.9, 1.9, 1.9, 2.0 times smaller than the HLL sketch. A similar pattern holds for $n\in \{10^7, 10^9\}$. 

We would like to point out that HLL requires 5 times less memory space than the FMS and FM sketch when using the same $m$. 
Under the same memory usage, our FMS is more accurate than the FM sketch but less accurate than the HLL sketch.
Despite this, as we discussed later, the memory usage for the sketch in our protocol is not the bottleneck.

\subsubsection{DP-DICE vs. MPC-FM}
We then demonstrate the accuracy of our proposed FMS sketch when adding an additive
noise variable for satisfying differential privacy, and then verify the efficiency of our DP-DICE protocol in comparison with MPC-FM.

\header{Accuracy}.
\cref{fig:central} shows the accuracy of our FMS sketch in comparison with the FM and HLL sketches
under the same differential privacy.
As we introduced in \cref{sec:preliminary}, the estimator of HLL has complex power operations, which limits its implementation on the SPDZ framework.
Therefore, we use the estimator given in the early version of HLL (i.e.,
LogLog sketch~\cite{DurandF03}), that is, $\tilde{n}=\tilde\alpha_m m2^ {Z^\#}$,
where $\tilde\alpha _m \approx 0.783$ and $Z^\#=\frac{1}{m}\sum_{i=1}^mR[i]$.
We directly add additive Gauss noise to HLL's variable $Z^\#$, FM's variable $Z^*$
(defined in \cref{eq:fmz}), and our FMS' variable $Z$ (defined in \cref{eq:Z}) to
achieve $(\varepsilon, \delta)$-differential privacy, where we set $\varepsilon\in \{0.1, 0.2, 0.3, 0.4, 0.5 \}$ and $\delta = 10^{-12}$ following the settings in~\cite{Hu0LGWGLD21}.

In \cref{fig:central}(a), we see that the AARE of our FMS is 15.5, 30.8, 44.5, 57.0, and 68.2 times
smaller than that of HLL when $n=10^3$ and $\varepsilon=0.1, 0.2, 0.3, 0.4, 0.5$, respectively.
Compared with FM, our FMS reduces the AARE by 155, 308, 454, 575, and 700 times when $\varepsilon= 0.1, 0.2, 0.3, 0.4, 0.5$, respectively. When $n=10^5$, as shown in \cref{fig:central}(b), the AARE of our FMS is 7.6, 7.2, 7.2, 6.8, and 7.2 times smaller than that of HLL and 963, 1337, 1461, 1498, and 1617 times smaller than that of FM. Similar patterns hold when $n=10^7$ and $n=10^9$. FM always gets high AAREs since its variable $Z^*$ has sensitivity $\Delta_{Z^*}=w m$ and would need significantly large noise to achieve the same level of differential privacy.

In \cref{fig:central}(e)--(h), we compare the accuracy of FMS, FM, and HLL for different $m$. When $n=10^3$, the AARE is more sensitive to inaccurate estimation and our FMS still has AAREs less than $0.08$, while FM and HLL have very large AAREs. HLL behaves badly for low cardinality scale, for instance, when $n=0$, HLL outputs about $0.7m$. When $n$ is set to $10^5$, $10^7$, and $10^9$, the AARE of FMS is roughly 10 to 15 times smaller than HLL for any $m$ and more than 300 times smaller than FM. 



As we mentioned, in the setting of our problem, it is not easy to directly add continuous noise to our FMS' variable $Z$ on the framework of SPDZ and there is no trusted central server. As a result, we evaluate the AAREs of FMS, FM, and HLL in distributed discrete Gaussian mechanism setting.
In \cref{fig:local}(a)--(d), we compare the AAREs of FMS, FM, and HLL for different values of $\varepsilon$ when using the distributed discrete Gaussian mechanism, where $\varepsilon\in \{0.1, 0.2, 0.3, 0.4, 0.5\}$.
In \cref{fig:local}(a), where $n=10^3$, the AARE of our FMS is 34.4, 63.6, 83.3, 100.6, and 111.1 times smaller than that of HLL and 310, 591, 758, 883, and 931 times smaller than that of FM for $\varepsilon= 0.1, 0.2, 0.3, 0.4, 0.5$, respectively. In \cref{fig:local}(b), where $n=10^5$, compared with HLL, our FMS reduces the AARE by 7.4, 7.2, 7.2, 7.3, and 7.3 times for $\varepsilon= 0.1, 0.2, 0.3, 0.4, 0.5$, respectively. Compared with FM, FMS reduces the AARE by 1198, 1321, 1288, 1425, and 1360 times for $\varepsilon= 0.1, 0.2, 0.3, 0.4, 0.5$, respectively. Also, similar patterns hold for $n=10^7$ and $n=10^9$.

In \cref{fig:local}(e)--(h), we vary $m$ from 1,024 to 8,192 to further study the effect of $m$. In \cref{fig:local}(e), when $n=10^3$, the AARE of FMS is always below 0.038, while FM is constantly larger than 9.
The AARE of HLL even gets larger when $m$ grows for the same reason in the continuous noise setting.
In \cref{fig:local}(f), the AARE of FMS is 8.1, 6.8, 7.4, and 9.8 times smaller than that of HLL and 453, 699, 1203, and 1810 times smaller than that of FM for $m = 1024, 2048, 4096, 8192$, respectively. Besides, the AAREs of FMS and HLL are roughly proportional to $1/\sqrt{m}$. The similar patterns hold for $n=10^7$ and $n=10^9$ in \cref{fig:local}(g)--(h).

In \cref{fig:local}(i)--(l), we vary the number of DHs from 5 to 25 to see whether the AARE of FMS is stable under different numbers of DHs. We set $m= 4096$ and $\varepsilon= 0.1$ by default. As shown in \cref{fig:local}(i), when $n=10^3$, the AARE of FMS reduces from 0.062 to 0.026 as the number of DHs grows from 5 to 25. While the AAREs of FM and HLL are constantly larger than 8 and 1.04 respectively. When $n=10^5$, the AARE of FM slightly reduces from 0.012 to 0.010 when the number of DHs grows from 5 to 25. While FMS reduces the AAREs of HLL and FM by more than 6 and 10 times respectively. Again, the similar patterns hold for $n=10^7$ and $n=10^9$ in \cref{fig:local}(k)--(l).



\header{Memory Cost}.
It is not difficult to find that our DP-DICE has the same memory complexity as the MPC-FM protocol.
Both protocols have online and offline phases,
therefore we record their memory overhead in the online and offline phases, respectively.
The experiments show that both DP-DICE and MPC-FM require a small amount of memory space.
The memory cost of DP-DICE in the offline phase is about 413MB per CP. The protocols in the online phase of DP-DICE are divided into CP execution and DH execution. Each CP execution protocol consumes about 302MB of memory, and each DH execution protocol consumes about 21MB of memory. In a real-world application, the CP is run on the server, and the memory cost of 413MB is not significant for the server. A large part of the protocols' memory overhead is consumed by the SPDZ protocols, while the FMS sketch consumes only a few hundred KB, which is almost negligible. 
Each DH only requires to allocate memory for the FMS sketch and so the memory overhead is also minimal.

\begin{figure}[t]
    \includegraphics[width=0.98\linewidth]{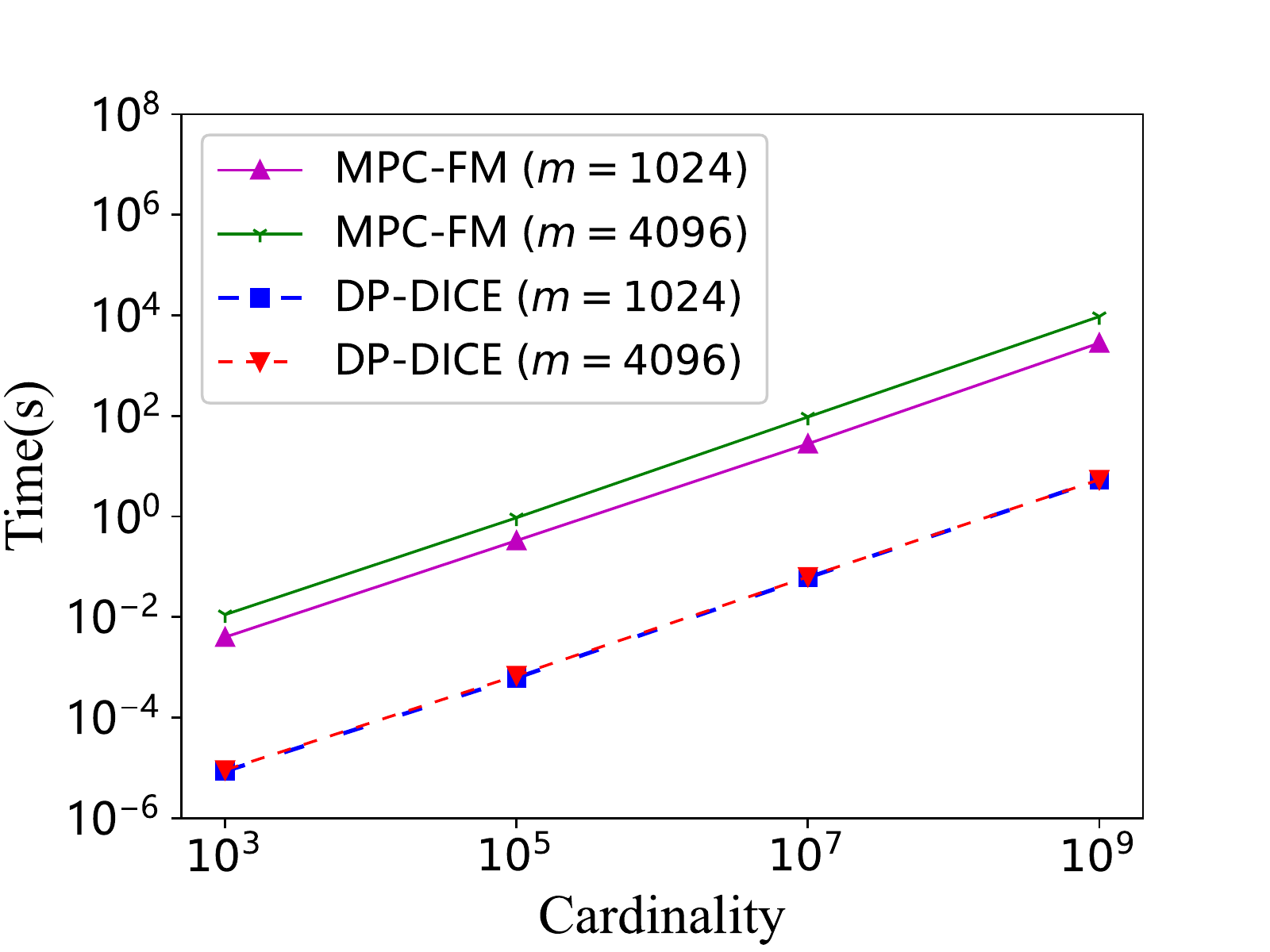}
\caption{CPs' sketch generation time for our DP-DICE.}
  \label{fig:sketchgeneration}
  \end{figure}

\begin{figure*}[htp]
  \subfloat[offline communication costs for different cardinalities]{%
    \includegraphics[width=.25\linewidth]{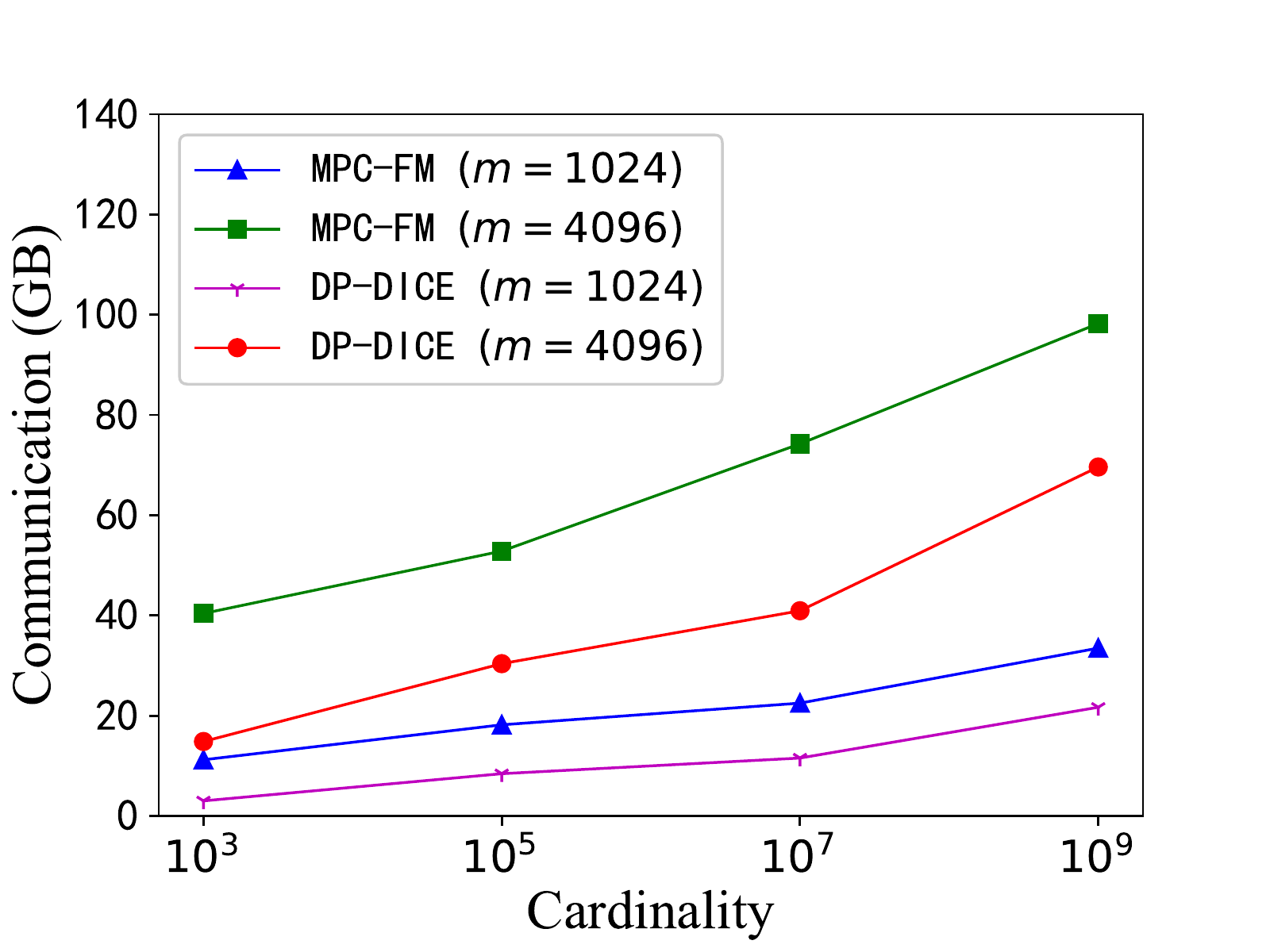}}
  \subfloat[offline communication costs for different numbers of CPs]{%
    \includegraphics[width=.25\linewidth]{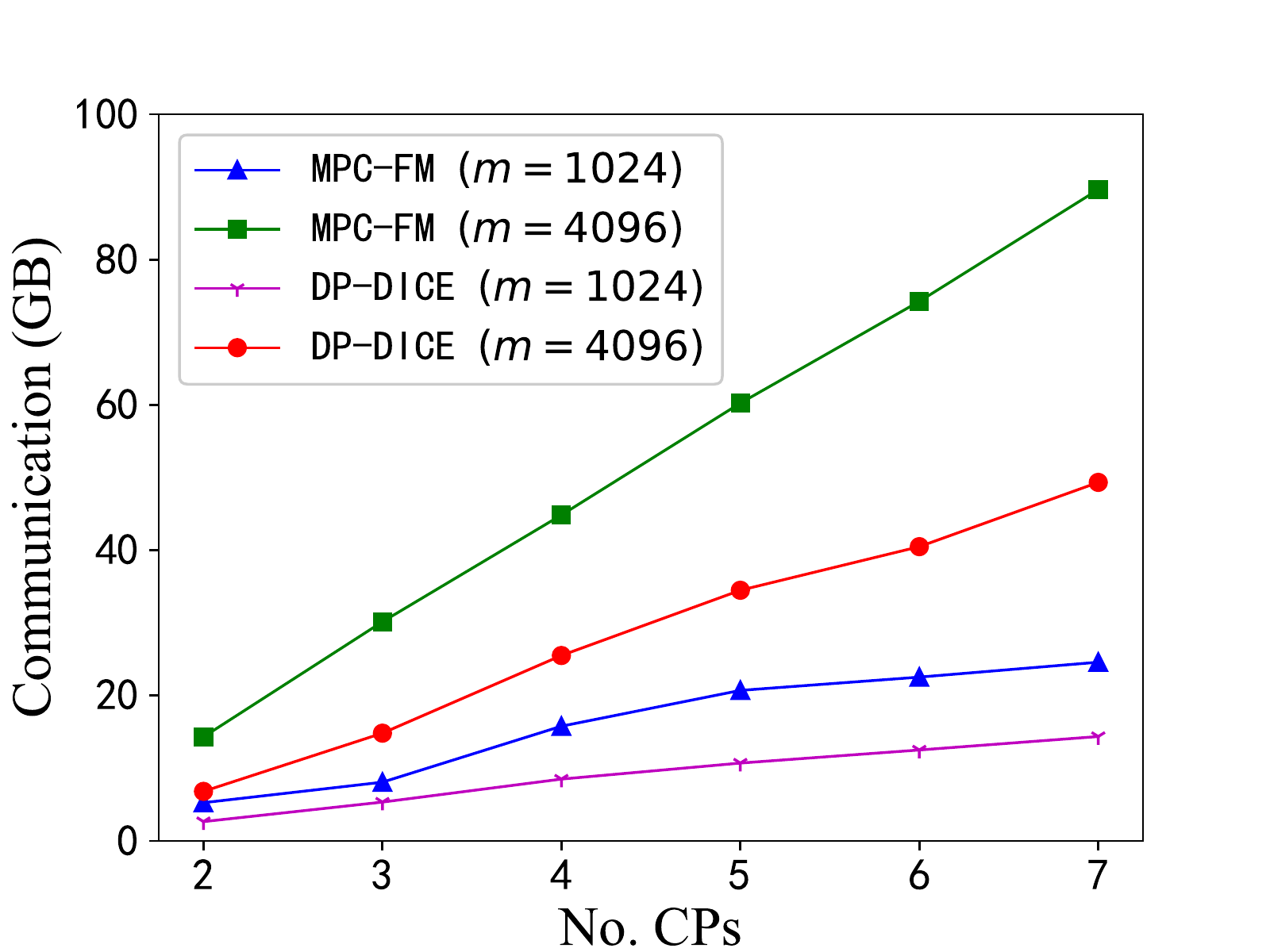}}
  \subfloat[online communication costs of different cardinalities]{%
    \includegraphics[width=.25\linewidth]{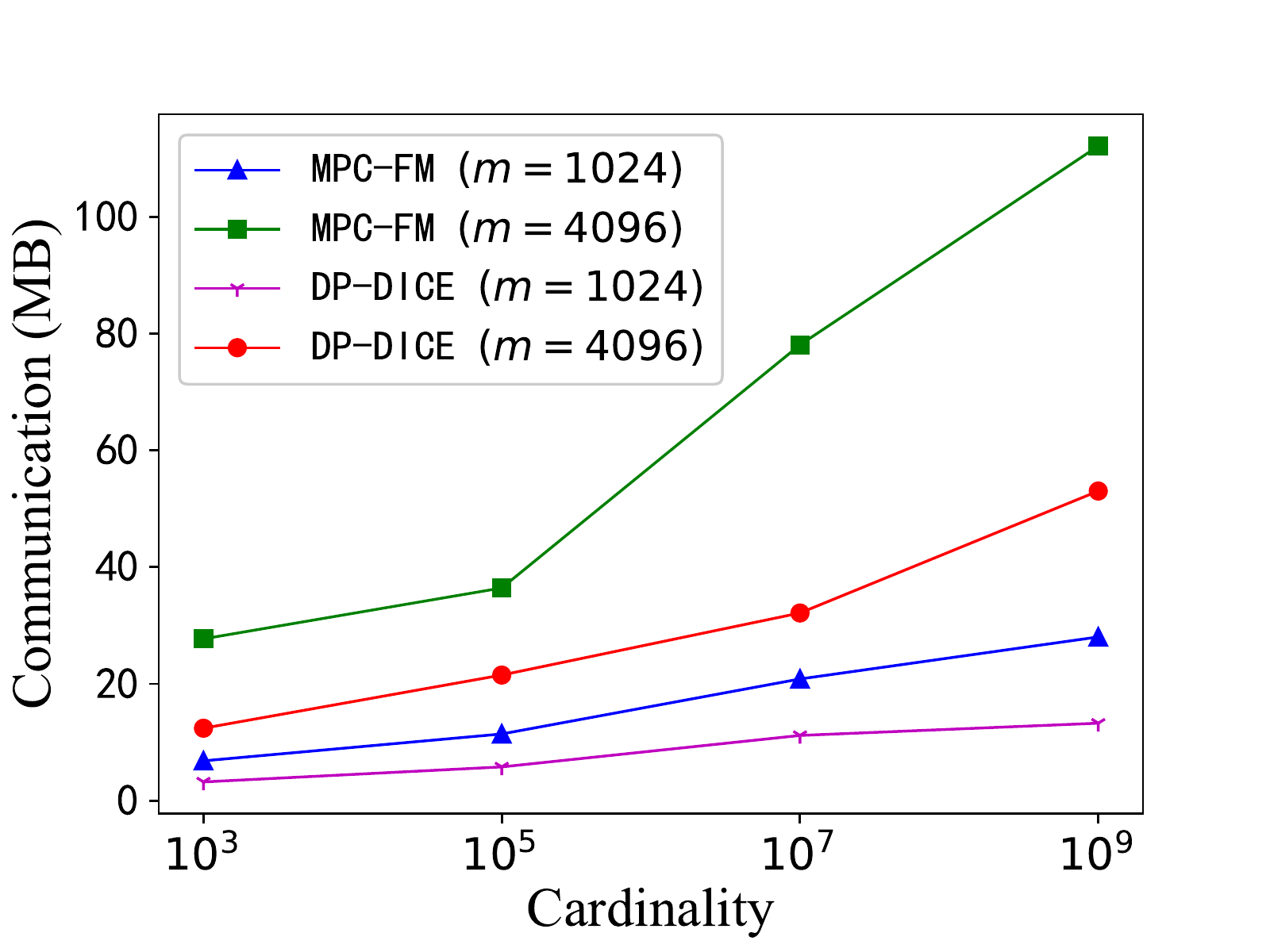}}
  \subfloat[online communication costs for different numbers of CPs]{%
    \includegraphics[width=.25\linewidth]{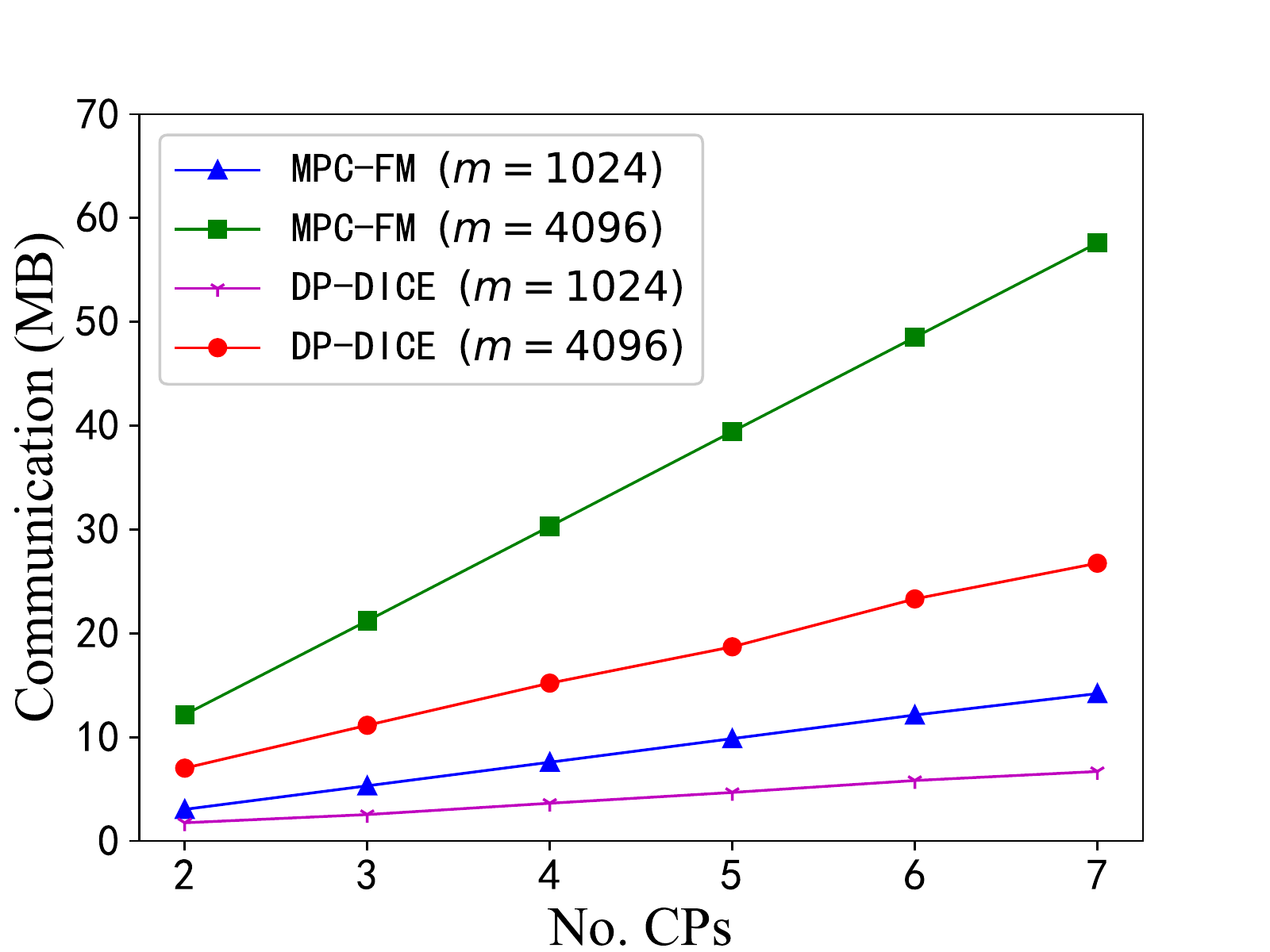}}\\
  \subfloat[(LAN) offline preparation time for different cardinalities]{%
    \includegraphics[width=.25\linewidth]{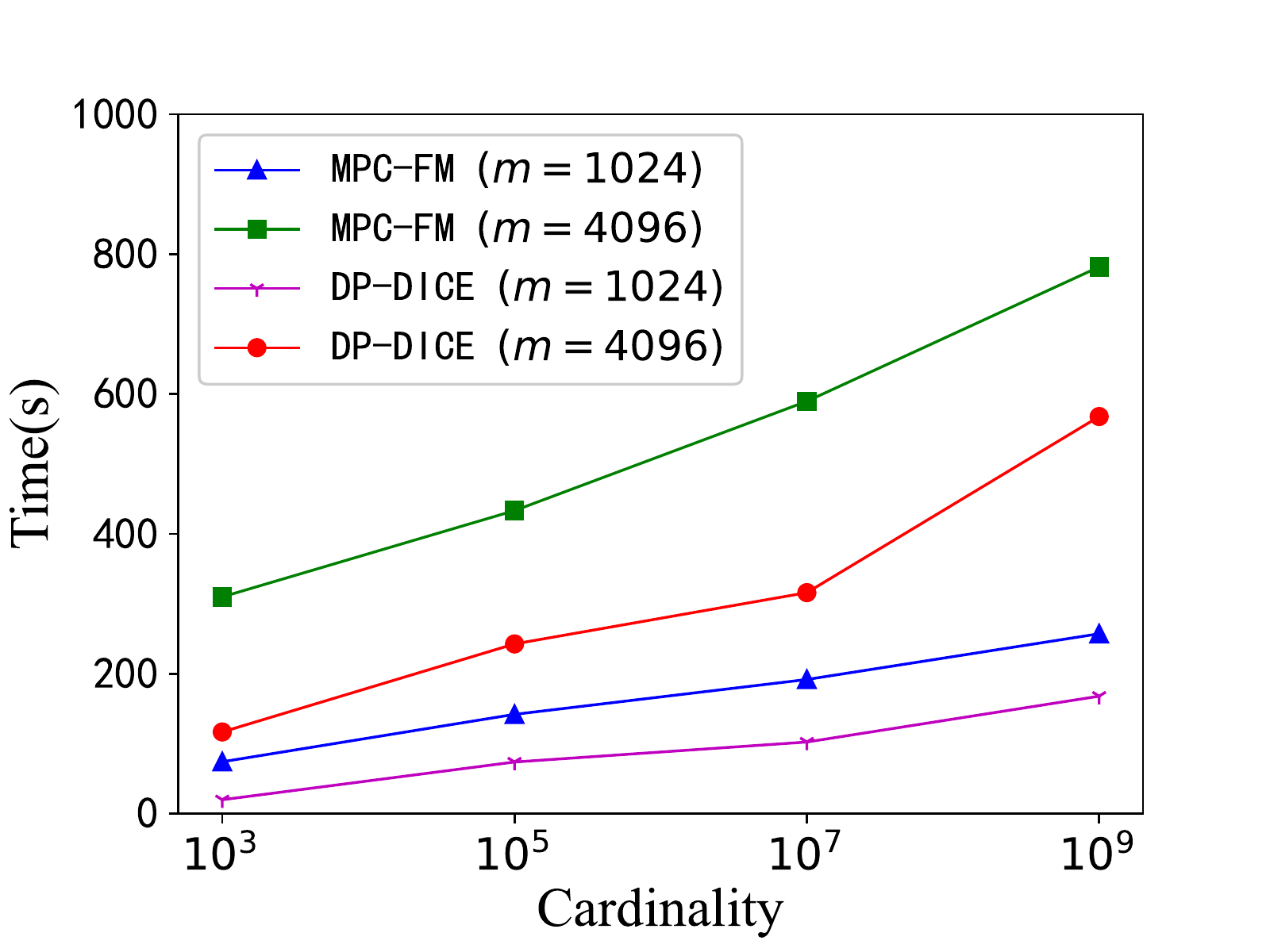}}
  \subfloat[(LAN) offline preparation time for different numbers of CPs]{%
    \includegraphics[width=.25\linewidth]{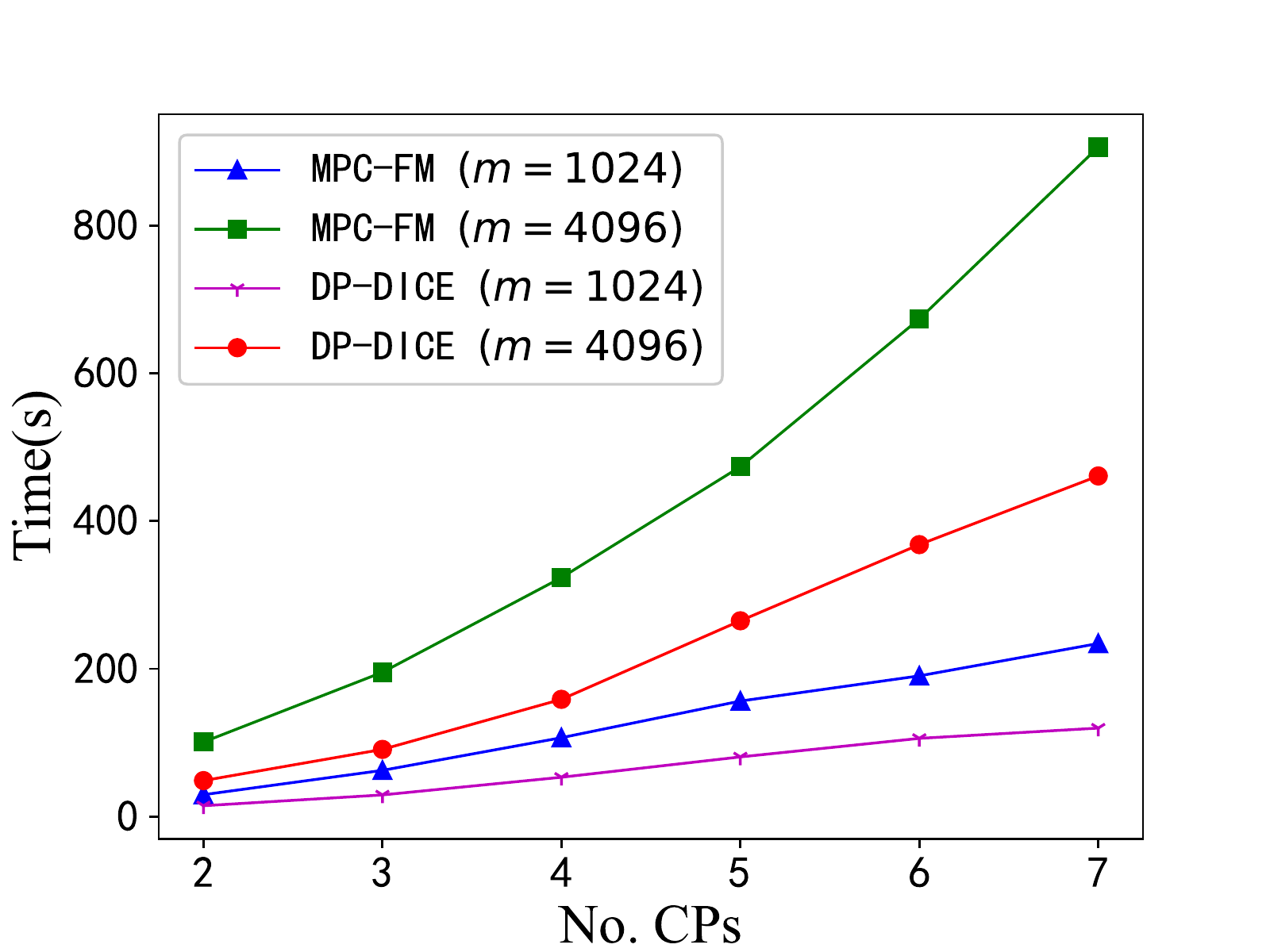}}
  \subfloat[(LAN) online running time for different cardinalities]{%
    \includegraphics[width=.25\linewidth]{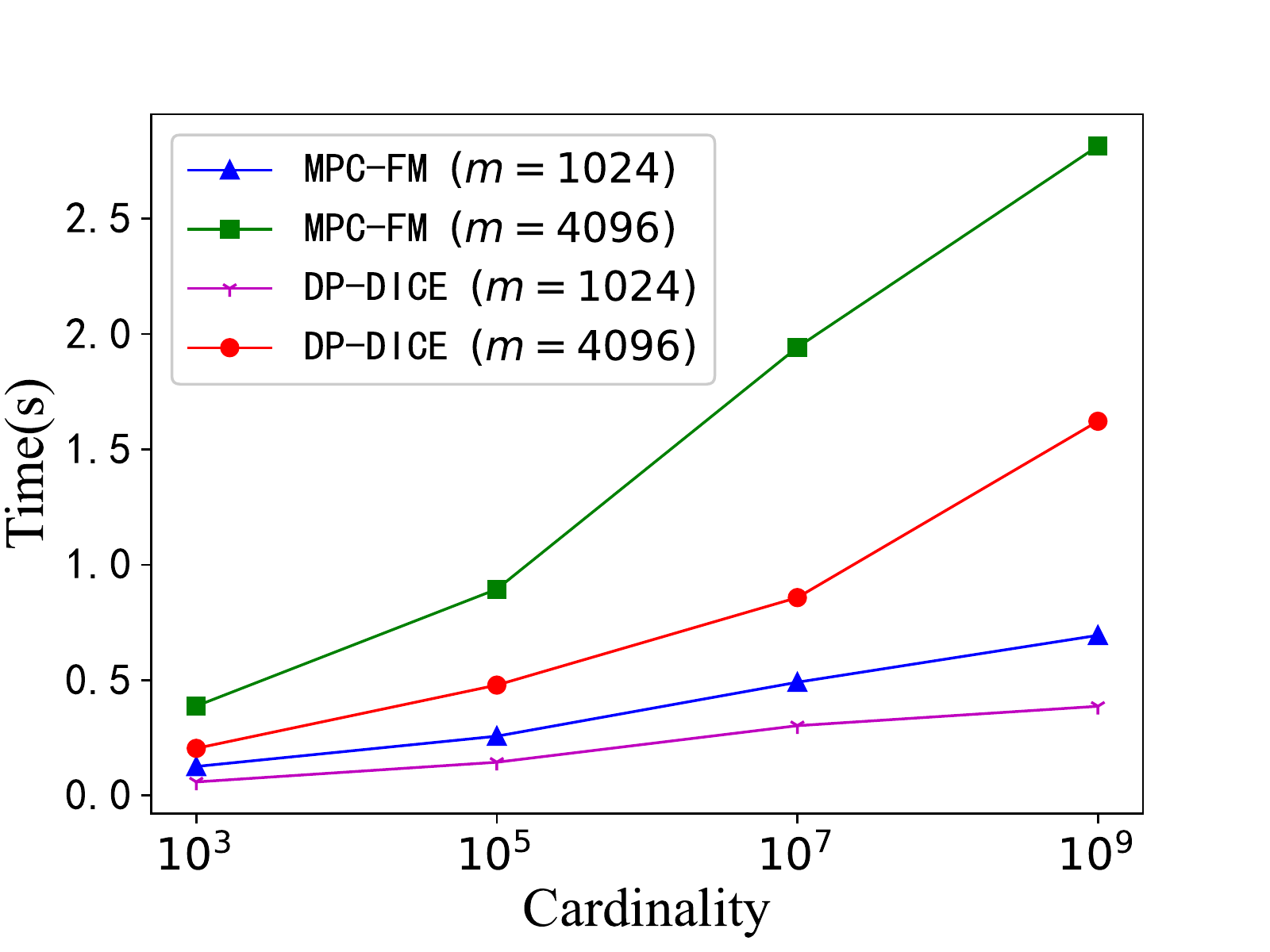}}
  \subfloat[(LAN) online running time for different numbers of CPs]{%
    \includegraphics[width=.25\linewidth]{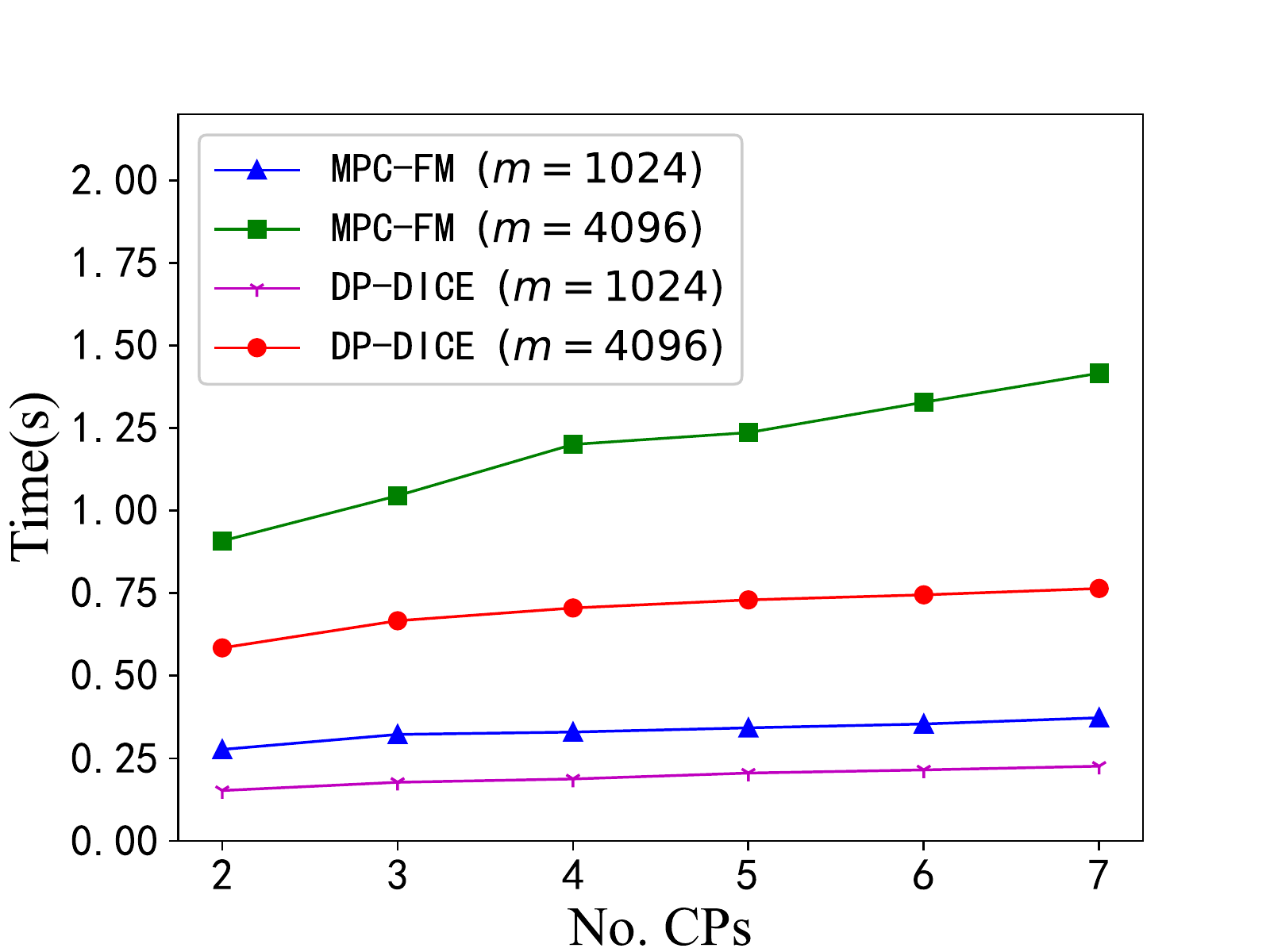}}\\
  \subfloat[(WAN) offline preparation time for different cardinalities]{%
    \includegraphics[width=.25\linewidth]{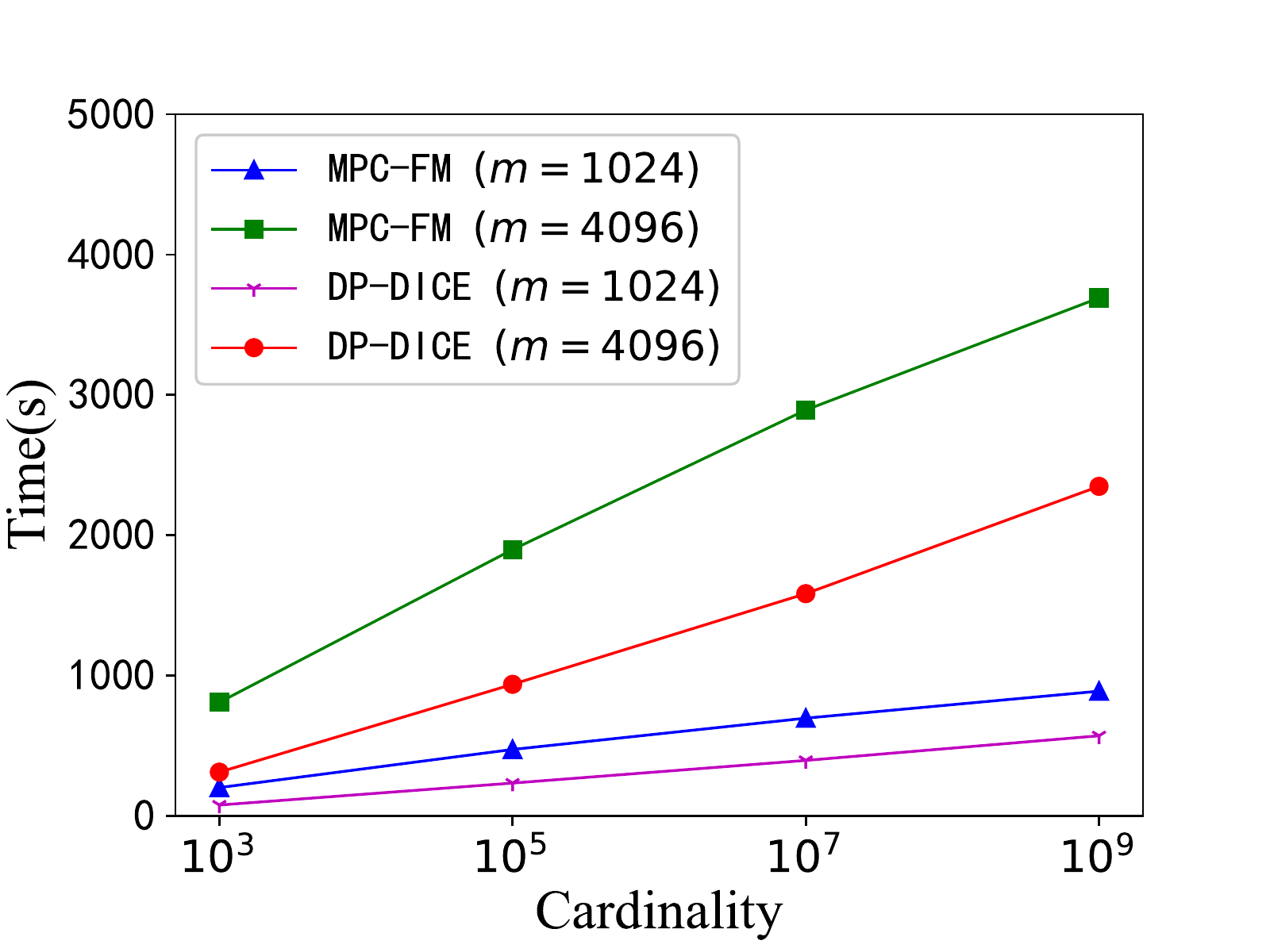}}
  \subfloat[(WAN) offline preparation time for different numbers of CPs]{%
    \includegraphics[width=.25\linewidth]{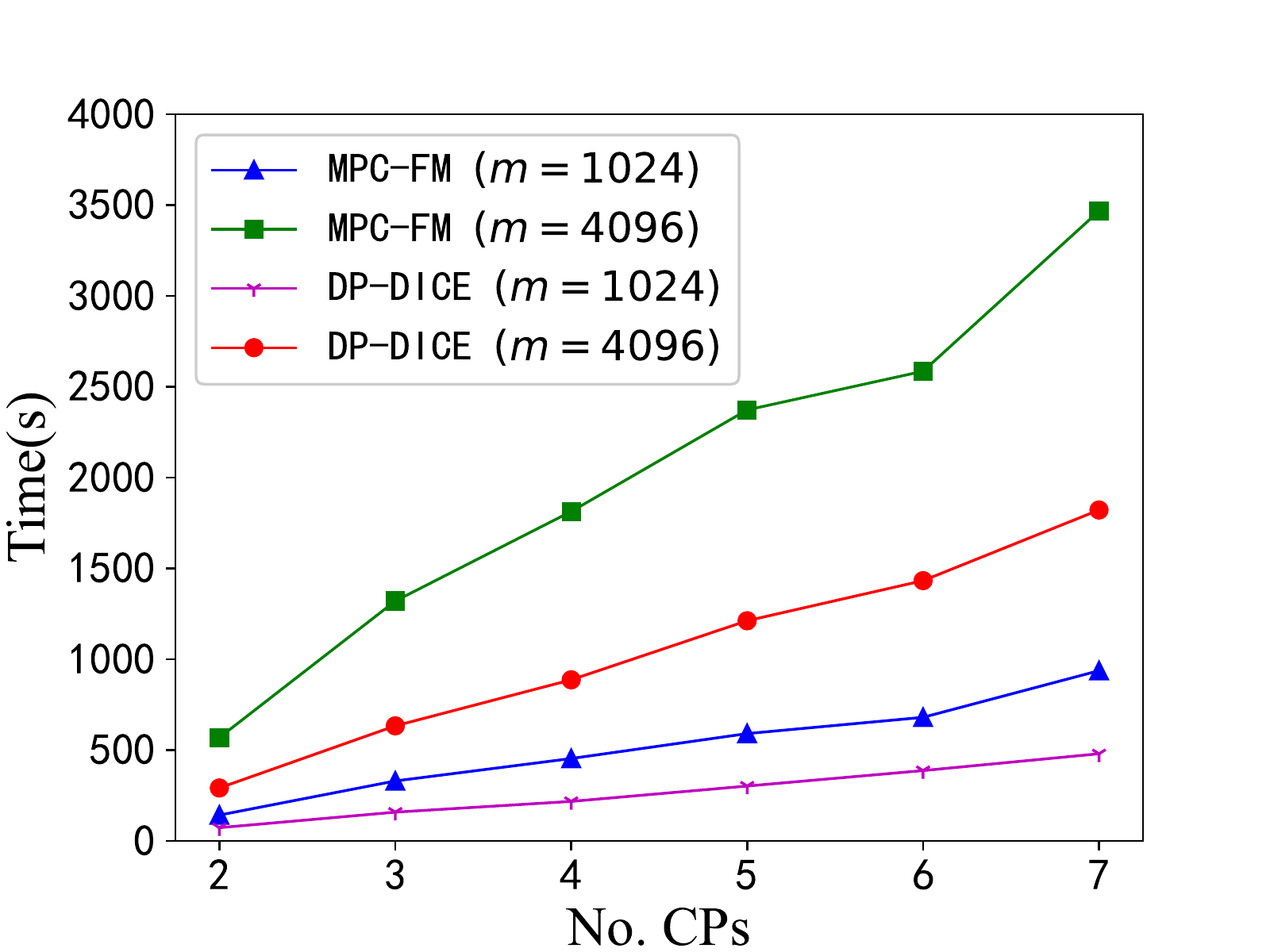}}
  \subfloat[(WAN) online running time for different cardinalities]{%
    \includegraphics[width=.25\linewidth]{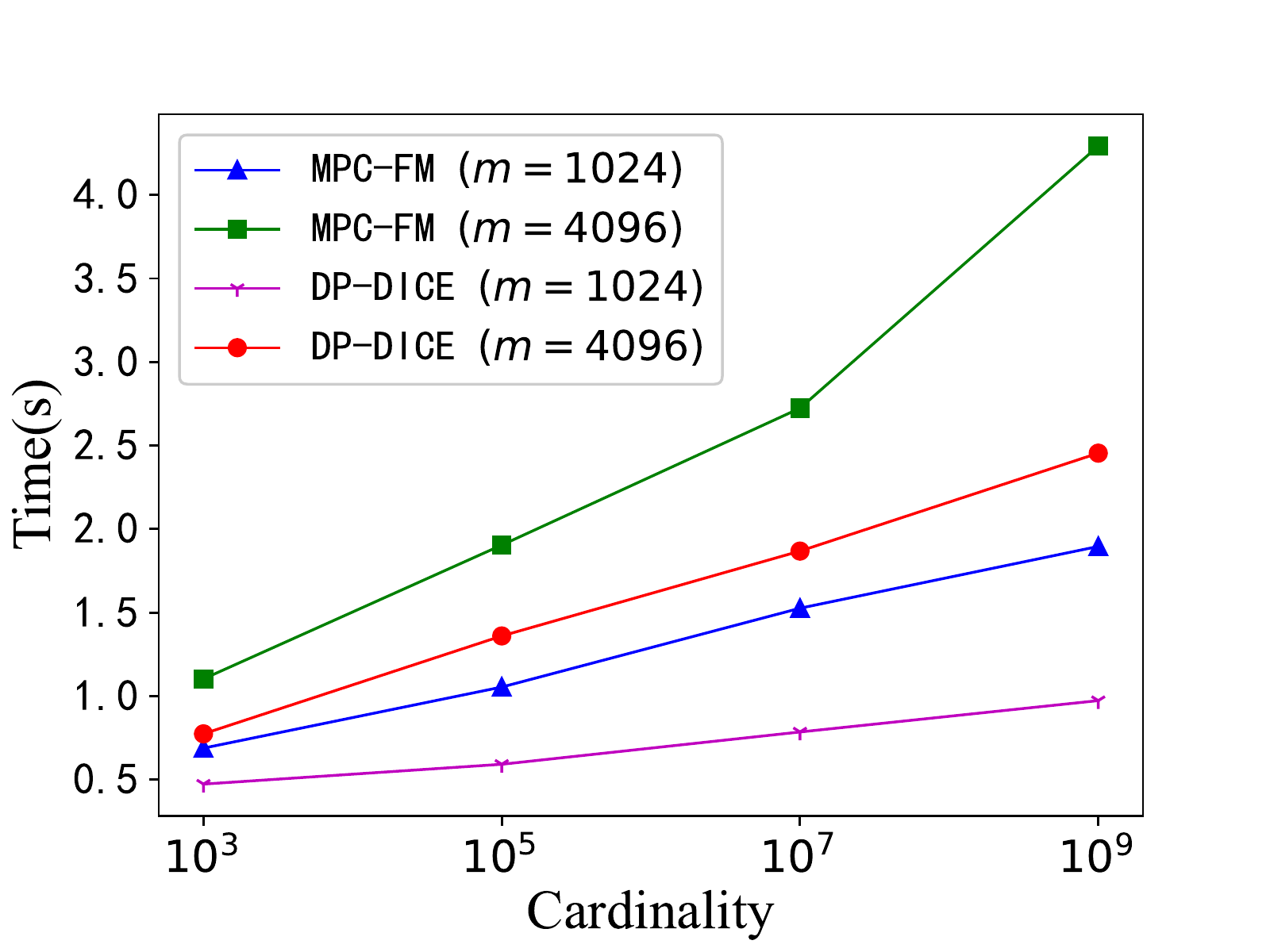}}
  \subfloat[(WAN) online running time for different numbers of CPs]{%
    \includegraphics[width=.25\linewidth]{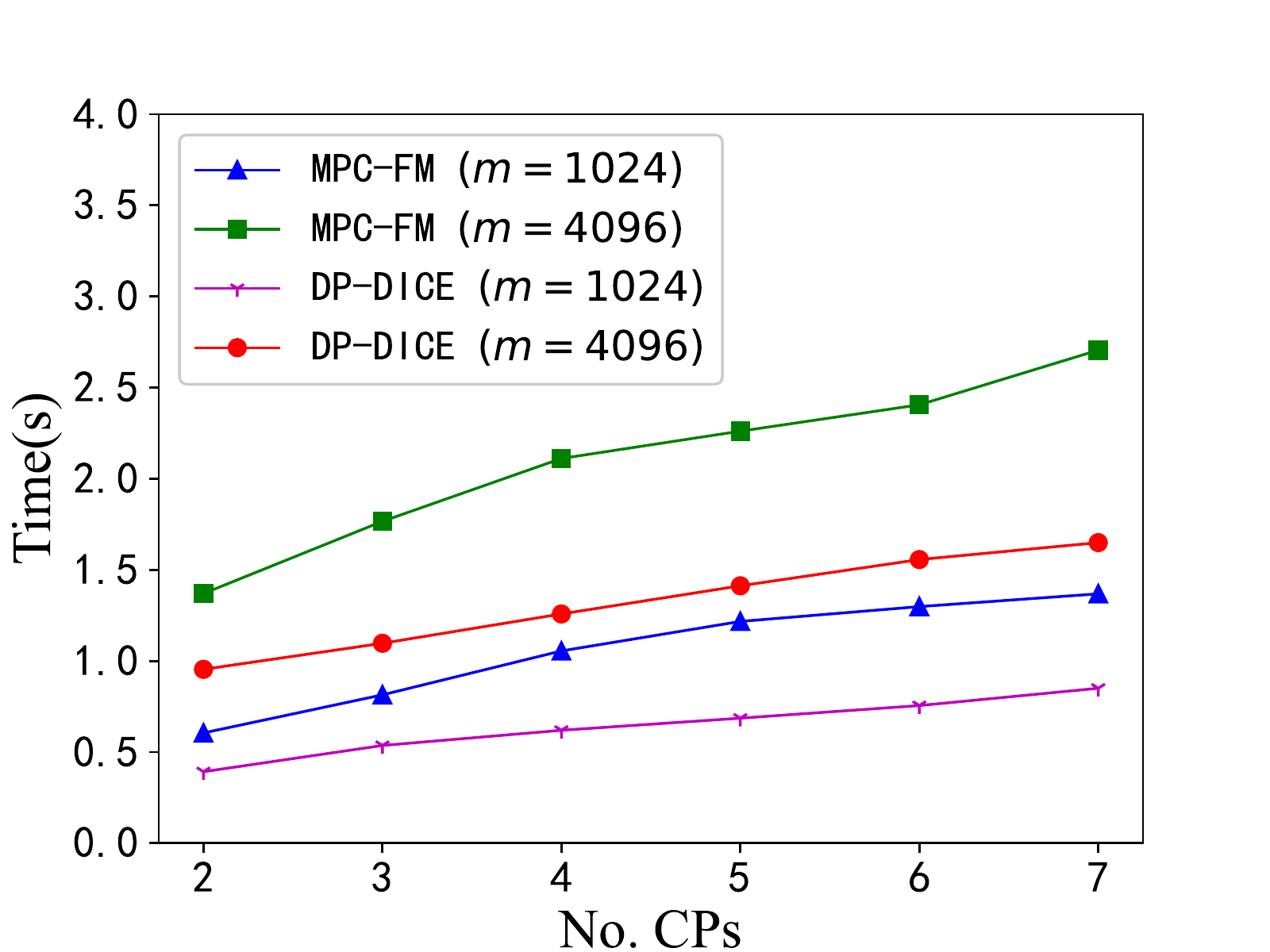}}     
    
\caption{Communication cost, offline preparation time, and online running time of our DP-DICE compared with
  MPC-FM.}
  \label{fig:efficiency}
\end{figure*}

\header{Running Time and Communication Cost}.
We compare the performance of our DP-DICE and MPC-FM under different settings $n\in\{10^3, 10^5, 10^7, 10^9\}$, $m\in \{1024, 4096\}$, and $c\in \{2,3, \ldots, 7\}$.
During the data collection phase,
each DP first computes the sketch of its holding set,
which is computationally extensive when the set is large.
In Fig.~\ref{fig:sketchgeneration}, when $m=1024$, we see that our DP-DICE reduces the sketch generation time of MPC-FM by 465.12, 536.13, 457.23, and 526.44 times for $n=10^3$, $10^5$, $10^7$, $10^9$, respectively. When $m=4096$, the advantage of DP-DICE is more significant.
From Fig.~\ref{fig:efficiency}(a)-(d),
we see that our DP-DICE reduces the online communication cost of MPC-FM by 1.7 to 2.4 times and reduces the offline communication cost by 1.4 to 3.7 times.

We evaluate the offline preparation time and online running time of DP-DICE and MPC-FM in two different network environments: LAN and WAN.
The results in the LAN environment are shown in Fig.~\ref{fig:efficiency}(e)-(h),
and the results in the WAN environment are shown in Fig.~\ref{fig:efficiency}(e)-(h).
In the LAN environment,
we see that our DP-DICE decreases the online running time by 1.6 to 2.3 times and the offline preparation time by 1.6 to 3.7 times.
In the WAN environment,
we see that our DP-DICE decreases the online running time by 1.4 to 2.0 times and the offline preparation time by 1.6 to 2.6 times.

\subsubsection{DP-DICE vs. PSC} 
Different from MPC-FM and DP-DICE,
the released code of PSC~\cite{fenske2017distributed} is in Go and
does not fully support multi-threading.
For a fair comparison,
Hu et al.~\cite{Hu0LGWGLD21} re-implemented the protocol of PSC in C++ and revealed that the performance
of the new implementation is much better than that reported in the original paper of PSC. 
While this C++ implementation is not publicly available,
we reuse the results of this advanced implementation of PSC given by Hu et al.~\cite{Hu0LGWGLD21}.
For cardinalities $n$ varying in $\{20,000, 30,000, 40,000, 50,000\}$,
the AARE of PSC is about 0.05 and 0.04 for $\epsilon=0.1$ and $\epsilon=0.3$, respectively (results from Fig.10 in the full version of MPC-FM\footnote{https://changyudong.site/assets/pdf/USENIX\_21\_FULL.pdf}).
For $n\in \{20,000, 30,000, 40,000, 50,000\}$,
the AAREs of our DP-DICE vary from 0.0079 to 0.0097 when $\epsilon=0.1$ and from 0.0064 to 0.0090 $\epsilon=0.3$.
We see that DP-DICE is several times more accurate than PSC.
Hu et al. demonstrate that MPC-FM reduces the online running time of PSC by orders of magnitude.
From the results in~\cref{fig:efficiency}, we see that DP-DICE is faster than MPC-FM, therefore it is also significantly more efficient than PSC.

\subsubsection{DP-DICE vs. MPC Sort\&Compare} 

A direct way to calculate the number of distinct elements under MPC is to sort the elements and  de-duplicate them. 
We compare this MPC Sort\&Compare method with our DP-DICE in the same experimental environment.
We evaluate the efficiency of the MPC-Sorting protocol for three CPs.
When there exist $n=10^2$ distinct elements, the MPC-Sorting protocol requires 561 seconds, which is 9 times slower than our DP-DICE.
When $n=10^5$, it takes about 20 hours, which is 468 times slower than our DP-DICE.
We easily see that the execution time of the MPC-Sorting protocol increases significantly with $n$.
The MPC-Sorting protocol takes more than a week when $n\ge 10^6$.
The above experiments demonstrate that our DP-DICE protocol is  significantly more efficient than the MPC Sort\&Compare protocol.

\pdfoutput=1
\section{Related Work}
\label{sec:related}
There are two different kinds of PDCE protocols: DH-PDCE and CP-PDCE.
For DH-PDCE protocols, the DHs collect data and also act as the CPs to participate in the computation.
Unlike DH-PDCE, the DHs in CP-PDCE protocols are only responsible for collecting data while the CPs compute the estimation.

\textbf{DH-PDCE Protocols.} 
To the best of our knowledge, Private Set Union Cardinality (PSU-CA)~\cite{CristofaroGT12} is the first DH-PDCE protocol.
Based on the principle of inclusion-exclusion, 
it reduces the original PDCE problem to the private computation of a series of set intersections' cardinalities,
which can be solved by well-studied Private Set Intersection (PSI) protocols.
The output of PSU-CA is the exact cardinality, which is not necessary for many real-world applications.
Davidson and Cid~\cite{DavidsonC17} exploit a membership query sketch method \emph{Bloom Filter} to privately estimate the cardinality of set intersections, which significantly accelerates the speed of PSU-CA.
However, the principle of inclusion-exclusion involves a number of summands that is exponential in the number of DHs, which is still computationally intensive when there exist many DHs.
To address this issue, ~\cite{TschorschS13,DongL17} propose protocols based on the FM sketch~\cite{flajolet1985probabilistic}, which is a well-known cardinality estimation method.
However, none of the above protocols are differential privacy. 
Recently, Chen et al.~\cite{ChenG0M21} proposed a DH-PDCE protocol for a particular scenario where each DH holds a single element,
Jia et al.~\cite{JiaSZDG22} propose a protocol to privately compute two sets' union, which is different from our problem.

\textbf{CP-PDCE Protocols.}
Ashok and Mukkamala~\cite{AshokM14} proposed a CP-PDCE protocol based on the Bloom Filter.
Egert et al.~\cite{EgertFGJST15} reveal that this protocol is not secure, and give a more secure variant of the protocol.
Unfortunately, both protocols do not support differential privacy.
For the protocol in~\cite{StanojevicNY17}, each of the DHs represents its local set as a bit vector, disturbs the vector using the randomized response method to achieve differential privacy, and then sends the vector to the CPs for estimating the union set's cardinality.
However, the protocol has large estimation errors (in the order
of the universal set's size), which does not meet the requirement of many applications. In ~\cite{choi2020differentially}, the authors use the LogLog sketch and the Laplace mechanism to estimate cardinality. However, it also exhibits large estimation errors for the LogLog algorithm holds high sensitivity and would need a large noise to achieve the requirement of differential privacy.
The protocol of PSC~\cite{fenske2017distributed,fenske2022accountable} maintains a hash table with a public
hash function for each DP.
For an element holding by multiple DHs,
PSC hashes it into the same bin of the hash table,
which facilitates eliminating duplicates.
To achieve reasonable accuracy, 
PSC needs to set the hash table's size much larger than the union set's cardinality,
which significantly limits its efficiency and scalability.
Hu et al.~\cite{Hu0LGWGLD21} proposed a CP-PDCE protocol that implements secure computation based on the FM sketch. They exploit the uncertainty imposed by the intrinsic estimation variance of the FM sketch to produce differentially private outputs~\cite{smith2020flajolet}. 
However, the protocol is not differential private (refer to~\cref{subsec:challenges} for details) when the DHs are honest but curious.
Ghazi et al.~\cite{GhaziKKMPSWW22} proposed to estimate both the union set's cardinality and the frequency histogram (i.e., the fraction of elements appearing a given number of times across all the DHs).
Their protocol is based on a variant of Bloom Filter and consists of multiple phases with noise injected to guarantee differential privacy.
The protocol is complex and involves many parameters to be set, which is not easy to be implemented.

\pdfoutput=1
\section{Conclusion and Future Work} \label{sec:conclusions}
In this paper, we propose a secure and efficient protocol DP-DICE for solving the problem of PDCE.
Experimental results demonstrate the efficiency and efficacy of our DP-DICE. Compared with the protocol MPC-FM, our DP-DICE speeds up the sketch generation by thousand times and the online running time by about 2 times and reduces the estimation error by more than a thousand times.
Compared with the protocol PSC, DP-DICE speeds up the online running time by orders of magnitude and reduces the estimation error by several times.
Similar to the recent proposed federated learning with secure aggregation protocol in~\cite{KairouzL021}, 
we notice that our DP-DICE also has the potential modular wrap-around when aggregating distributed discrete Gaussian noise variables on the SPDZ framework, which does not compromise security and precision but accuracy.
Fortunately, this modular clipping happens with a very small chance because a very large modulus $p$ is typically used in our protocol.
In the future, we plan to completely address this issue and extend our protocol to handle more complicated cases.
For example, some applications may have a few malicious DHs. In addition, the MPC framework used by our DP-DICE is SPDZ, which is not robust. In other words, the DP-DICE protocol will terminate and return no result if any CP aborts. Robust MPC is an active research area and we will migrate our DP-DICE protocol to a robust MPC framework when it is ready for use.


%

\bibliographystyle{unsrt}

\end{document}